\documentclass[useAMS,usenatbib,aps,floatdix,nofootinbib]{mn2e}

\usepackage{graphics} 
\usepackage{amssymb,amsmath}
\usepackage{amsmath}
\usepackage{psfrag}
\usepackage{graphicx}

\def\ep{\epsilon}

\def\si{\sigma}

\def\be{\begin{equation}} 
\def\ee{\end{equation}} 
\def\bea{\begin{eqnarray}} 
\def\eea{\end{eqnarray}}  
\def\bean{\begin{eqnarray*}} 
\def\eean{\end{eqnarray*}}

\def\bk{{\bf k}}  
\def\bx{{\bf x}}  

\def\br{{\bf r}}  
\def\ba{{\bf a}}

\def\bn{{\bf n}} 
\def\bu{{\bf u}}
\def\bR{{\bf R}}

\def\bK{{\bf K}}

\def\tbu{{\tilde\bu}}

\def\bq{{\mathbf q}}
\def\bse{\begin{subequations}}
\def\ese{\end{subequations}}

\def\da{{\dot a}}

\def\bF{{\mathbf F}}

\def\bq{\mathbf{q}}

\def\mP{\mathcal{P}}

\def\lsim{\raise 0.4ex\hbox{$<$}\kern -0.8em\lower 0.62ex\hbox{$\sim$}} 
\def\gsim{\raise 0.4ex\hbox{$>$}\kern -0.7em\lower 0.62ex\hbox{$\sim$}} 
\def\mP{\mathcal{P}}
\def\mQ{\mathcal{Q}}

\def\bk{\mathbf{k}}

\def\f0N{f_0^{(N)}}
\def\bec{\begin{center}}
\def\eec{\end{center}}

\title[Discreteness error in the non-linear regime of cosmological N body simulations]
{Towards quantitative control on discreteness error in the non-linear regime of cosmological N body simulations}
\author[M. Joyce, B. Marcos and T. Baertschiger]
{M. Joyce$^1$, B. Marcos${^{2,}}{^3}$ and T. Baertschiger$^4$\\
$^1$Laboratoire de Physique Nucl\'eaire et de Hautes Energies,  
UMR 7585 \\ Universit\'e Pierre et Marie Curie --- Paris 6, 
75252 Paris Cedex 05, France\\
$^{2}$``E. Fermi'' Center, Via Panisperna 89 A, Compendio del 
Viminale, I-00184 Rome, Italy\\
$^3$ISC-CNR, Via dei Taurini 19, I-00185 Rome, Italy\\
$^4$Dipartimento di Fisica, 
Universit\`a ``La Sapienza'',
P.le A. Moro 2,
I-00185 Rome,
Italy}

\begin{document} 

\date{\today}

\maketitle

\begin{abstract}   
  
The effects of discreteness arising from the use of the $N$-body 
method on the accuracy of simulations of cosmological structure 
formation are not currently well understood. In the first part 
of this paper we discuss the essential question
of how the relevant parameters introduced by this discretisation
should be extrapolated in convergence studies if the goal is to
recover the Vlasov-Poisson limit. 
In the second part of the paper we study numerically, and
with analytical methods we have developed recently, the central 
issue of how finite particle density affects the precision of 
results above the force smoothing scale. In particular we focus
on the precision of results for the power spectrum at wavenumbers
around and above the Nyquist wavenumber, in simulations in which 
the force resolution is taken smaller than the initial
interparticle spacing. Using simulations of identical theoretical 
initial conditions sampled on four different  ``pre-initial''  
configurations (three different Bravais lattices, and a glass)
we obtain a {\it lower bound} on the real discreteness error.
With the guidance of our analytical results, which match 
extremely well this measured dispersion into the weakly 
non-linear regime, and of further controlled tests for 
dependences on the relevant discreteness parameters,
we establish with confidence that the measured dispersion
is not contaminated either by finite box size effects or
by subtle numerical effects. Our results show notably
that, at wavenumbers {\it below} the Nyquist wavenumber,
the dispersion increases monotonically in time throughout 
the simulation, while the same is true above the Nyquist
wavenumber once non-linearity sets in. For normalizations 
typical of cosmological simulations, we find
lower bounds on errors at the Nyquist wavenumber of order of a percent,
and larger above this scale. Our main conclusion is that the 
only way this error may be reduced below these levels at 
these physical scales, and indeed convergence to the physical 
limit firmly established, is by extrapolation, at fixed values of the 
other relevant parameters, to the regime in which the mean comoving 
interparticle distance becomes less than the force smoothing scale. 
\end{abstract}    


\section{introduction}
Dissipationless cosmological $N$-body simulations aim to reproduce the
clustering of dark matter in the universe, assumed to be in the form of
a microscopic particle with extremely weak non-gravitational
interactions (for reviews 
see e.g. \cite{bertschinger_98, bagla_review, dolag_etal_2008}). 
In the absence of an analytical treatment of the strongly non-linear 
regime, these
simulations have become increasingly central in extrapolating 
the predictions of the current ``standard model'' of cosmology 
to the corresponding scales. Many kinds of observations 
now probe directly or indirectly
the distribution of dark matter at these scales, and will do so
with greater precision in the coming years. The resultant need for
precision in the theoretical results 
makes more necessary than ever a better understanding of 
these simulations. This paper concerns one potentially important
source of error which is currently still poorly understood: rather 
than evolving numerically the theoretical Vlasov-Poisson equations 
describing the self-gravitating dark matter, simulations
employ the $N$-body method in which the matter is sampled
by ``macro-particles''. 
The errors introduced, i.e., the
difference between the results of the finite $N$ simulation and
those in the theoretical model (which corresponds to an appropriate
$N \rightarrow \infty$ limit), are not understood. This is the
discreteness problem in cosmological $N$-body simulation. 
It is a problem which has received, given its potential
importance, a very modest amount of attention (see references below). 
Further the existing literature on the issue is marked by a 
considerable diversity in its conclusions, both 
qualitatively {\it and} quantitatively. Given the 
ever more pressing need for robust control on the very considerable 
precision required of simulations --- a goal of one percent 
precision is now typically 
considered (see e.g. \cite{Huterer:2004tr, mcdonald05}) 
---  it is an issue which deserves attention.

In this paper we first give a brief review of the problem of
discreteness in cosmological $N$-body simulation. We both
describe briefly previous work by other authors on the issue, as well
as some recent work by ourselves and our collaborators, in which
we have developed new analytical approaches to describe discreteness
effects both in the initial conditions of simulations and in their
early time evolution. As a starting point we attempt here to give a 
precise explicit formulation of the problem of discreteness. This 
distinguishes 
notably the problem from strictly numerical issues (e.g. about 
the agreement of codes using different summation techniques).
We emphasize in particular the necessity to establish, before
any discussion about the quantification of errors,  precisely how 
numerical simulations should be extrapolated to approach 
the desired theoretical limit. Our conclusion is simply that 
an appropriate such extrapolation is one which takes 
the interparticle spacing $\ell$ to zero, {\it at fixed values 
of the other relevant discreteness parameters}. Further
such extrapolation should be done keeping fixed the initial 
conditions, which --- given that simulations are performed in 
a finite periodic system --- means using the same realization
(and modes) of the initial theoretical power spectrum.

After this introductory discussion we turn to a numerical and
analytical study of the issue. The goal of this study is to answer the
more practical question of {\it how small} the interparticle distance
$\ell$ must be to attain convergence of physically relevant quantities
to a desired precision. In particular we focus on the issue, which is
at the centre of some controversy in the literature, of the accuracy
of results, at scales around and below the initial
interparticle distance, of simulations which use a force smoothing 
scale smaller than this latter scale. To attempt to resolve the 
question we use here a
simple numerical method to isolate errors which manifestly must arise
from discreteness, and which therefore give a lower bound on the
discreteness errors. We focus here on the two point correlation
properties of clustering, but the method we use can be extended to any
other quantity (e.g. mass function, merger rates). The essential 
difference between our study and the few
previous such attempts of this kind (see references below) is that we
use, as mentioned above, also an analytical formalism which describes
fully the measured discreteness effects at sufficiently early times.
This allows us to ``calibrate'' the errors, in the sense that it
allows us to establish without doubt that the measured quantities
arise from physical discreteness effects, and not from the other
possible sources of dispersion in our results (poor numerical
convergence, or finite size effects). We then study the further
evolution of these errors into the non-linear regime, allowing us to
place with confidence a lower bound on the true discreteness error in
this regime.

More specifically our tests, and principal conclusions,
are as follows. We consider
$N$-body simulations in an EdS universe of identical 
theoretical initial conditions, given by a random Gaussian 
realization of a power spectrum $P(k) \propto k^{-2}$.
In our central test the simulations differ only in the
choice of ``pre-initial'' configuration, i.e.,  
the point distribution chosen to represent the uniform universe prior 
to the application of the perturbations corresponding to the 
given theoretical  initial conditions. The canonical choices 
in the literature are ``grid'' (a simple cubic lattice), or ``glass''
\citep{white_leshouches}. Here we consider a wider class of such configurations, 
employing also two different Bravais lattice configurations 
(body-centred cubic and face-centred cubic). The reason for our
choice of these configurations is that they allow us to apply
in a very powerful way our analytical treatment. This formalism 
gives a very accurate description for the early time 
evolution of simulations starting, in principle, from any perturbed 
Bravais lattice. The differences in the evolved power spectra, starting from
the same realization of the theoretical model discretised on 
these different distributions, measured in our simulations are 
in extremely good agreement with these analytical predictions 
for {\it all wavenumbers} at 
early times, and progressively deteriorates, as anticipated,
as we go into the strongly non-linear regime. In this latter
regime we observe that these differences show similar 
dependences on the discreteness parameters as in the 
regime where we can fit them analytically. Notably at a given 
physical scale, they decrease as $\ell$ decreases, and
increase monotonically as a function of time at fixed
$\ell$. 

These tests give us a robust non-trivial lower bound on the size 
of systematic discreteness errors. For the power spectrum, which is
the quantity we focus on, these lower bounds are of order 
a few percent for wavenumbers comparable to and larger than
the Nyquist frequency for a starting red-shift equal to $2^5$,
and then decrease monotonically at smaller wavenumbers. While
the precise bounds for any given cosmological model (and choice
of other relevant simulation parameters) will differ, they will be 
of this order (or larger, as the bounds monotonically increase
with the starting redshift). Our results allow us then 
to draw conclusions about the question of how far 
$\ell$ must be extrapolated to attain errors smaller than of this
order. Specifically we conclude that the common practice of using results 
considerably below the scale
$\ell$ (or $\pi/\ell$ in reciprocal space) {\it may} be justified,
but only with a discreteness error bar which is, for the
power spectrum, and for normalisations typical of cosmological 
simulations, of order of these lower bounds, i.e., several percent. 
Precision greater than this for a given wavenumber $k$, e.g., 
down to below the one percent level for the power spectrum now often 
cited as necessary, can be achieved only by using particle 
densities such that $k \ell <1$.
Indeed results of simulations do not converge
to the continuum limit until this parameter range is reached,
and thus one cannot have real confidence in results 
without performing such an extrapolation. While this conclusion
has been argued for in several studies by some authors
(see references and discussion below), it is a much more 
stringent requirement than that assumed in much of the 
literature, and formulates a considerable challenge to simulation.

\section{The problem of discreteness}
\label{The problem of discreteness}

The problem of discreteness in cosmological simulation arises from
the fact that the numerical simulations are not a direct discretisation
of the equations of motion of the theoretical model. 
The latter 
is (usually) 
assumed to be described, on the 
physical scales of relevance,  by Vlasov-Poisson (VP) equations 
(or ``collisionless Boltzmann equations'')  which give the 
evolution of the (smooth) phase space mass density.
%
%
%
$N$-body simulations, on the other hand, are numerical integrations
of the equations of motion of $N$ self-gravitating particles,
i.e., 
\be
\label{Nbody-eom}
\ddot\bx_i +2\frac{\da}{a} \dot \bx_i =
-\frac{Gm}{a^3}\sum_{j\ne i}\frac{\bx_i-\bx_j}{|\bx_i-\bx_j|^3} 
\, W_\varepsilon(|\bx_i-\bx_j|)
\ee
where dots denote derivatives with respect to time, $m$ is the mass of 
the particles, $\bx_i$ is the (comoving) position of the $i$-th particle, and 
$W_\varepsilon(|\bx_i-\bx_j|)$ is a function which regularizes the 
divergence in the gravitational force at $|\bx_i-\bx_j|=0$ below a
characteristic scale $\varepsilon$. 
These unphysical ``macro-particles'' are artefacts of the $N$-body 
simulation technique, with a mass many orders of magnitude (typically $\sim
10^{70}$) larger than those of the theoretical dark matter particles.

As the VP equations may, in principle, be obtained as an appropriate 
$N \rightarrow \infty$ limit of the particle system, the problem
of discreteness is in practice that of determining the discrepancy 
between the solution of the $N$-body equations for some finite
$N$ and their solution for a much much larger $N$, representative
of the VP limit. It is therefore 
evidently essential  to specify precisely how to extrapolate
cosmological $N$-body simulations to this limit. This is
the point we first discuss.  

\subsection{Discreteness parameters}

In the case of the $N$-body method
employed to solve the cosmological problem, the unphysical parameters characterising the numerical solution
can be clearly divided into two. Firstly there are those
required, {\it in addition to the parameters of the input 
theoretical model}, to characterise the equations 
(\ref{Nbody-eom}) and their initial conditions.
Secondly there are the parameters introduced to then solve these 
well posed equations numerically (e.g. time step, parameters 
controlling the precision of the calculation of the force).
It is only the former, which we will refer to as the 
{\it discreteness parameters} and denote by $\{ D_\alpha \}$,
which are the subject of study here. The latter set of 
parameters, which we will refer to as the {\it numerical parameters} 
of a simulation, control the accuracy with which the set of 
equations (\ref{Nbody-eom}), with well defined initial conditions, are 
solved. They therefore have no relevance to the problem of discreteness
which we are focussing on: we wish to understand
the relation between the results of a ``perfect'' $N$-body 
simulation, i.e., an arbitrarily precise numerical solution
of the equations (\ref{Nbody-eom}) from well specified initial
conditions, and the evolution of the
theoretical model from its corresponding 
initial conditions\footnote{The sensitivity 
of results to this second set is, of course, essential to understand 
in order to characterise 
the precision of results of simulations (e.g. using different codes),
and indeed considerable effort to improve control has been made
in the last few years (see e.g. \cite{heitmann07,lukic08}). We note that
the distinction we make here between the two kinds of parameters is not
usually made in the literature on the ``convergence'' of simulations 
(see e.g. \cite{power_03, lukic08}).}.  

The set of discreteness parameters $\{ D_\alpha \}$ 
we consider is the following:

\begin{itemize}

\item 1. The mass of the macro-particles (referred to as {\it mass
      resolution} in the literature), or equivalently (since the mean
      mass density is specified) their mean (comoving) number density
      $n_0$. We will parametrize this by $\ell \equiv n_0^{-1/3}$, which we
      refer to as the {\it mean interparticle spacing}.

\item 2. The {\it smoothing parameter} $\varepsilon$ characterising the
       regularisation of the force (known as the {\it force resolution} 
       in the literature\footnote{It is often referred to simply as the
``spatial resolution''. We will not use this nomenclature here as the
central issue we discuss is whether such an identification of the 
force smoothing scale with that of the spatial resolution is
valid.}). 

\item 3. The {\it pre-initial configuration}, i.e., choice of grid, glass, or
      other distribution. We will denote this discrete variable $preIC$.

\item 4. The {\it initial red-shift}, $z_i$.

\end{itemize}

Some remarks on this list are appropriate:
 
\begin{itemize}

\item  Discreteness effects  depend 
on the number of particles $N$ used in a simulation only through the
{\it density} of these particles. Change in results when $N$ varies, at 
fixed $\ell$, is {\it not} a discreteness effect: to pass 
from the VP equations to the $N$-body equations 
(\ref{Nbody-eom}), and their initial conditions,
we do not need to introduce the side $L$ of the cubic box 
(which, with periodic replicas, is canonically used to approximate the 
infinite universe) as both sets of equations are well defined in 
infinite space. The box size $L$ thus belongs to the second set 
of parameters, as it is introduced to solve the Eqs.~(\ref{Nbody-eom})
in a (finite) numerical simulation. The dependences on it, i.e.
on the variation of $N$ at fixed particle density, are 
{\it finite-size effects}. We do not study these effects 
here, and will always work at fixed $L$ in our numerical
study below. For studies of them see e.g. 
\cite{pen, sirko, bagla_finitesize06, bagla_finitesize08}.

\item  The smoothing parameter $\varepsilon$, on the other hand, cannot,
       in modern cosmological simulation, be considered as belonging to
       the numerical parameters: it is not, in this context, a parameter introduced
       to facilitate the numerical solution\footnote{Indeed the 
       equations (\ref{Nbody-eom}) may be solved 
       numerically without any such smoothing (and often are in
       other contexts e.g. galactic dynamics).}. Rather, as we will 
       discuss further below, it is used with the aim of reducing
       effects 
       of two-body collisionality, i.e., to try to make the $N$-body 
       solution approach better the theoretical collisionless behaviour
       corresponding to the VP equations. 

\item  That the initial red-shift $z_i$ is a discreteness parameter in
the sense we have defined above has been shown explicitly in 
\cite{joyce_05, marcos_06, discreteness2_mjbm} (and summarised also
briefly in Sect.~\ref{PLT-summary} below). Put simply, the 
treatment of the evolution of Eqs.~(\ref{Nbody-eom}) in this work
shows, analytically, that the initial conditions for the $N$-body 
system derived for a given input power spectrum (PS)  
at a redshift $z_1$ (using the canonical method
based on the Zeldovich approximation), do not evolve exactly under 
Eqs.~(\ref{Nbody-eom}) to those set up from the same PS at
a different red-shift $z_2$. This is true  {\it in the limit of 
arbitrarily small
initial relative displacements to the lattice} where non-linear 
(fluid) corrections to the Zeldovich approximation can be neglected.  
      

\end{itemize}
This list of discreteness parameters is a minimal one, appropriate for, say, a 
       standard $P^3M$ type code. Even in this case it could be
      elaborated to be more precise. For example, the regularisation 
      involves the choice of a function which is not always the same,
      and $\varepsilon$ can vary in time\footnote{We have implicitly 
      assumed it to be 
      fixed in comoving length units, which is usually the case,
      although many other variants can be found in the literature.}.
      Different choices of the sampled modes may also be made in 
      setting up initial conditions.  
      The list is adequate 
      also for a simple $PM$
      code, but evidently it would need to be expanded to 
      describe adaptive codes in which the particle number changes
      in time and space according to some criteria. We will not consider such 
      complexities here, apart from a few further comments on this point
       in our conclusions: it is sufficiently ambitious to hope, at least
       as a first step, to control fully the effects of discreteness for 
      these simpler cases.

\subsection{Convergence to Vlasov-Poisson limit}

Let us now denote by ${\cal Q} (r, z ; \{ D_\alpha \})$ the measured value of
any physically relevant quantity in an $N$-body simulation, 
at red-shift $z$, with values of the discreteness parameters 
$D_\alpha$, e.g., a two point correlation function or a PS
(where the variable $r$ is given the appropriate interpretation, and 
could equally well represent a set of vector separations for a higher
order statistic). 

The discreteness problem can be schematically represented then as that 
of determining an estimate of the  difference  
\be
\Delta{\cal Q} (r, z ; \{ D_\alpha \}) \equiv
{\cal Q} (r, z ; \{ D_\alpha\}) 
- {\cal Q}_{VP} (r, z)
\ee
where ${\cal Q} (r, z ; \{ D_\alpha \})$ is the result of a 
``perfect'' $N$-body simulation, and  ${\cal Q}_{VP} (r, z )$ is
the result of the same quantity in the VP equations 
evolved from the same initial conditions. 

By construction ${\cal Q}_{VP} (r, z )$ is, in general,
unknown. Indeed it is because we cannot determine it analytically that
we turn to $N$-body simulation.  To estimate $\Delta{\cal Q} (r, z ;
\{ D_\alpha \})$ the best one can do is thus to study, numerically,
the convergence of ${\cal Q} (r, z ; \{ D_\alpha \})$ towards some
fixed value as the $\{ D_\alpha\}$ are appropriately extrapolated.
If the goal is to
approach as closely as possible the evolution of the VP equations, one
should evidently extrapolate the relevant parameters in a way which indeed
gives convergence to this limit of the $N$-body system.  While it is
evident that the interparticle distance $\ell$ should be decreased,
how the other parameters should be varied (or not) is
not. Indeed, as we will discuss further in our conclusions, most of
the few convergence studies of cosmological simulations in the
literature do not adopt an extrapolation which converges directly to
the VP limit\footnote{A notable exception is the study reported in
\cite{splinter}, which will we discuss below, as well as a recent
paper by \cite{romeo08}.}.

There is in fact no rigorous treatment in the literature on
cosmological $N$-body simulations, or more broadly in the
cosmology literature, establishing the existence of the
VP limit: derivations of the VP equations
(see e.g. \cite{peebles, saslaw}) are limited to showing 
that these equations may be obtained by a truncation
to the leading term of a BBGKY hierarchy of equations,
but do not rigorously establish the conditions under which
the required truncation may be made\footnote{For an alternative
derivation of the VP equations using a coarse-graining of 
the microscopic equations for the particle system, 
see \cite{buchert_dominguez}.}. Formal proofs establishing 
the validity of the Vlasov mean field approximation for long-range
interacting systems can, however, be found 
in the mathematical physics literature (for a discussion
see e.g. \cite{spohn}). Notably \cite{braun+hepp} have proved 
that in a {\it finite} system of particles interacting through $1/r^2$ pair
forces, {\it regularized} so that the potential is bounded below at $r=0$, the Vlasov limit 
corresponds to $N \rightarrow \infty$. In taking this limit the volume,
mass and time of evolution are kept fixed\footnote{Formally the coupling
in the interaction (i.e. $Gm^2$ for gravity, where $m$ is the particle
mass) scales in proportion to $1/N^2$.}. We will assume, without 
rigorous proof, the evident extension of this result to the 
infinite volume case of cosmological simulations: we take 
the VP limit as $\ell \rightarrow 0$ (i.e. particle number in any
finite volume goes to infinity) at fixed mass density, {\it followed by} 
$\varepsilon \rightarrow 0$\footnote{As noted, the proof of 
\cite{braun+hepp} is for finite non-zero smoothing, and 
the existence of the exact VP limit $\varepsilon \rightarrow 0$ 
has not in fact been proven. We neglect this mathematical subtlety 
here, which one would expect to be relevant, at most, to the 
asymptotically long time behaviour of the system (which does not 
interest us in this context)}. The convergence at fixed 
temporal duration corresponds to keeping also the initial 
red-shift $z_i$ fixed, and as the limit should clearly not 
depend on $preIC$ (the choice of pre-initial configuration), 
we also keep this fixed.

In summary, applied to an $N$-body simulation, this tells us 
that {\it an} appropriate extrapolation is given by 
{\it decreasing $\ell$ (i.e. increasing the particle density) 
while keeping the other discreteness parameters $\{ D_\alpha \}$ fixed}. 
The limit is taken at {\it fixed} $\varepsilon \neq 0$,
which means that the spatial resolution (for unsmoothed gravity)
is limited to above this scale. In other words this extrapolation
converges to a smoothed version of the VP equations, which then
(we assume) would converge to VP as $\varepsilon \rightarrow 0$.
This extrapolation is not necessarily unique --- convergence
may in principle be obtained while allowing $z_i$ and/or 
$\varepsilon$ to vary in various manners as a function of
$\ell$ --- but it is certainly simple. The use of any alternative
({\it if}, we emphasize, the goal is to obtain direct convergence to
the VP limit) should, however, be carefully considered to establish
(at least as rigorously as here) that it gives convergence to
the VP limit. 

It is important to note the specific order of the limits in $\ell$ and
$\varepsilon$. Beyond the necessity to introduce a regularisation in
rigorous proofs of the VP limit mentioned above, the reason for this
can be understood easily on physical grounds: the scale $\varepsilon
\neq 0$ provides a characteristic scale which is clearly necessary to
give physical significance to the limit $\ell \rightarrow 0$. Indeed
taking $\varepsilon=0$ and initial conditions specified by a pure
power law input PS, $\ell$ is the sole characteristic scale of the
discrete system (in the limit $L\rightarrow \infty$), and defines
itself the unit of length. Varying $\ell$ gives, up to a trivial
rescaling, a system with exactly the same dynamics, which is
manifestly not that of the VP limit (as it includes explicitly
non-mean field effects such as two body collisionality). When, on the
other hand, the same system is treated, but now with $\varepsilon \neq
0$, the limit $\ell \rightarrow 0$ has a non-trivial meaning if it is
taken at constant $\epsilon$. Indeed non mean-field effects such as
two body scattering, and other ones we will describe below, are
explicitly no longer present in the dynamics in this limit.

Finally we make one important remark about initial conditions.
Evidently to establish convergence to the evolution of a given 
continuum model, one should keep the initial conditions in this 
limit fixed. While we have stated explicitly that, in extrapolating, 
the initial redshift $z_i$ should be held fixed, this does not 
prescribe unambiguously the initial conditions at any finite $\ell$:
given that simulations are performed in a finite volume, the
number of modes in any interval of wavenumber is finite, and
thus different realizations of the same initial conditions 
introduce intrinsic statistical fluctuations in the initial
conditions compared to the average theoretical behaviour.
For convergence studies of discreteness effects specifically
it is simplest to keep also the realization fixed, although of course 
such effects can in principle be averaged out by a sufficiently 
large number of realizations\footnote{Further, when 
$\ell >\varepsilon$, one needs to specify whether 
power should be added (if present in the theoretical model) in 
the larger range of wavenumbers which can be sampled in the 
initial conditions as $\ell$ decreases. While this 
is not the source of ambiguity in our extrapolation
(as power should evidently not be added in the range 
that $\ell < \varepsilon$), for studies in the range
$\ell >\varepsilon$, such as that we will report below,
it is relevant. We will use the prescription that 
the realization of the input displacement field is kept 
fixed, as this allows us most clearly to identify
effects of discreteness.}.

\subsection{How far must one extrapolate?}

While the above discussion simply tells us how to extrapolate 
towards the VP limit, the practical form
of the discreteness problem is more detailed: {\it How small do we 
need $\ell$ to be, given certain fixed values 
of other parameters in the set $\{D_\alpha\}$, to attain a desired 
precision $\Delta{\cal Q} (r, z ; \{ D_\alpha \})$ 
on the theoretical quantity ${\cal Q}_{VP} (r, z)$ ?} 

\subsubsection{Common wisdom}

Current practice in interpreting the results of cosmological
simulations appears to repose on an approximate answer to this
question, which we now attempt to summarize (see e.g. \cite{smith, power_03}).
It is supposed that there are essentially two ways in which 
discreteness can play a significant role in making an $N$-body 
simulation deviate from the desired VP evolution:
\begin{itemize}
\item D1. Through the {\it limits placed on the accurate representation of 
      the initial conditions}. Indeed, to avoid aliasing effects, 
      only modes of the input theoretical PS up to the Nyquist 
      frequency $k_N=\pi/\ell$ should be sampled. Unless 
      $k_c \ll k_N$ this means that there is ``missing power''.
      In simulations of CDM type models, notably, this is always the
      case. Further, for {\it any} initial PS, there 
      is always additional power in the initial conditions,
      predominantly at $k > k_N$, generated purely by the 
      discretisation\footnote{Analytical expressions for the full 
      initial conditions 
      are given in \cite{discreteness1_mjbm}. On a lattice, at linear
      order in an expansion in the amplitude of the input spectrum, the
      discrete power is non-zero only for $k>k_N$. In a glass there
      is also a contribution, $\propto k^2$ at small $k$, for $k<k_N$.}. 

\item D2. Through {\it two body collisions in the course of the dynamical
      evolution} which cause deviations from the desired mean field 
      behaviour of the VP limit.

\end{itemize}

The first point is believed not to place, in practice, an important
limitation on the accuracy of simulations once they are evolved.
The reason is that gravitational clustering, from CDM cosmological 
initial conditions, is understood to develop essentially by the 
transfer of power from large to small scales, non-linear structures 
being formed by the evolution of fluctuations at initially larger 
scales. The spatial resolution thus improves rapidly as time 
goes on, essentially following the forming non-linear structures
which depend only on the presence of the initial fluctuations 
which seeded them\footnote{A numerical study which nicely 
illustrating this may be found in \cite{Little+weinberg+park_1991}.
See also \cite{bagla+padmanabhan_1997} and \cite{bagla_discreteness08}.}. Small residual effects are 
envisaged, arising from 
the ``spurious'' power generated by the sampling
on a specific pre-initial configuration (grid or glass), but
they are usually assumed to be negligible and of no practical
importance\footnote{Some works (e.g. \cite{smith}) attempt to 
correct for the associated effects by subtracting this 
power which can be measured in the initial conditions. 
This procedure assumes that this spurious power does not evolve,
an assumption which we have shown analytically to be incorrect 
in  \cite{discreteness1_mjbm}.}.
An exception is in the case of hot (or warm) dark matter spectra. 
In this case one may have $k_c \ll k_N$ so that all the initial 
power is well represented, but the small scale power generated 
by discreteness 
can evolve to form structures which may not be ``wiped out'' 
sufficiently rapidly by the structures forming at larger
scales. Recently interest in this case has been regenerated
in the context of simulation of ``warm dark matter'' models, and 
it has been shown explicitly in numerical studies 
\citep{gotz+sommerlarsen_WDM, wang+white_HDM}, using different 
$preIC$ (grid or glass) that such effects may be important,
leading to gross discreteness effects in such simulations.

The effects of two body collisionality (D2) are understood to be taken 
care of by the smoothing $\varepsilon$. Indeed it is explicitly for 
this reason that such a smoothing is introduced, its value being chosen 
ideally large enough to suppress the related effects, but small enough
so that too much spatial resolution is not lost. Since, according to 
simple estimates \citep{binney}, one expects such effects to be largest 
in regions of highest density, $\varepsilon $ is chosen just large
enough to suppress them, over the relevant cosmological time scales, 
in such regions. 

In summary, these physical arguments may be formulated as qualitative
answers to the question posed above, as follows. For 
typical quantities measured in simulations (e.g. two point correlation 
function, PS, halo masses and profiles), the errors 
$\Delta{\cal Q} (r, z ; \{ D_\alpha \})$  due to discreteness,
for any $r$ a little larger than $\varepsilon$ are negligible, i.e., so 
small as to be of no practical interest (compared to attainable 
numerical errors, notably), if:

\begin{itemize}

\item A1.  $\ell$ is sufficiently small so that, at red-shift $z$, 
the fluctuations at scale $r$ may be formed by the collapse of 
fluctuations initially at scales $k < k_N$.    

\item A2.  $\ell$ is sufficiently small so that the collisional
relaxation time scale in the densest resolved regions (i.e. 
the highest density in a region of radius $\sim \varepsilon$) 
is large compared to the age of the universe. 

\end{itemize}

Both of the answers can be converted, by making use of known phenomenological 
models describing the results of simulations (notably halo models, or 
the model of \cite{peacock}), into approximate criterion 
for the necessary $\ell$ (i.e. particle density)
expressed in terms of the parameters of the theoretical model, and
the scales $r$ and $\varepsilon$, and of the red-shift $z$.
In \cite{discreteness-hamana} one can find, for example, approximate
criteria derived using halo models, while \cite{knebe_etal2000}, 
\cite{power_03} and \cite{diemandetal_convergence} present extensive numerical
studies\footnote{In these latter papers the question of discreteness 
is not separated from the question of numerical convergence of 
the $N$-body equation. Thus particle density and the smoothing $\varepsilon$
are considered on the same footing as choice of time step, and
parameters for force precision etc.. The sets of simulations 
studied do not in fact define a convergence study to the 
VP limit, as we have discussed above: power is added as the 
particle density is increased, and the initial red-shift also 
changes. We will return to this point in our conclusions.}.
A series of other articles focus specifically on the effects
of two body relaxation in placing limits on the accuracy
of density profiles in halos, using mostly numerical approaches
(\cite{knebe}, \cite{diemandetal_2body}, \cite{elzant_2006}).
A recent paper by \cite{bagla_discreteness08} concludes, 
on the basis of some simple numerical tests on different 
theoretical initial conditions, that discreteness effects may be 
neglected once the non-linearity scale has evolved to be larger 
than the mean interparticle separation.  

\subsubsection{Dissenting views}

While these answers may be correct, they are certainly 
not in any way rigorous. The essential problem is that 
they {\it assume that the physical effects of discreteness are known}, 
or, at least, that those which play any significant role in 
simulations are known.
While the latter may a posteriori prove to be true, the former 
certainly is not. Indeed understanding of the role of discreteness
in the highly non-linear evolution of these systems is extremely
limited.  

One of the surprising aspects, at least at first sight, of the
standard criteria just discussed is that they allow the resolution
scale of a simulation (at $z=0$) to be very much smaller than
the scale $\ell$.  Indeed in practice the spatial resolution is 
usually taken to be fixed by $\varepsilon$, with 
$\varepsilon \ll \ell$ \footnote{In the ``Millenium'' simulation 
\cite{springel_05}, for example, $\varepsilon \approx \ell/50$.}. 
If one considers that this smoothing is introduced to make
the ``macro-particles'' behave like fluid elements, moving under the
effect of the mean field, it would appear to be necessary to
have, at least, $\varepsilon \sim \ell$. This point has been
forcefully argued by Melott et al. in a series of papers
during the nineties \citep{melott_alone, kuhlman, melott_all, splinter},
and restated in a recent comment \citep{melott_comment2007}.
In \cite{melott_alone}, \cite{kuhlman} and \cite{melott_all} 
specific non-Vlasov
effects are explicitly shown to be present in numerical 
experiments, and of much greater importance once the
regime $\varepsilon < \ell$ is attained. One of the few
controlled studies of the issue of discreteness in the 
literature in a spirit resembling that advocated above 
is given in \cite{splinter}. The paper focuses on the
difference between results of simulation using $PM$ codes 
and $P^3M$ codes at different resolutions. Its conclusion
is that results of the latter codes in the regime
$\varepsilon < \ell$, {\it do not} agree well, most notably for 
phase sensitive statistics, with those obtained from 
higher resolution $PM$ codes (for which $\varepsilon=\ell$). 
These results, which place a question mark over the
reliability of results below the scale $\ell$, have 
been largely ignored and addressed only very incompletely 
in subsequent works (see, notably, \cite{knebe_etal2000, discreteness-hamana}) 
which support, broadly, the ``common wisdom''  which we have 
outlined above \footnote{We say ``incompletely'' because no other 
published work has, to our knowledge, reported similar precise 
tests measuring the same quantities. \cite{knebe_etal2000} ascribe 
the differences seen in the two point correlation properties 
by \cite{splinter} to  ``erroneous evolution in high resolution runs'', 
but without any proof (their own numerical tests, unlike those 
of \cite{splinter}, are not tests for discreteness effects but
for the coherence of results produced by different codes).
The analysis of \cite{discreteness-hamana}, which explicitly 
calculates how resolution improves with time as foreseen by
the ``common wisdom'' described above, suffers from the
weakness, underlined by the authors themselves, that it 
is based on the use of a halo model description of non-linear
clustering, itself drawn from numerical simulations.}.

The common wisdom has also been questioned by several other groups
of authors (\cite{suisalu+saar,Baertschiger:2002tk,xiao_etal, romeo08}), all 
placing in question (like Melott et al.) the use of a smoothing 
$\varepsilon < \ell$ on the 
basis of numerical results. In particular we note that the role 
played by interactions of particles with their nearest 
neighbours --- which give physical effects clearly not representative 
of the mean field Vlasov-Poisson limit --- in the
evolution of clustering at early times in simulations has
been highlighted in cosmological simulations 
in \cite{Baertschiger:2002tk}, and in a simplified class of
gravitational N-body simulations in \cite{sl1, sl2, sl3}.
In a very recent study \cite{romeo08} conclude, on the basis of a study using
wavelet techniques to analyse a set of $\Lambda$CDM 
simulations, also that results below the scale $\ell$
are unreliable.  
We note also the discussion of discreteness effects
in \cite{binney_discreteness}, which illustrates with a study 
of a one-dimensional sheet model that discreteness may 
induce effects prior to virialisation (and distinct from
two body effects) by artificially bounding above the 
growth of the phase space density.

\subsection{Analytical results}
\label{PLT-summary}

In recent work by ourselves and our collaborators
\citep{joyce_05, marcos_06, discreteness2_mjbm, marcos07}, 
we have used a perturbative treatment of cosmological 
$N$-body simulations to treat discreteness effects analytically. 
While the method is limited by its range of application (to
sufficiently early times) it has the advantage of
providing an exact quantification of these effects
in that range, as well as an understanding of the physical
mechanism at play. In this section we will briefly review this
formalism, which we will employ in the next section in the
analysis of our numerical results. 

The treatment can be understood as a generalization to discrete
distributions of the standard linearization of the equations of a
self-gravitating fluid, in the Lagrangian formalism (see
e.g. \cite{buchert2}). At linear order we thus refer to it as
``particle linear theory'' (PLT).  The full details can be found in
these publications, and we will limit ourselves to a short summary of
the essential idea, and the salient results. We note that while
\cite{discreteness2_mjbm} presents the details of the use of PLT to
quantify discreteness effects in the usual case of a simple cubic (SC)
lattice as $preIC$, \cite{marcos07} develops fully its generalisation
to the cases that $preIC$ is a body centred cubic (BCC) or face
centred cubic (FCC) lattice. We will exploit fully this latter
generalisation in the next section.

The principle of this approach is very simple: it consists simply in
Taylor expanding the force on each particle due to any other in 
their relative (vector) displacement from the lattice 
configuration\footnote{The treatment is analogous to one used
standardly in solid state physics (see e.g. \cite{pines}) to treat 
perturbations about a crystal, both for the case of short range 
two-body interactions (e.g. Lennard-Jones) and Coulomb 
interactions. See \cite{marcos_06} for further discussion.}.
Since the force is zero in the unperturbed lattice the force 
$\bF (\bR)$ on a particle originally at lattice site $\bR$
can be written, at linear order in the displacements 
${\bf u}({\bf R},t)$, as  
\be
\bF (\bR)= -\sum_{{\bf R}'} 
{\cal D} ({\bf R}- {\bf R}') {\bf u}({\bf R}',t)\,,
\label{linearised-force}
\ee
where the sum is over all the lattice sites, and
the matrix ${\cal D}$ is 
\bea
{\cal D}_{\mu \nu} ({\bf R} \neq {\bf 0})&=&
Gm\left(\frac{\delta_{\mu \nu}}{R^3}
-3\frac{R_\mu R_\nu}{R^5}\right) 
\nonumber \\
{\cal D}_{\mu \nu} ({\bf 0})&=& 
-\sum_{{\bf R} \neq {\bf 0}} {\cal D}_{\mu \nu} ({\bf R})
\eea
where $\delta_{\mu \nu}$ is the Kronecker delta, and the 
subscripts are the cartesian indices\footnote{A sum over 
the copies, due to the periodic boundary conditions, 
is left implicit in these expressions.}. With this approximation
to the force, the equations of motion
for the particles Eq.~(\ref{Nbody-eom}) may then be written as
\be
{\bf {\ddot u}}({\bf R},t) 
+2 H {\bf {\dot u}}({\bf R},t) 
= -\frac{1}{a^3} \sum_{{\bf R}'} 
{\cal D} ({\bf R}- {\bf R}') {\bf u}({\bf R}',t)\,. 
\label{linearised-eom}
\ee
Defining the discrete Fourier transform on the lattice and its inverse 
by 
\bse
\label{def-discreteFT}
\begin{align}
\label{def-discreteFT-tok}
{\bf {\tilde u}}({\bf k},t)&= \sum_{{\bf R}} e^{-i {\bf k}\cdot{\bf R}}
{\bf u}({\bf R},t) \\
\label{def-discreteFT-tor}
{\bf u}({\bf R},t)&= \frac{1}{N} \sum_{{\bf k}} e^{i {\bf k}\cdot{\bf R}}
 {\bf {\tilde u}}({\bf k},t)\,,
\end{align}
\ese 
where the sum in Eq.~\eqref{def-discreteFT-tor} is over the first
Brillouin zone (FBZ) of the lattice, i.e., the set of $N$ 
non-equivalent reciprocal lattice vectors closest to the
origin $\bk=\mathbf 0$\footnote{For a SC lattice the vectors
of the FBZ are thus $\bk = \bn (2\pi/L)$, where $\bn$ is a vector 
of integers of which each component $n_i$ ($i=1,2,3$) 
takes all integer values in the range $-N^{1/3}/2 < n_i \leq N^{1/3}/2$.
See \cite{marcos07} for the explicit expressions for the FBZ vectors
of a FCC and BCC lattice.},  Eq.~(\ref{linearised-eom}) can be
written in reciprocal space as
\be 
{\bf \ddot{{\tilde u}}} ({\bf k},t) + 2 H (t) {\bf
\dot{{\tilde u}}} ({\bf k},t) = -\frac{1}{a^3} {\cal {\tilde D}} ({\bf
k}) {{\bf {\tilde u}}}({\bf k},t) 
\ee 
where ${\cal {\tilde D}} (\bk)$, the Fourier transform (FT) 
of ${\cal D} ({\bf R})$, is a symmetric $3 \times 3$ matrix 
for each ${\bf k}$.

The solution of the dynamical problem now reduces simply to the
diagonalisation of the ${\cal {\tilde D}} ({\bf k})$, which is
straighforward (and inexpensive) numerically..  For 
each ${\bf k}$ this gives three orthonormal eigenvectors 
${\bf e}_n ({\bf k})$ and their eigenvalues 
$\omega_n^2({\bf k})$ ($n=1,2,3$). The evolved displacements 
from {\it any} initial perturbed lattice configuration, specified
at a time $t_0$, may then be written as
\be
\label{eigen_evol}
\bu(\bR,t)=\frac{1}{N}\sum_{\bk}\left[\mathcal{P}(\bk,t)\tbu(\bk,t_0)+\mathcal{Q}(\bk,t)\dot\tbu(\bk,t_0)\right]e^{i\bk\cdot\bR}
\ee
where the matrix elements of the ``evolution operator'' $\mathcal{P}$ and $\mathcal{Q}$ are
\bse
\label{evol_operators}
\begin{align}
\mP_{\mu\nu}(\bk,t)=&\sum_{n=1}^3 U_n(\bk,t)(\mathbf{e}_n(\bk))_\mu(\mathbf{e}_n(\bk))_\nu\\
\mQ_{\mu\nu}(\bk,t)=&\sum_{n=1}^3 V_n(\bk,t)(\mathbf{e}_n(\bk))_\mu(\mathbf{e}_n(\bk))_\nu\,.
\end{align}
\ese
The functions $U_n({\bf k},t)$ and $V_n({\bf k},t)$ are linearly
independent solutions of the mode equations \be {\ddot{f}} + 2 H
{\dot{f}}= -\frac{\omega_n^2({\bf k})}{a^3} f
\label{mode-equation}
\ee
chosen such that $U_n({\bf k},t_0)=1$, $\dot{U}_n({\bf k},t_0)=0$,
$V_n({\bf k},t_0)=0$ and $\dot{V}_n({\bf k},t_0)=1$.


The expression Eq.~(\ref{eigen_evol}) for the evolution is, up to
the validity of the linearized approximation to the force, {\it exact for
the discrete system}. To use it to determine discreteness effects we
must, as we have discussed at length in the previous subsections,
first identify unambiguously the correct continuum limit, and how
it is obtained by extrapolation of the discreteness parameters.
We have shown in \cite{joyce_05,marcos_06} that this may be done
straighforwardly directly from Eq.~(\ref{eigen_evol}): taking
the limit $\ell \rightarrow 0$, this expression for the
evolution converges exactly to that obtained \citep{buchert2}
by linearizing the equations for a self-gravitating fluid in
the Lagrangian formalism, which reduces asymptotically 
to the Zeldovich approximation. The latter represents the appropriate
analogous treatment of the Vlasov-Poisson limit. Note again that the 
limit $\ell \rightarrow 0$ in Eq.~(\ref{eigen_evol}) is taken at
fixed $t_0$ (i.e. fixed initial red-shift $z_i$) and for
a fixed input spectrum of displacements $\tbu (\bk)$ 
(and velocities $\dot{\tbu} (\bk)$) \footnote{The convergence 
to the VP limit is shown most easily keeping the initial 
conditions fixed, as described above, i.e., by introducing 
an ultraviolet cut-off in the initial spectrum so that 
the initial modes remains the same as $\ell \rightarrow 0$.
Alternatively, and in line with the general 
prescription given above, the 
limit may be recovered with a finite smoothing 
$\varepsilon$ which is kept fixed as $\ell \rightarrow 0$.
To do so one exploits the fact that the PLT formalism 
can be applied to any two body 
interaction potential, and specifically a softened 
gravitational potential (see \cite{discreteness2_mjbm} for
details).}.

The differences between the evolution given by Eq.~(\ref{eigen_evol})
and this continuum evolution may then be computed exactly for any given
initial conditions, yielding the discreteness effects, in the regime
of validity of the PLT approximation. In \cite{marcos_06} we have studied
the domain of validity of PLT numerically for a range of different 
initial conditions, and conclude that, for some simple statistical 
quantities, it describes the evolution up to the time when many 
particles approach their nearest neighbours, which
corresponds approximately to shell crossing in the corresponding 
fluid limit\footnote{We will see below that, for the quantities which
we will study in the next section  --- the differences in the
evolution of the same initial conditions sampled on  different 
$preIC$ --- the PLT approximation actually hold for much longer 
times, apparently following well the evolution of any mode until
it goes non-linear.}. In \cite{discreteness2_mjbm} we have presented
a quantitative analysis of the discreteness effects in this
corresponding regime, for some basic quantities in typical 
cosmological simulations. Some of the essential results are 
the following:

\begin{itemize}
\item The modification at shell crossing (up to which the PLT 
treatment described fully the discreteness effects) of the 
evolution of any given mode of the displacement field grows
monotonically with time.  Indeed taking the limit 
$z_i \rightarrow \infty$ at fixed particle density (i.e fixed $\ell$)
the evolution of the N-body system {\it diverges from the continuum
VP limit}. 

\item The modification due to discreteness at shell crossing,
for a fixed $z_i$, depends approximately on the ratio $k\ell$,
increasing as $k\ell$ does. This is physically very reasonable:
the longer the wavelength of a mode compared to the interparticle
scale, the less affected is the evolution by discreteness. 
For the typical values of $z_i$ in cosmological simulations,
the effect is typically to reduce the power in modes, by
up to about $50\%$ at the Nyquist frequency, and by
about $10\%$ at half this value.

 
\end{itemize}

Given this treatment of discreteness effects --- exhaustive
and analytical, but with a limited domain of validity in time --- 
what can we conclude about the questions raised
in the previous subsections? Concerning the formal extrapolation 
to the VP limit, we have already noted that 
PLT indeed converges to the theoretical VP behaviour
when the parameters are extrapolated as prescribed above.
With respect to the question of 
how far we must extrapolate 
in $\ell$ in order to converge with some required precision
to the VP evolution, the treatment also gives a clear answer,
{\it at shell crossing}. The answer depends of course on
the quantity considered, and then also on $z_i$ and 
on the cosmological model (which determined the red-shift
of shell crossing given $\ell$). In \cite{discreteness2_mjbm}
we have shown, for example, that, for $z_i$ a factor of five
larger than the redshift of shell crossing, errors of five
percent in the PS are achieved only for $k < k_N/4$.
Thus if we want, at shell crossing, an accuracy of less
than this on the power, we can use only results in
this range. 

This second conclusion is strikingly different from what
one might expect given the ``common wisdom'': the evolution
of the simulation makes the range of scale over which the
continuum model is accurately represented (to some given
precision) {\it decrease}, rather than increase. It 
in fact suggests that the view that $\ell$
should be a lower limit for spatial resolution may even be
too optimistic. These findings, however, only apply at 
shell crossing, and the ``common
wisdom'' above may still apply later on. Indeed, as we
described, the justification for this common wisdom is 
that when the transfer of power, characteristic of 
gravitational clustering in these systems, sets in, 
differences at smaller scales are wiped out. These
results at shell-crossing show, however, that
between $z_i$ and shell crossing errors develop in 
the long-wavelength modes (below $k_N$) which were not 
present in the initial conditions. As a result modes
at later time which depend on this power will necessarily
inherit this error.

An important point which we emphasize is that the fundamental reason
why the discreteness errors determined using PLT do not behave
as expected by the common wisdom is that {\it they arise from physical effects
of discreteness which are not usually envisaged}.
Indeed the physical effects described by PLT compared with the VP
limit are different from the two effects envisaged usually 
which we listed above. Firstly, they are {\it dynamical} 
effects which modify the evolution of any given mode in 
a way which is independent of the initial conditions.
Secondly, they are clearly not two body collisional effects
\footnote{To make this very explicit we have shown 
in \cite{discreteness2_mjbm}
that the inclusion of a simple Plummer smoothing in the force actually increases 
the difference between PLT and the VP limit for unsmoothed gravity.}. 
The effect they describe can be characterised 
physically as a  {\it dynamical sparse sampling effect}: PLT
compared with its VP limit tells us how the evolution of a 
fluctuation depends on the spatial density of the sampling
particles. An important question is then evidently to understand how this
physical effect --- which there is no reason to believe should
go away when we pass to the non-linear regime ---quantitatively
affects results in the latter regime. We will return to 
this point in our conclusions.
 
\section{A calibrated numerical study of discreteness effects}

We return now to the practical question of how 
small $\ell$ needs to be for a measured 
quantity to have converged to a desired precision. Since the force 
smoothing $\varepsilon$ places a lower bound on the spatial 
resolution, a simplified, more specific, form of the 
question is: {\it how small does $\ell$ have to be in order that, at 
any given red-shift, the effects
of discreteness are negligible down to scales of order} $\varepsilon$?
The answer provided by the ``common wisdom'' above is that 
$\ell$ is sufficiently small, in typical simulations, if 
$\ell/\varepsilon$ is less than about one hundred (see e.g.
\cite{knebe_etal2000}). According to 
the ``dissenting views'' $\ell$ must be at least as small as 
$\varepsilon$. 

One way to determine, in principle,  which view is correct is 
evidently to compare results from simulations with large 
$\ell_1/\varepsilon$ in the range $\ell_1 > r > \varepsilon$ with
those obtained in much higher resolution simulations, with
$\ell_2 \leq \varepsilon \ll \ell_1$. This is indeed 
the strategy advocated in \cite{splinter}, which reports a study
of this type down to a resolution $\ell_2=\varepsilon$. It 
concludes, as noted above, that there are significant differences 
in results, i.e., no evidence for convergence, in the range 
$\ell_1 > r > \ell_2$. Other authors \citep{knebe_etal2000}
argue however that these differences are ascribable 
to ``erroneous evolution in high resolution runs''. The 
difficulty in reaching a convincing conclusion is that 
the questions of discreteness effects are intertwined
with numerical and finite size effects. While such differences should 
be resolvable by further numerical tests, this would require 
considerable investment of resources which, apparently
because of the wide acceptance of the ``common wisdom'', 
has not been made\footnote{See, however, the recent paper
by \cite{romeo08}, which we will comment on in our conclusions.}.

Instead of undertaking such a numerical study --- which, given the
modest numerical resources at our disposition, would not in any case
likely to be any more conclusive than that reported by \cite{splinter}
--- we focus in the rest of this paper on another kind of test.  We
will see that this will allow us to reach conclusions, with modest
sized (but very well numerically converged) simulations, about the
central issue: the validity/precision of results in the range of
scales around or below $\ell$, in simulations with $\ell \gg
\varepsilon$.  The aim is to provide a method which gives a
non-trivial lower bound on discreteness error in such simulations.  To
do so we simply compare the results of simulations from identical
theoretical initial conditions, changing only the choice of the
discreteness parameter $preIC$, i.e., the pre-initial
configuration. We can then study how this error depends on time and
scale. Although the measured effects are quite small --- at most
of order of five percent in the PS for the times and scales relevant
to cosmological simulations --- we can establish clearly, using the
analytical PLT formalism combined with numerical tests of their
dependence on $\ell$ and $\varepsilon$, that they are indeed
discreteness effects. We can then address in a controlled way the
question of how far $\ell$ needs to be extrapolated so that one can be
confident that the true systematic errors due to discreteness have
converged to significantly less than this lower bound (e.g. to less
than one per cent).

Rather than considering a specific cosmological model, we consider a
simple power law PS with exponent n=-2, evolved in an EdS
universe. This choice is both suitable for our study as it is simple
--- introducing no characteristic scale in the input model --- and yet
close to the currently favoured CDM-like cosmological model, which has
an initial PS with effective exponent ranging between $n \approx -1$
and $n \approx -3$ over the relevant range of scales.  In particular
we note that this PS is, like these cosmological models,
long-wavelength dominated so that the very efficient transfer of power
from long to short wavelengths which, as we have discussed above, is
believed to play a role in wiping out discreteness effects, should be
well represented.  We will comment further in our conclusions on the
generalization to other initial conditions, and specifically to those
of currently favoured cosmological models.

All our simulations have been performed using the publically available
parallel tree-mesh code GADGET2 \citep{gadget_paper}.  We use this
single (widely used and highly tested) code for our study for the
reasons we discussed above: the discreteness effects we are trying to
understand and control for are distinct from differences arising
between different codes, and indeed distinct from any dependence of
results on the numerical parameters of a given code.  The
``calibration'' of our results with our analytic tools here provide
in fact a robust check that the GADGET2 code's integration of the $N$-body
equations of motion is sufficiently precise that this is indeed the
case. Comparison with other codes would be, in the relevant regime, a
check on the accuracy of these codes, rather than a check on our
results.  In the regime where our analytic results do not apply, we
can have, of course, less confidence in the identification of our
measured effects as physical discreteness effects, and a comparison
with other codes could be instructive. We will address this issue
below, where we give details of the detailed checks of numerical
convergence of our results which we have performed using GADGET2.

\subsection{Initial conditions}
\label{Initial conditions}

\begin{figure}
\includegraphics[width=0.15\textwidth]{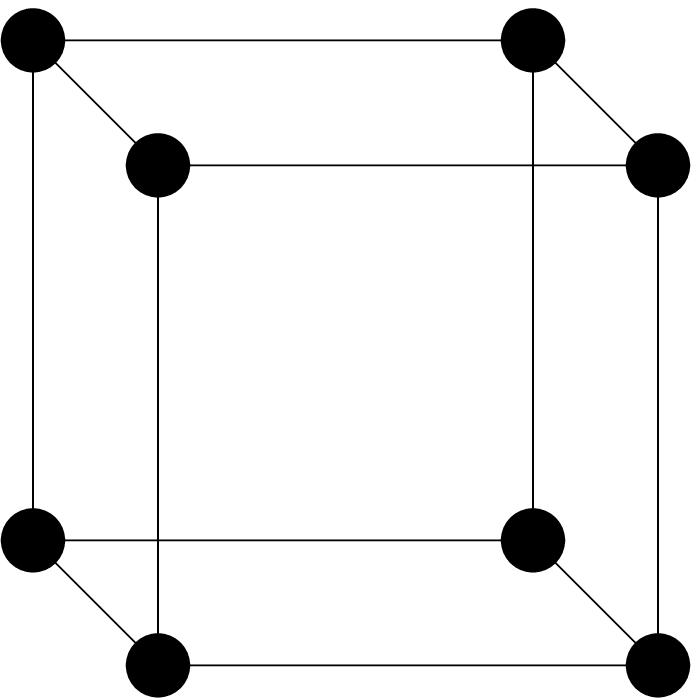}
\includegraphics[width=0.15\textwidth]{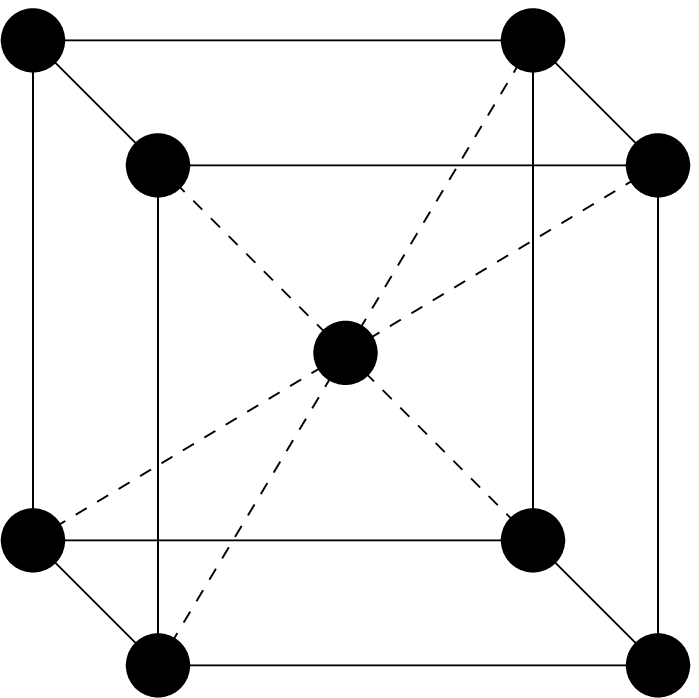}
\includegraphics[width=0.15\textwidth]{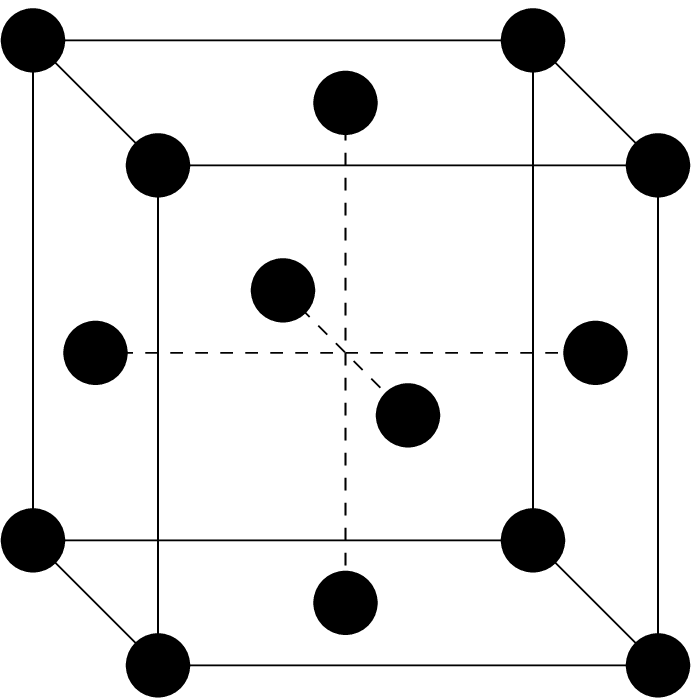}
\caption{From left to right, unit cell of the SC, BCC and FCC lattices.
 \label{unitcells}}
\end{figure}

We use the standard method, based on the Zeldovich approximation,
to set up initial conditions by applying appropriate displacements
to four different $preIC$: a simple cubic (SC) lattice, a body centred
cubic (BCC) lattice, a face centred cubic (FCC) lattice, and a glass 
configuration, shown in Fig.~\ref{unitcells}.

Our reference set of simulations, which we denote S1, have 
the number of particles shown in Table \ref{table1}.
The numbers for the BCC and FCC configurations have been chosen 
to be as close as possible to those of the SC and glass 
configurations\footnote{As can be seen in Fig.~\ref{unitcells},
there are two particles per elementary cell
of a BCC lattice, four per cell in a FCC lattice. Thus in a 
cubic box we have $2 M^3$ in a BCC, $4M^3$ in an FCC, lattice,
$M$ is an integer.}. The glass is generated, starting from 
Poisson distributed points with zero velocity, using an option 
in the GADGET2 package which evolves the particles under 
Coulomb forces (without expansion) and with a damping implemented 
by setting the velocities to zero at each time step.   
In what follows our results will always be given in 
{\it units of length in which the box size 
is equal to unity}. In these units the value of $\ell$ in 
the four different $preIC$ varies by less than one percent.

\begin{table}
\begin{center}
\begin{tabular}{@{}lr}
\hline
PI configuration & N \\
\hline
SC& $64^3=262144$ \\
BCC & $2\times 51^3 = 265302$ \\
FCC & $4\times 40^3 = 256000$ \\
glass & $64^3 = 262144$ \\
\hline
\end{tabular}
\end{center}
\caption{Number of particles in the four PreIC of the reference
set of simulations S1 \label{table1} }
\end{table} 

\begin{figure}
\psfrag{X}[c]{\large $k/k_N$}
\psfrag{Y}[c]{\large $P(k)$}
\includegraphics[width=0.45\textwidth]{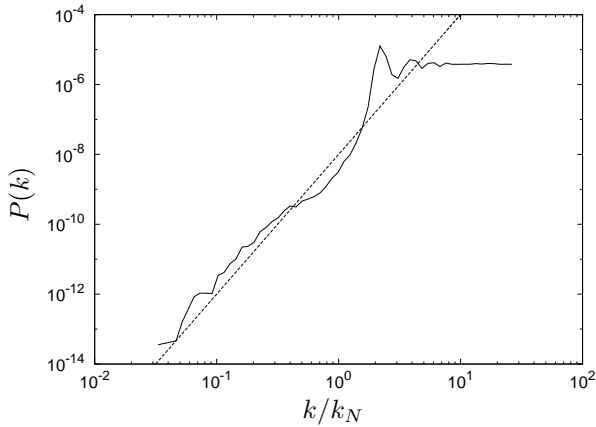}
\caption{The power spectrum(PS) of the pre-initial glass. The dashed line 
indicates the behaviour $\propto k^4$.
 \label{glassPS}}
\end{figure}

The three lattices have PS which can be written
\begin{equation}
P(\bk)=(2 \pi)^3 \sum_{\bK \neq 0} \delta (\bk - \bK)
\label{lattice-ps}
\end{equation}
where $\delta(\bk)$ denotes the Dirac delta function and 
the sum runs over non-zero vectors $\bK$ which are an
infinite subset of the full reciprocal lattice appropriately
defined for a given lattice\footnote{For the SC lattice 
we have simply $\bK=\bn (2 \pi/\ell_{sc})$ where $\bn$
is any non-zero vector of integers; see \cite{marcos07}
for the more general definition for any Bravais lattice.
Note that when we use the term ``reciprocal lattice'' here,
we are refering to that defined for the periodic box of
side $L$, i.e., for the SC lattice $\bK=\bn (2 \pi/L)$ 
where $\bn$ is any vector of integers.}. The delta function structure
of these PS is a result of the translational symmetries of the 
lattices. The PS of the glass is, in contrast, a continuous function.
Indeed, up to finite size effects, it is a function only of 
$k=|\bk|$ because of its statistical isotropy. 
It is shown in Fig.~\ref{glassPS}, along with
a line indicating its approximate small $k$ behaviour,
$P(k) \propto k^4$. 

Given an input theoretical PS $P_{th} (k)$, we generate 
a realisation of the displacement field $\bu(\bx)$ 
to be applied to the particles at spatial positions
$\bx$ of the four $preIC$ in a cubic box taking 
\be
\label{disp_gen}
\tbu(\bx)=\sum_{\bk}\left[a_{\bk}\sin(\bk\cdot\bx)\bk -b_{\bk}\cos(\bk\cdot\bx)\bk\right],
\ee
with
\be
a_{\bk}=R_1 \frac{\sqrt{P_{th}(\bk)}}{k^2},\qquad b_{\bk}=R_2
\frac{\sqrt{P_{th}(\bk)}}{k^2} \,,
\ee
where $R_1$ and $R_2$ are two independent Gaussian random numbers with
dispersion equal to unity. In writing the displacement field as a 
Fourier sum we use the fact that the $preIC$ are set up on a periodic
cube, and the sum over the vectors $\bk$ extends then over the 
appropriate reciprocal lattice. Further, if the input PS itself 
does not have a cut-off at a wavenumber significantly smaller than 
the Nyquist wavenumber of the sampling $preIC$ distribution, such a cut-off 
must be imposed to avoid aliasing effects. Here, where we consider
a simple power-law PS without a cut-off, we will take the 
sum in $\bk$ to extend over the first Brillouin zone 
{\it of the SC lattice}, i.e., the reciprocal 
vectors $\bk=\bn (2\pi/L)$ with each integer component $n_i \in [-N/2,
N/2[$. As we will discuss further below, this is the choice which minimises
aliasing effects for the SC lattice, but not for the other $preIC$  
configurations. We will measure the associated very small aliasing effects 
in the initial conditions and keep track of their role in generating
differences in the evolved distributions.

The only other parameter which needs to be fixed is the
normalization of the input PS $P_{th} (k)$ (which is equivalent 
to the choice of the initial red-shift $z_i$). In the set 
S1 we have taken, for all preIC, 
\be
k_N^3 P(k_N)=0.6,
\ee
where, in our units, $k_N=64\pi$. We have made this choice for our
reference simulations because it is close to that chosen
for such initial conditions, on a SC lattice, by the widely used  
GRAPHICS package \citep{bertschingercode}\footnote{This package 
determines the starting red-shift $z_i$
by normalizing so that the {\it maximal} value of the density 
fluctuation field at any point of the lattice is unity. This
gives a mass variance at this scale considerably less than
unity, sufficiently small that non-linear corrections to
the Zeldovich approximation should be small.}. We will discuss
below the effect of modifying this choice. 

Before turning to the evolution from this set of initial
conditions, let us consider more precisely their correlation 
properties, and in particular the effects of aliasing we
have mentioned above. To do so we make use of the detailed
analysis of initial conditions of N-body simulations reported 
in \cite{discreteness1_mjbm}. The PS of the perturbed $preIC$ 
distribution can be written conveniently in the form
\be
\label{discrete-continuous}
P(\bk)= P_{\mathrm {c}} (k)+P_{\mathrm {d}}(\bk).
\ee
where $P_{\mathrm {c}}(k)$ is the ``continuous'' part, independent
of the $preIC$ distribution, and $P_{\mathrm d}(\bk)$ is a
``discrete'' term which depends on the latter.
The full analytic expression for both these quantities
can be expanded order by order
in the amplitude of the input theoretical PS
$P_{\mathrm {th}}(k)$. At leading order one obtains
\bea
P_{\mathrm c} (k) &=& P_{\mathrm {th}}(k) 
\nonumber \,,\\
P_{\mathrm d} (k) &=& P_{\mathrm {PI}} (\bk)+P_{\mathrm {al}} (\bk) 
\label{PSdiscrete}
\eea
where $P_{\mathrm {PI}}(\bk)$ is the PS of the unperturbed $preIC$ configuration
(i.e. lattice or glass) and 
\be
\label{iofk}
P_{\mathrm {al}}(\bk)=\frac{k^2}{(2\pi)^3} \int d^3 q (\hat \bq\cdot\hat\bk)^2
\frac{P_{\mathrm {th}} (q)}{q^2}\left[P_{\mathrm {PI}}(\bq+\bk)-
P_{\mathrm {PI}}(\bk)\right]
\ee
is a contribution to the PS which, if non-zero at small $k$, 
describes an aliasing of the input PS.  If the $preIC$ is a perfect lattice, 
$P_{\mathrm {PI}} (\bk)$ is given by Eq.~(\ref{lattice-ps}) and 
therefore, for $\bk \neq \bK$,
we have 
\be
\label{iofk-lattice}
P_{\mathrm {al}}(\bk)= k^2 \sum_{\bK\ne\mathbf 0} 
\frac{\left[(\bK- \bk)\cdot \hat\bk\right]^2}{(\bK- \bk)^4}
P_{\mathrm {th}}(\bK- \bk)\,,
\ee
where $\bK$ are the appropriate subset of reciprocal lattice vectors
for each lattice. For the SC lattice the vectors $\bK$ are given
by $\bK= \frac{2\pi}{\ell}\bn \equiv 2k_N \bn$, where $\bn$ is any
non-zero vector of integers. It can then be verified easily\footnote{See
\cite{discreteness2_mjbm} for a more detailed discussion.} 
from the expression in Eq.~(\ref{iofk-lattice}) that $P_{\mathrm{al}}(\bk)$ 
is zero inside the first Brillouin zone of this lattice, i.e., 
for all $\bk$ with each component $k_i \in ]-k_N, k_N]$, 
{\it if} we impose a cut-off by making $P_{\mathrm{th}}(\bk)$ zero
{\it outside} the same region. This choice, which is the 
one we have used (and which
is that standardly used in this context \cite{bertschingercode, couchman}) 
is optimal, in the sense that it maximizes the size of the region
about $\bk=0$ in reciprocal place where the representation of the 
input power is exact, at linear order in the amplitude of the input PS.   

For other lattices an analagous, but different, optimal choice can be
made, taking the input PS non-zero only inside the given lattice's
first Brillouin zone. Here we have not done so as such a procedure
would require sampling the input PS at different wavevectors, which is
incompatible with the requirement that we use an identical
realization of the theoretical initial conditions. As a result we will 
have for the non-sc lattices a small
contribution coming from the aliasing term as given in
Eq.~(\ref{iofk-lattice}). For the glass, on the other hand, a more
significant contribution from this term [as given in Eq.~(\ref{iofk})]
is expected, as it is always non-zero and proportional to $k^2$ at
sufficiently small $k$ (see \cite{discreteness1_mjbm} for further
detail). In what follows we will study carefully in simulations the
evolution of these residual differences at small $k$ power in the
initial conditions, showing that they can in fact be neglected in
understanding the differences in the evolved power which emerge at
these scales.
 
\subsection{Numerical Evolution of S1}

We evolve\footnote{The details on the numerical parameters we have
used for the results reported are given in Appendix~\ref{app-gadget}.}
these four initial conditions in an EdS cosmology, 
from a scale factor $a=1$ to $a=2^7$. At this final time, as we
will see below, the scale of non-linearity has reached the box size
and finite size effects dominate. GADGET
implements a smoothing which modifies the force from exactly
Newtonian only below a scale $\varepsilon$.
We take here $\varepsilon=\ell_{\mathrm{sc}}/15$,
where $\ell_{\mathrm{sc}}$ is the interparticle spacing
of the SC lattice. This is, according to the ``common wisdom'',
a conservative choice for the  final resolution 
scale\footnote{For comparison
we note that, if our comoving particle density is assumed
equal to that in the Millenium simulation (\cite{springel_05}), 
the comoving size of our box is then approximately 
$15$ h $^{-1}$ Mpc. The ratio $\ell/\varepsilon$
in \cite{springel_05} is approximately fifty.}. 

\begin{figure*}

\includegraphics[width=0.45\textwidth]{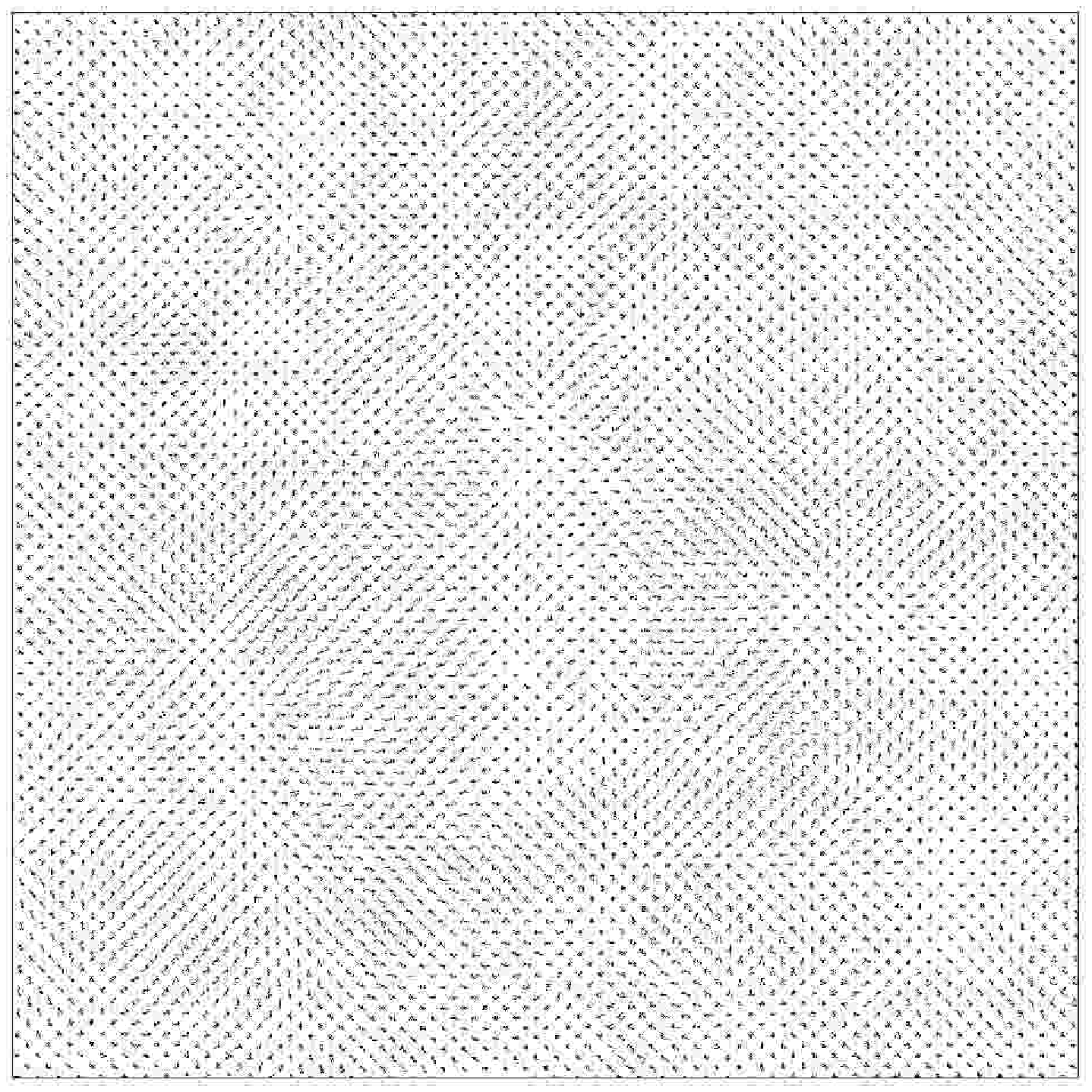}
\includegraphics[width=0.45\textwidth]{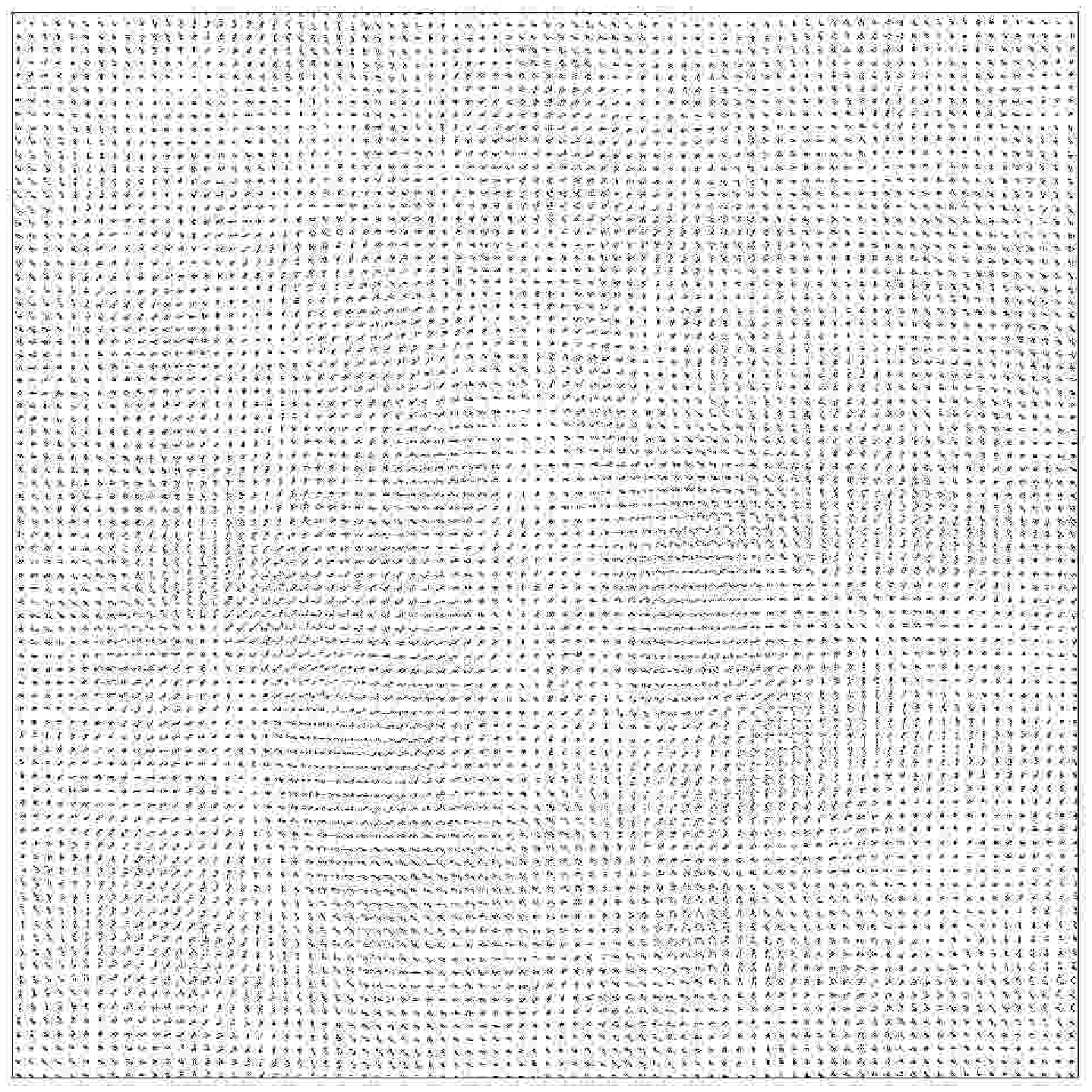}
\\
\includegraphics[width=0.45\textwidth]{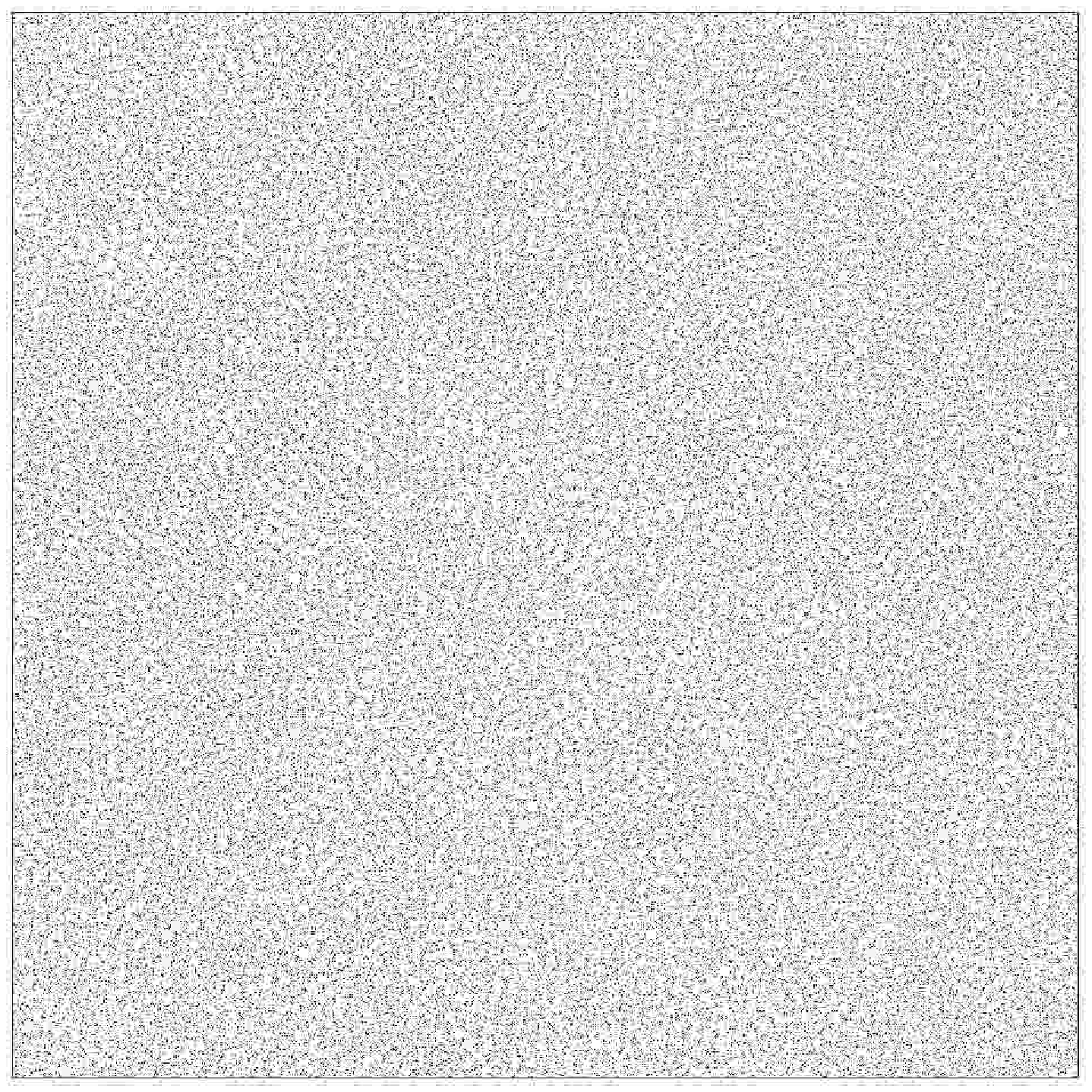}
\includegraphics[width=0.45\textwidth]{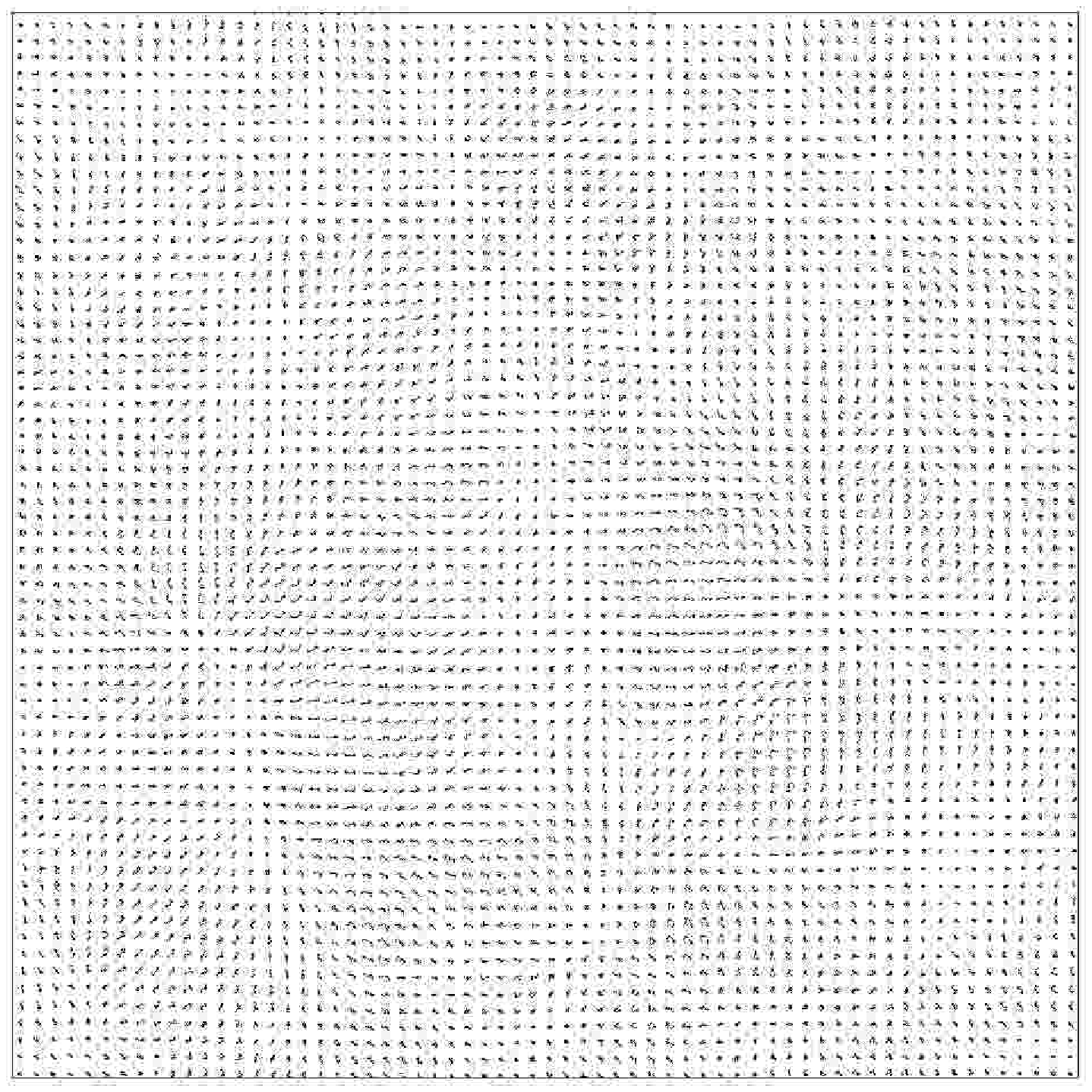}
\\
\includegraphics[width=0.45\textwidth]{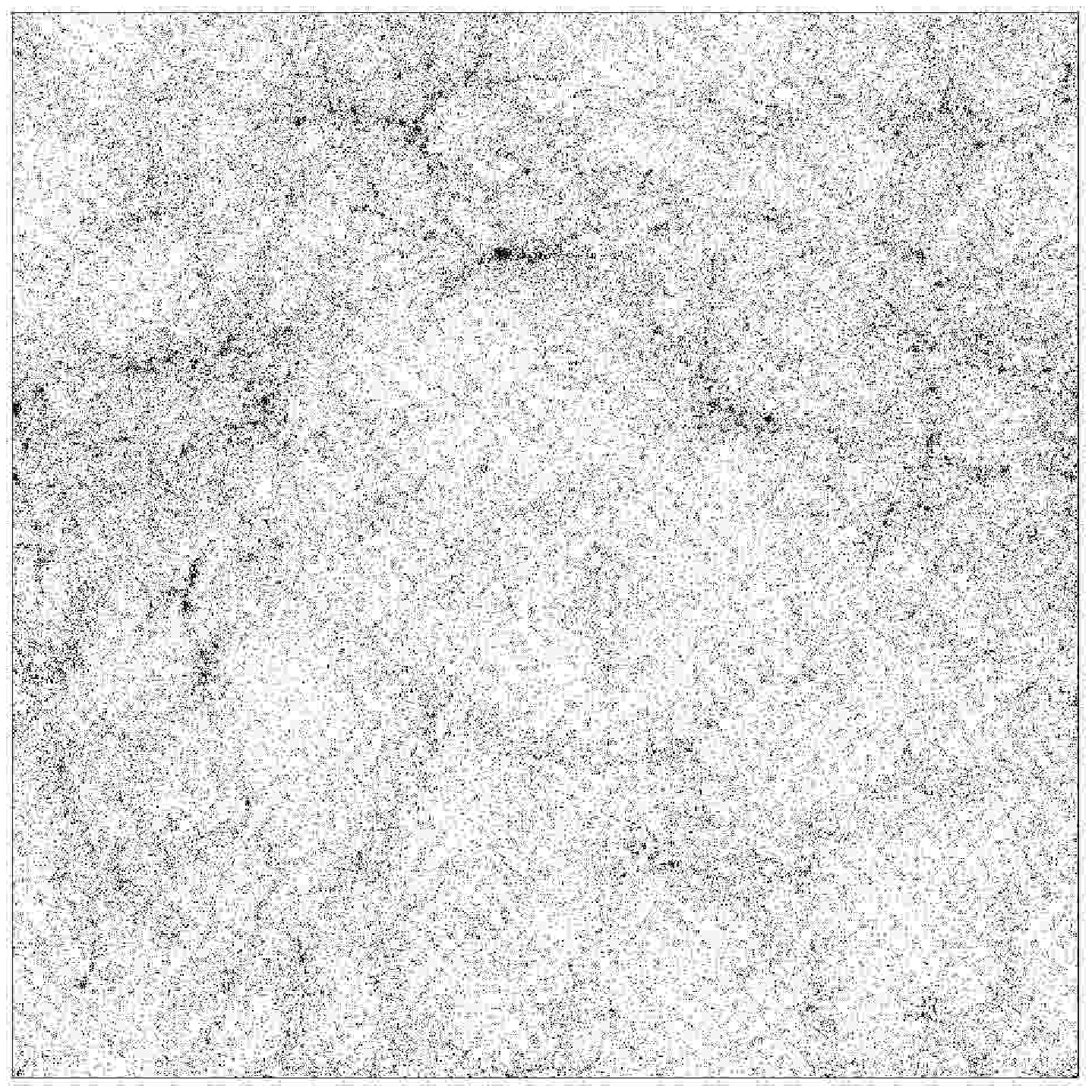}
\includegraphics[width=0.45\textwidth]{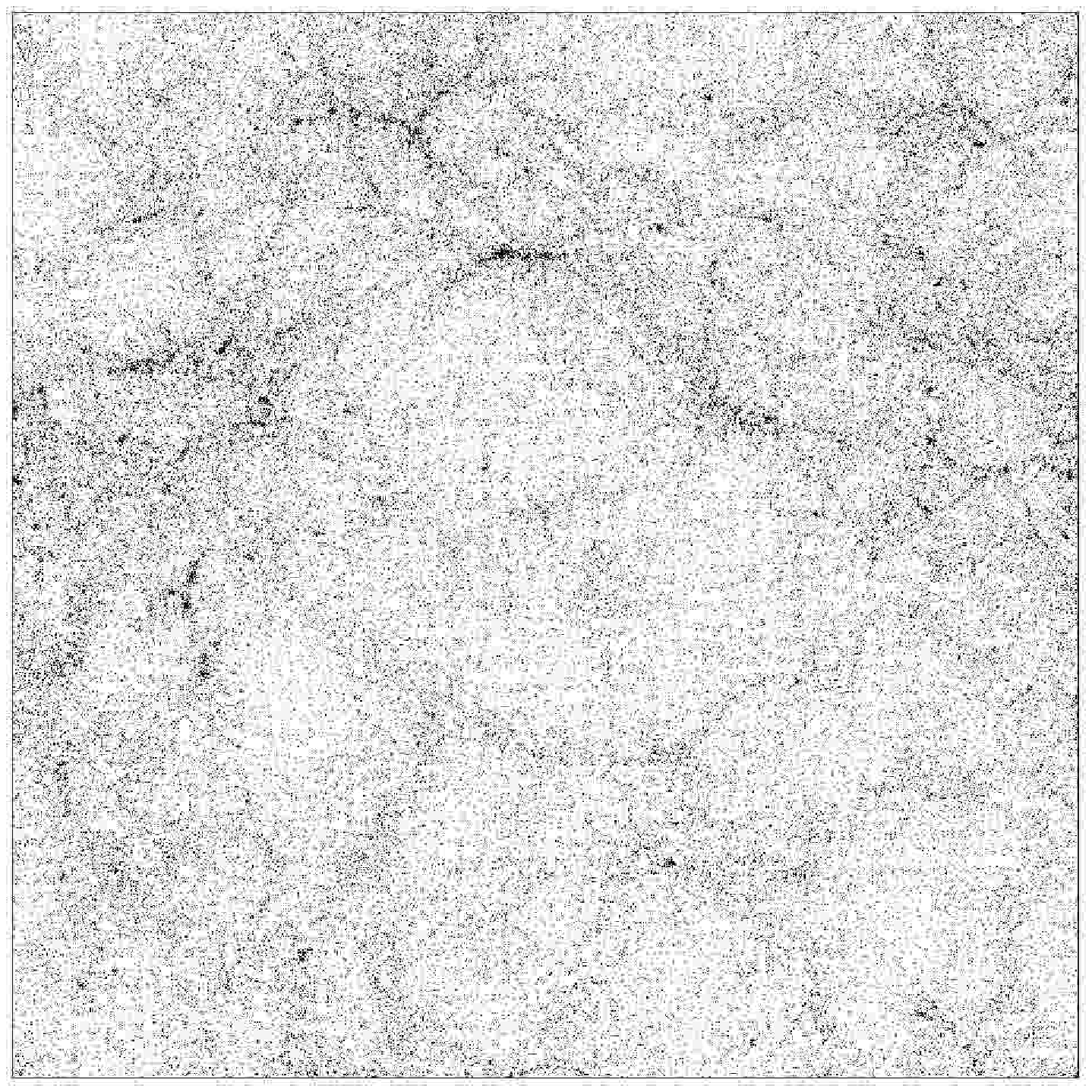}
\\
\includegraphics[width=0.45\textwidth]{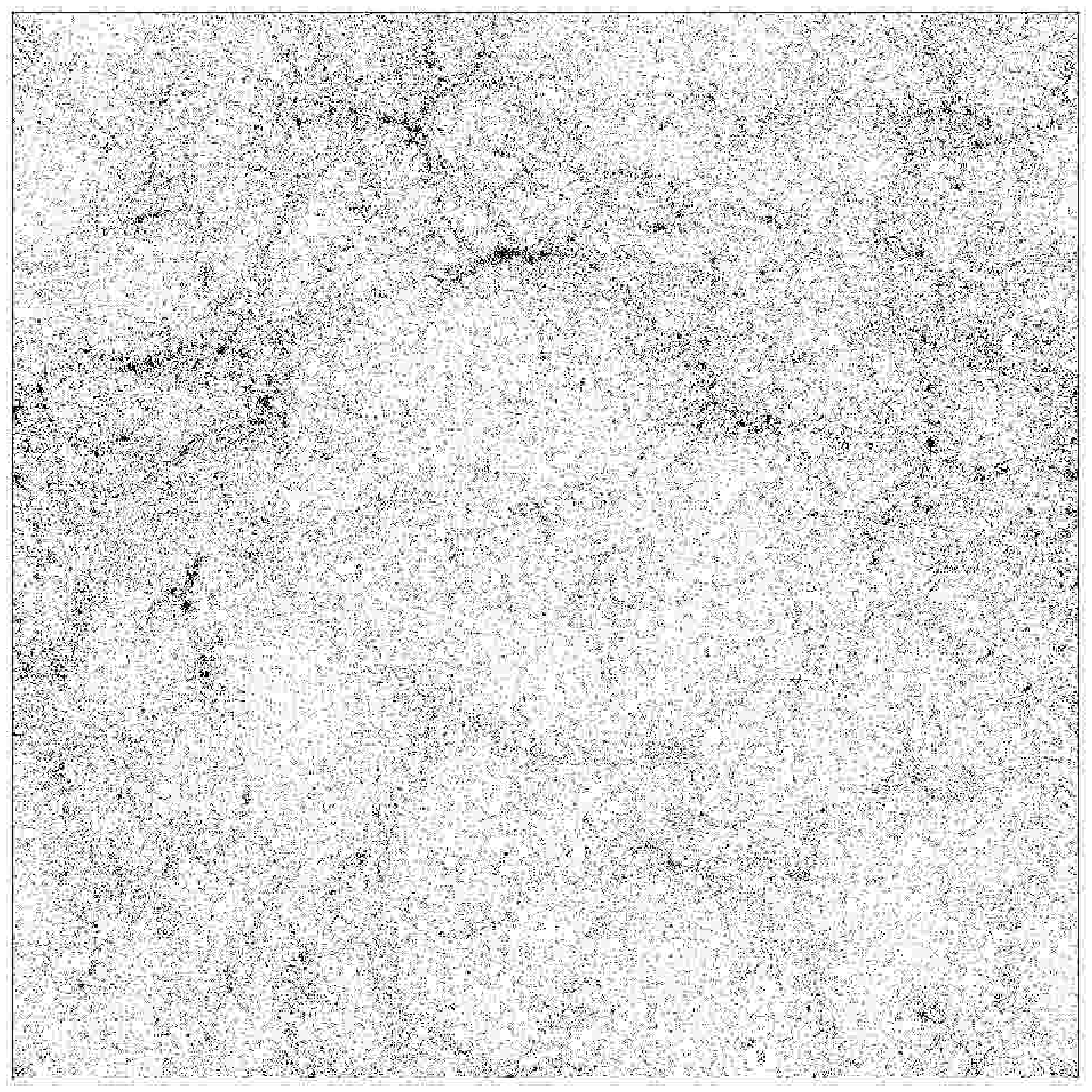}
\includegraphics[width=0.45\textwidth]{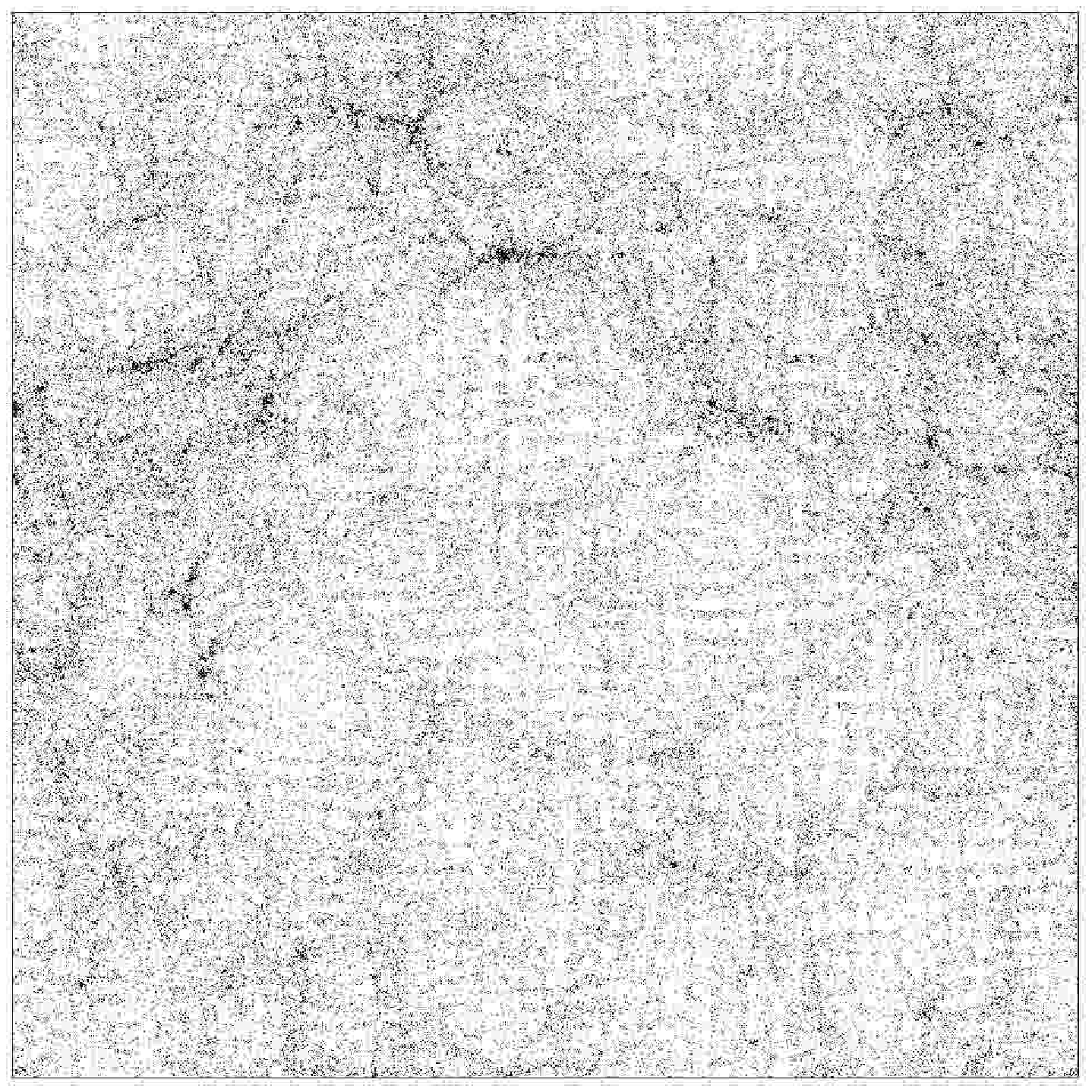}
\caption{Snapshots of a slice of the system for the different ICs 
(BCC, top left; FCC, top right; SC, bottom right; glass, bottom left)
at $a=1$ (upper four panels) and $a=2^3$ (bottom four panels). 
 \label{snapshot0}}
\end{figure*}

\begin{figure*}
\includegraphics[width=0.45\textwidth]{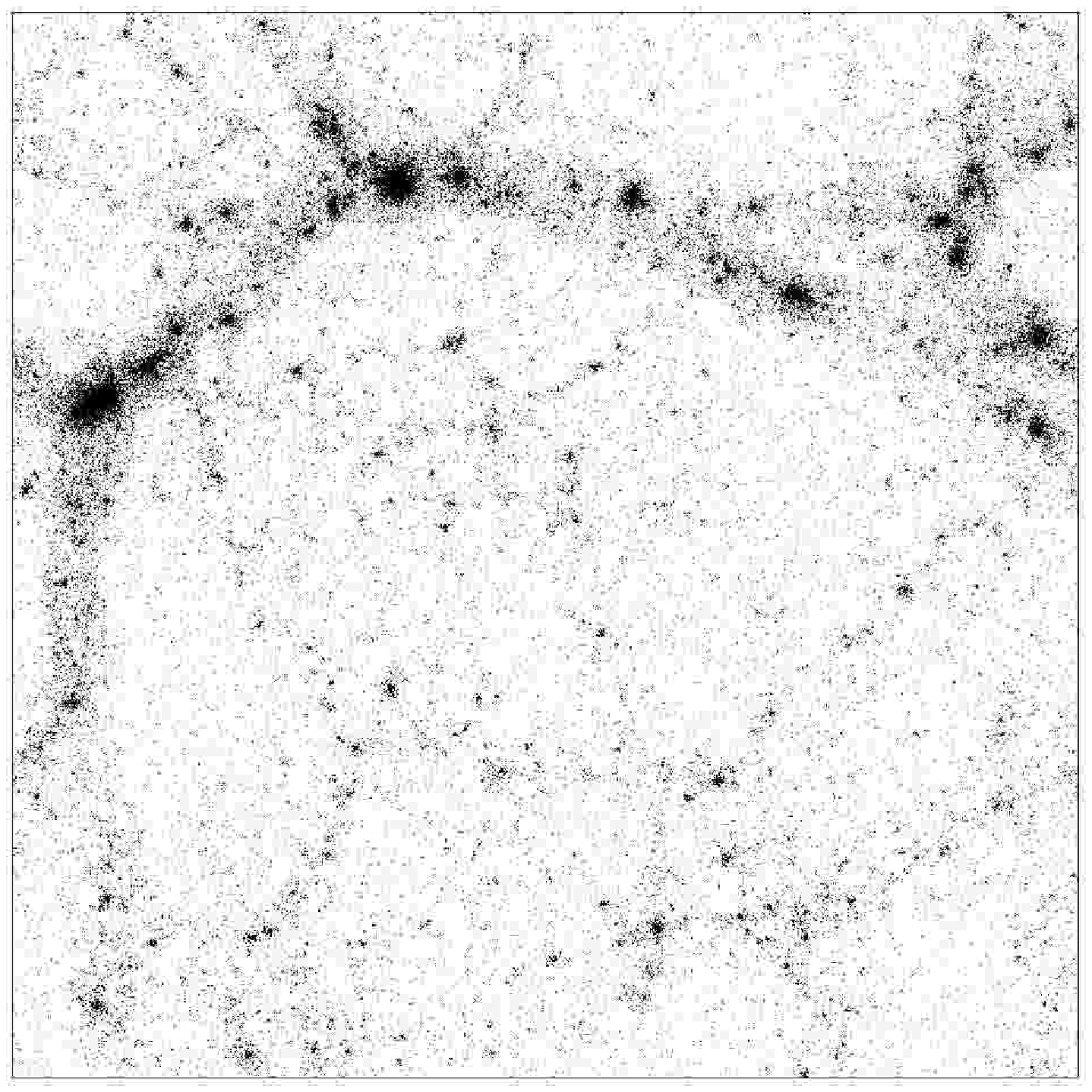}
\includegraphics[width=0.45\textwidth]{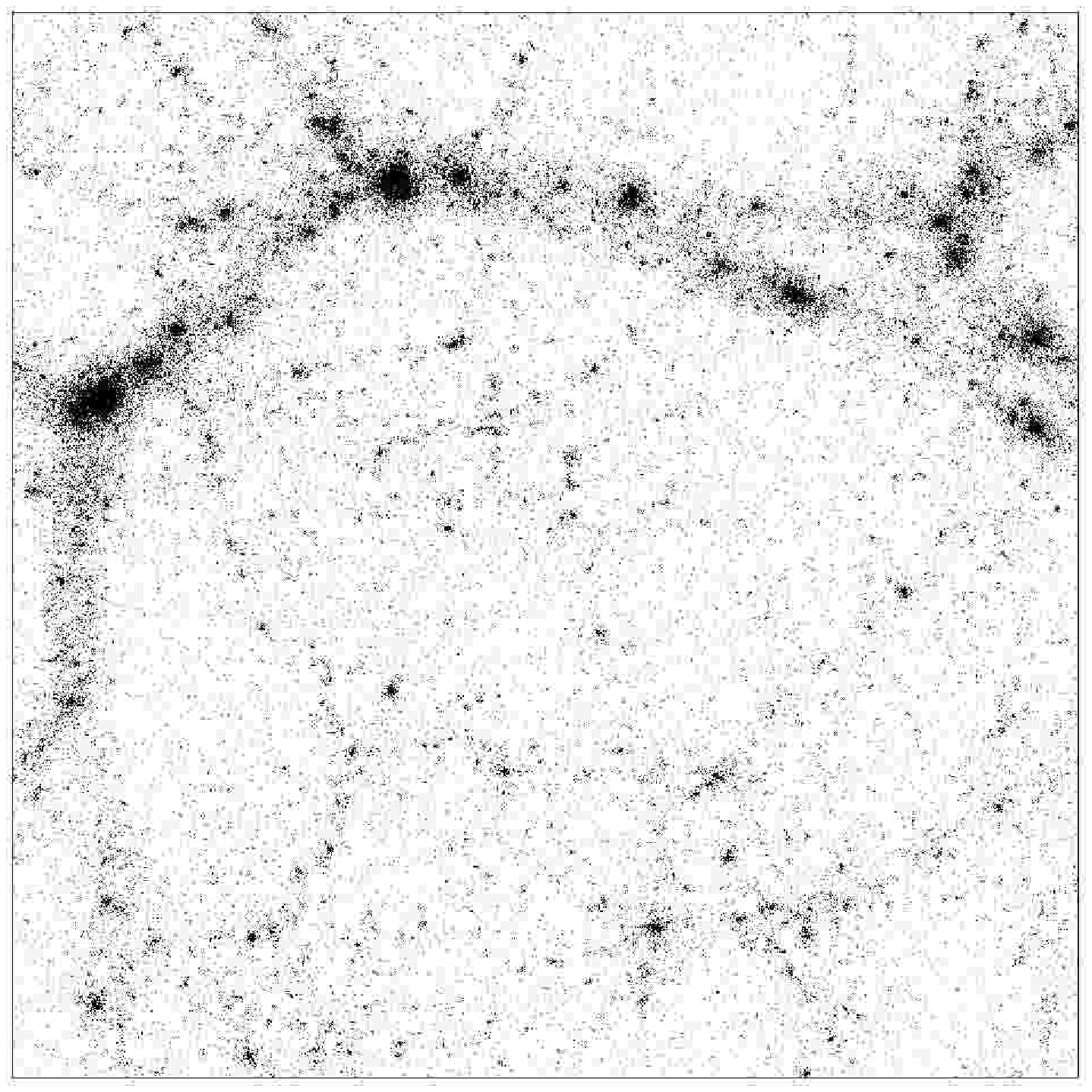}
\\
\includegraphics[width=0.45\textwidth]{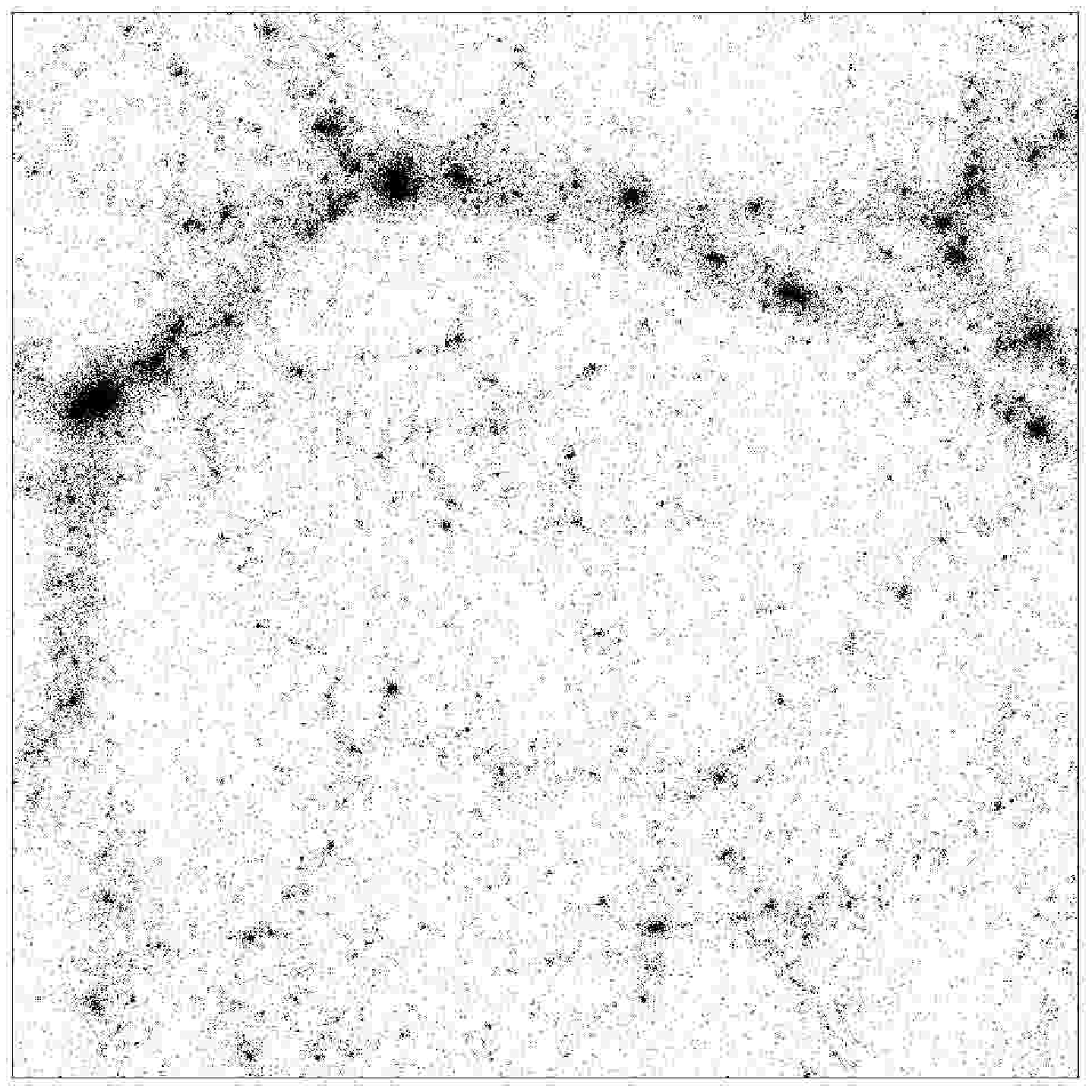}
\includegraphics[width=0.45\textwidth]{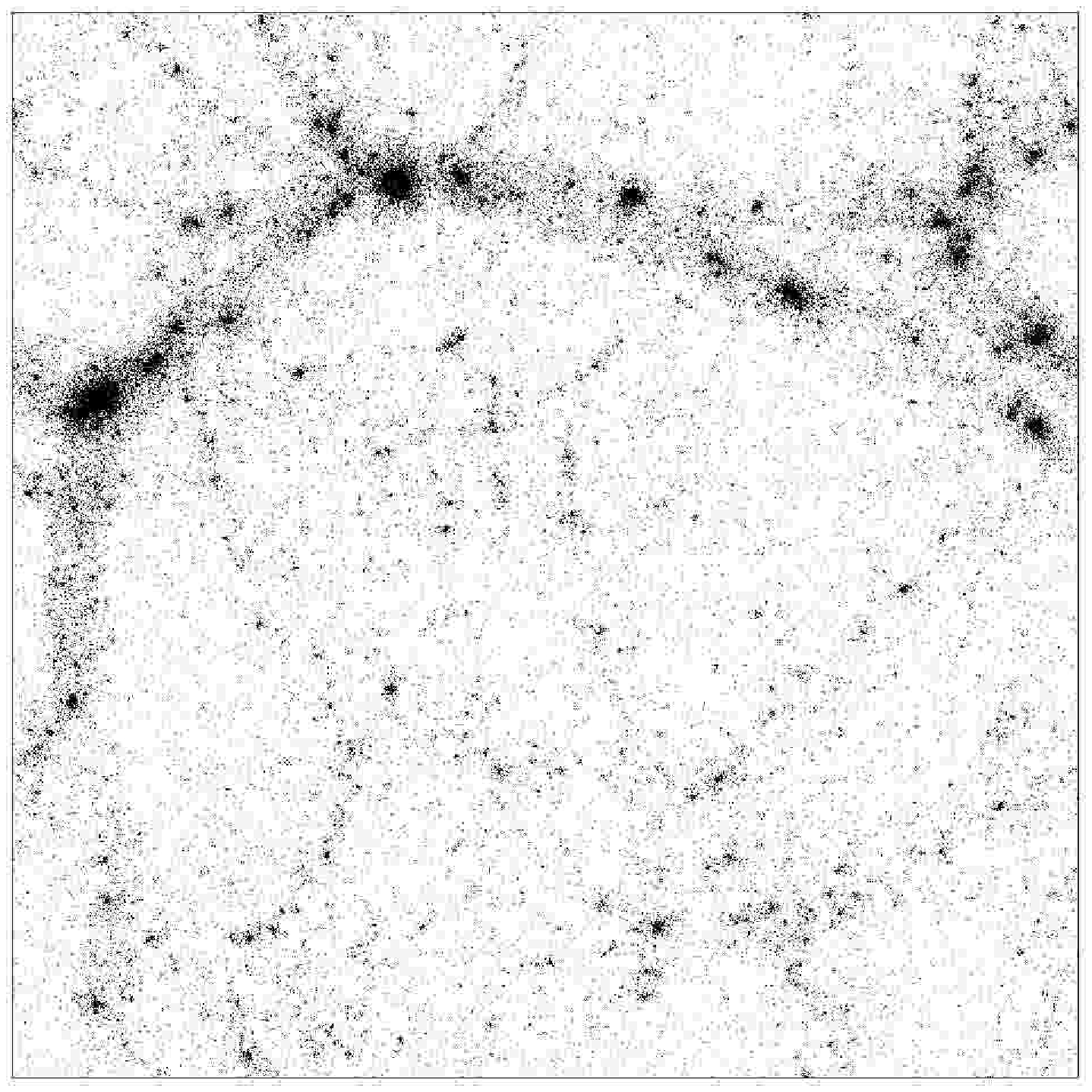}
\\
\includegraphics[width=0.45\textwidth]{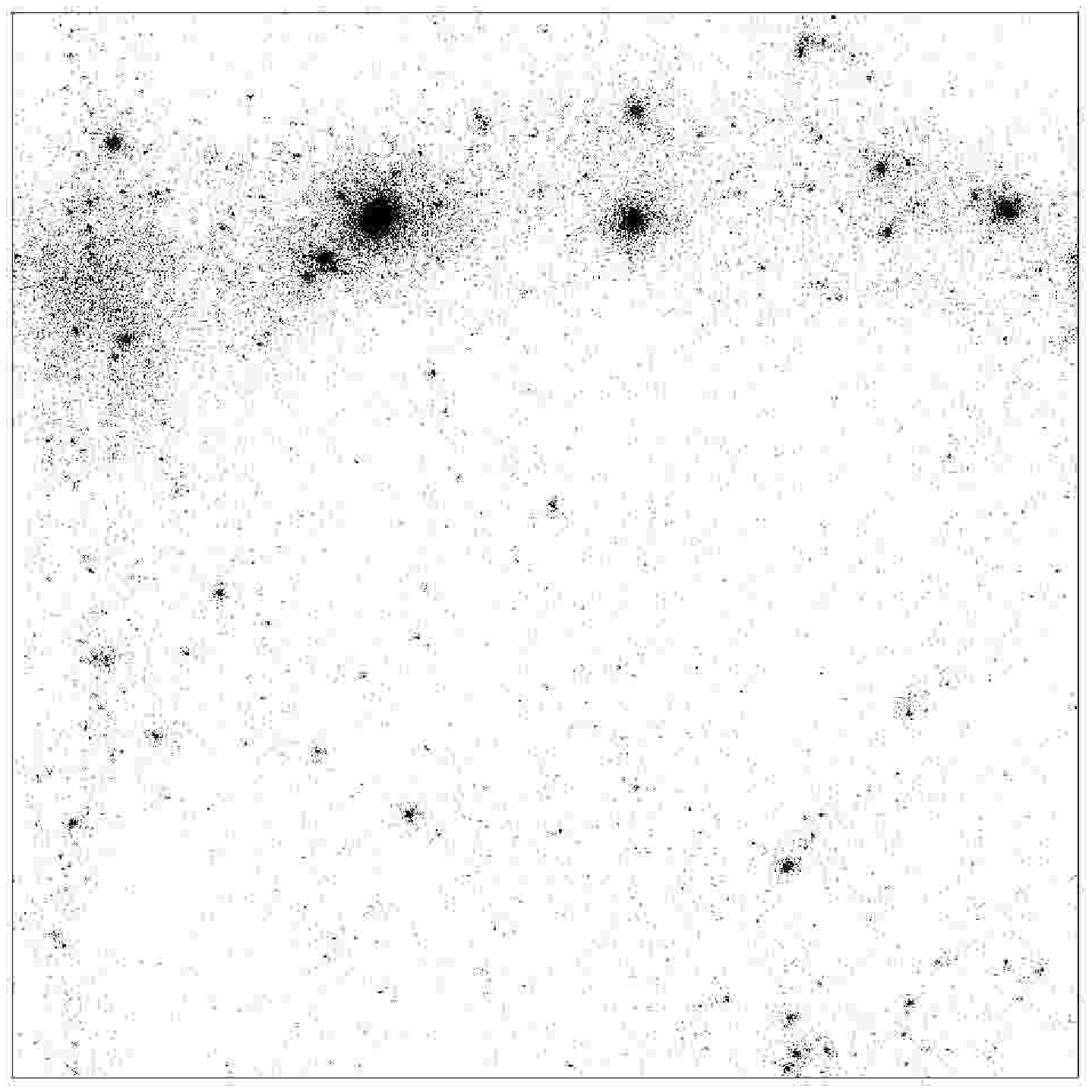}
\includegraphics[width=0.45\textwidth]{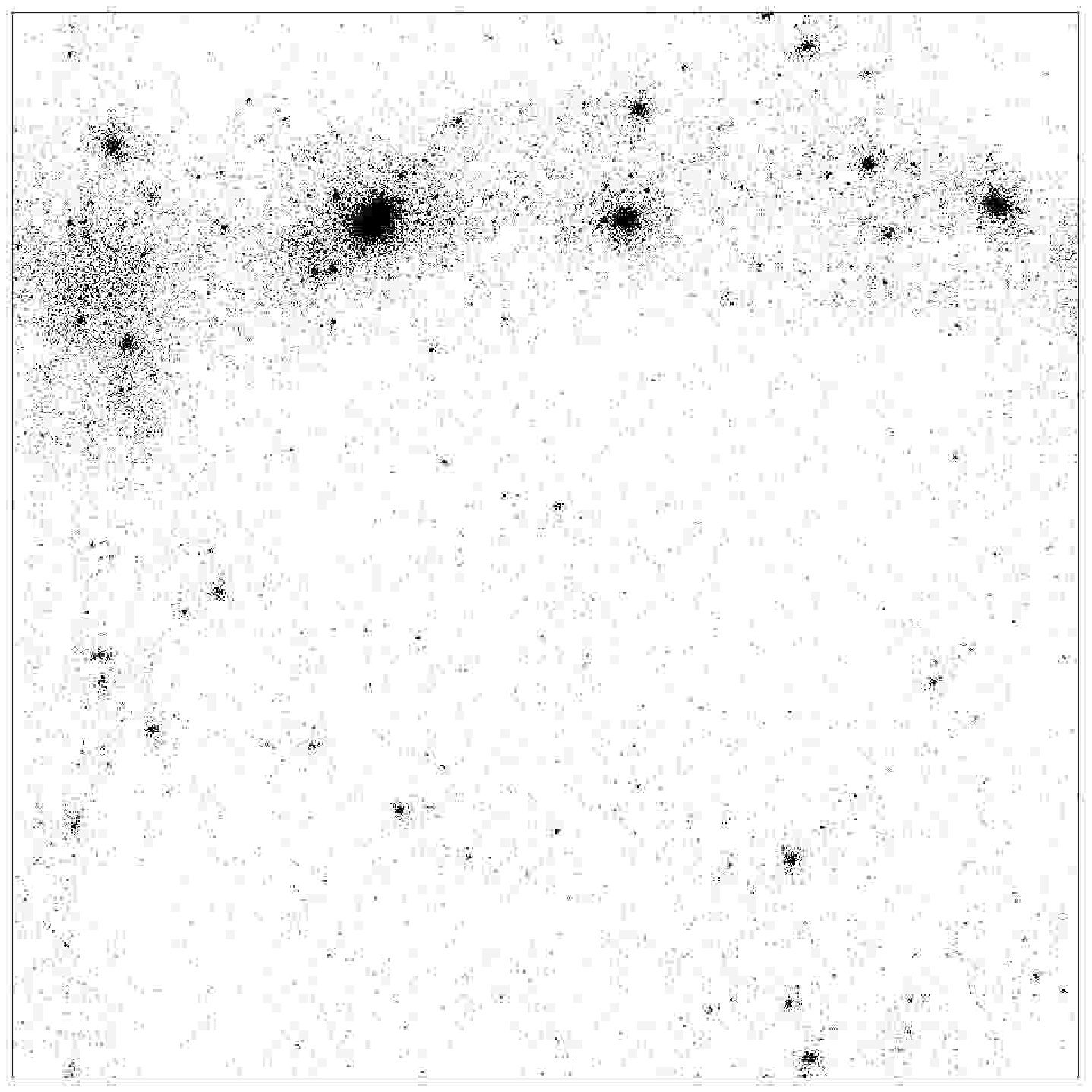}
\\
\includegraphics[width=0.45\textwidth]{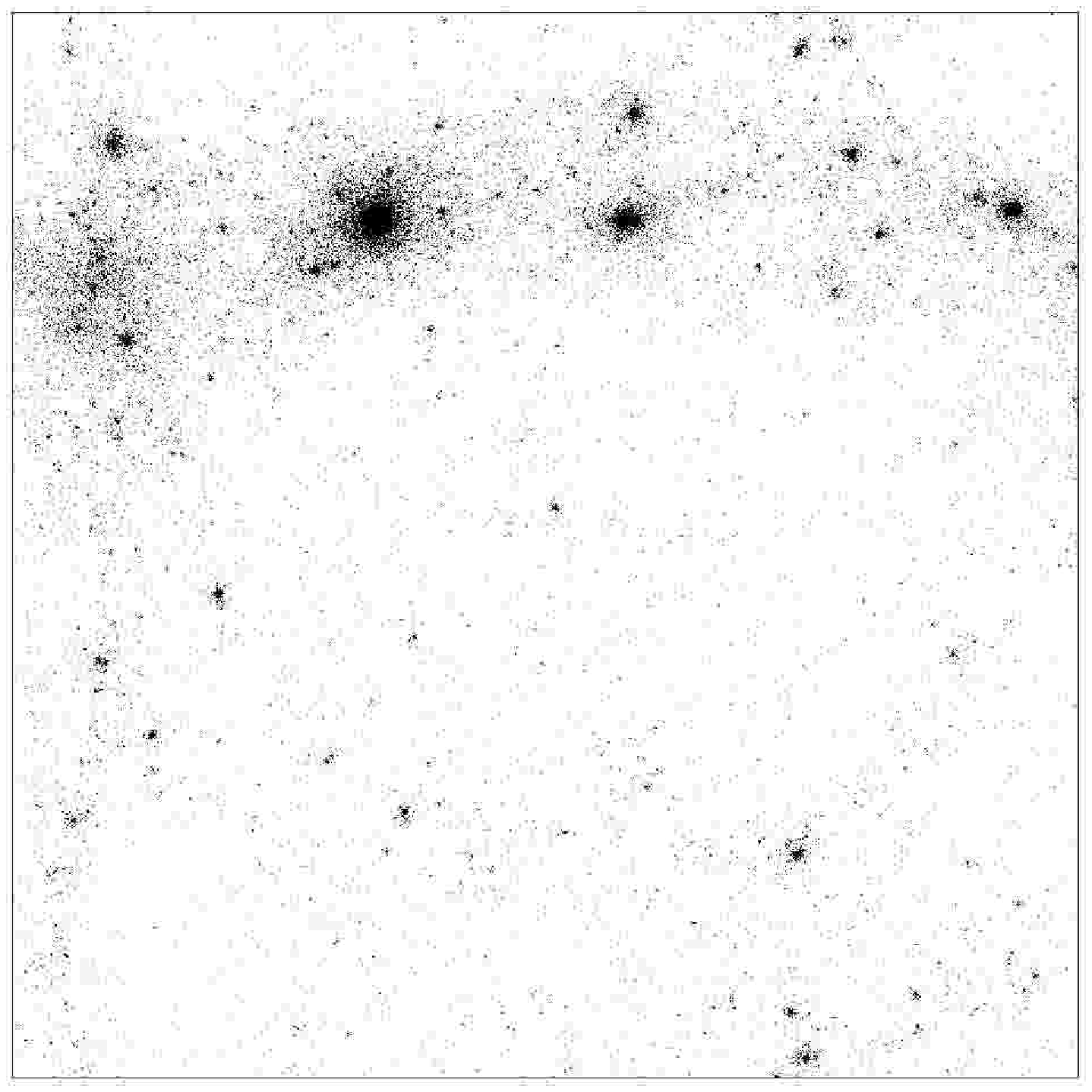}
\includegraphics[width=0.45\textwidth]{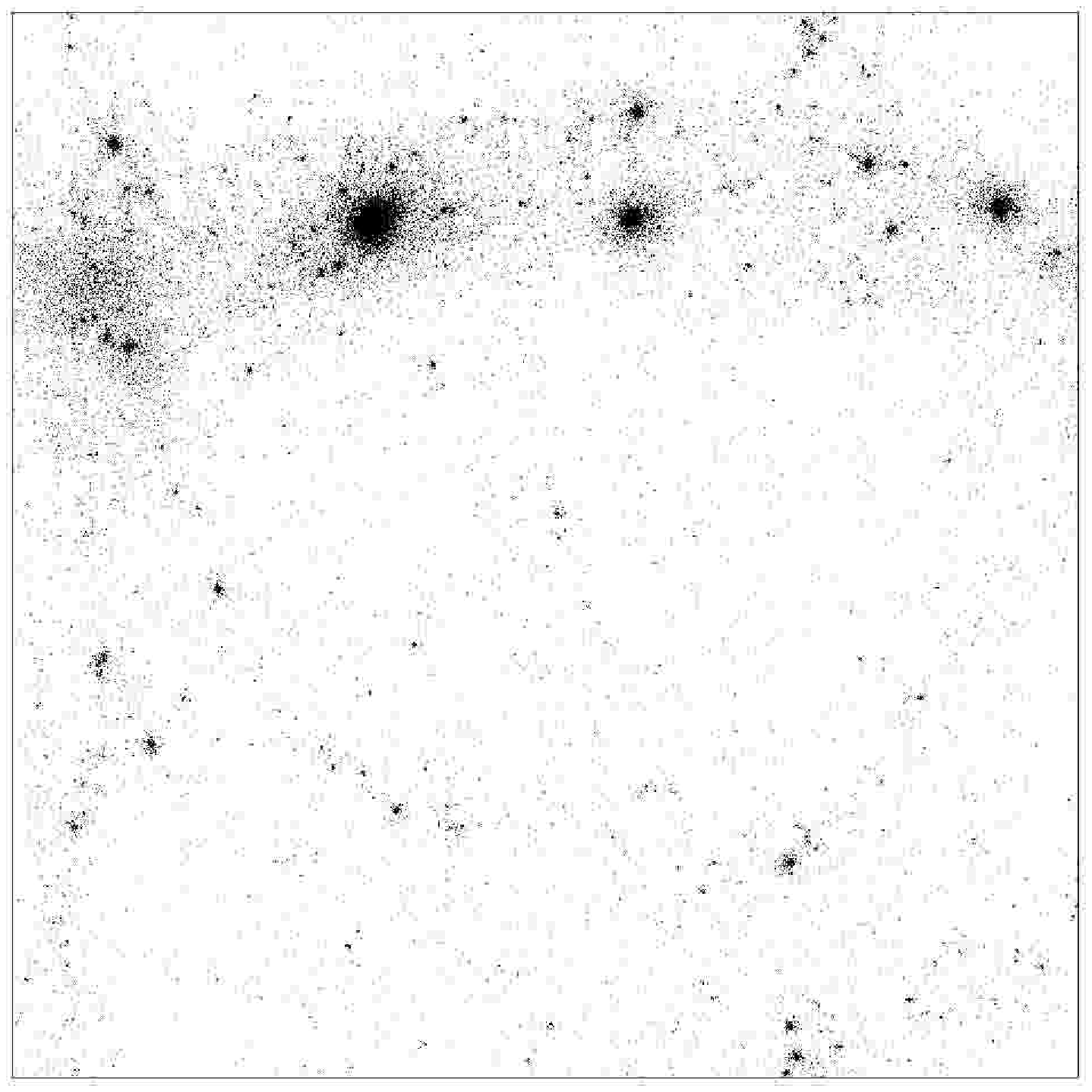}
\\
\caption{Snapshots of the evolved configurations at $a=2^5$ and $a=2^7$,
 in the same arrangement as in previous figure.
\label{snapshot_evol}}
\end{figure*}

\subsection{Snapshot inspection}

In Figs~\ref{snapshot0} and \ref{snapshot_evol}, we show snapshots,
for each of the four initial conditions, of a slice of depth $0.3L$
of the simulation box. The four snapshots correspond to $a=1$,
$a=2^3$, $a=2^5$ and $a=2^7$. In the initial conditions, at $a=1$, the
distributions look very different, reflecting the different small
scale properties, and long-range order, of the $preIC$
configurations. Blurring slightly one's vision, however, one can make
out clearly in the lattice configurations the very similar
superimposed fluctuations at larger scales. The glass looks very
different because it does not have the deterministic long range order
of the lattice, which makes the projection appear considerably
denser\footnote{In passing we underline that, contrary to what is
sometimes stated (e.g. \cite{wang+white_HDM}), the glass {\it is} a
long-range ordered distribution. In fact it has the property that
$P(k=0)=0$, which imposes the {\it global} constraint that the
integral of the two point correlation function is zero. Discussion of
the very particular stochastic long-range order of such
``superhomogeneous'' (or ``hyperuniform'' ) distributions, with
$P(k=0)=0$, may be found in \cite{glasslike,lebo_and_all,to03}. To
generate them starting from a Poisson distribution, as here, one
requires long-range correlation in the displacements of the particles,
provided here by the dynamics under Coulomb force (which rearranges
the points so that the fluctuations in any volume are proportional to
the surface i.e. sub-Poissonian).}. In the second slice, at $a=2^3$,
the first non-linear structures have formed, and already now the
visual impression is of a very strong resemblance in the
clustering. Distinct differences are however still evident. In
particular alignments inherited from the lattice configurations are
clearly visible, most evidently in the SC lattice.  In the next slice
at $a=2^5$ (which, as we will see below, is about the time at which
the largest modes included in the box go non-linear) the first visual
impression is of an even greater resemblance of the configurations,
but again closer inspection reveals differences at smaller
scales. Likewise in the last slice, when almost all the mass is in
just a few halos, the broad features at large scales are impressively
similar, while the spatial organisation of smaller structures reveals
evident differences.

\begin{figure*}
\psfrag{X}[c]{\large $k/k_N$}
\psfrag{Y}[c]{\large $P(k)$}
\psfrag{Y2}[c]{\large $\delta P^I(k)$}
\includegraphics[width=0.45\textwidth]{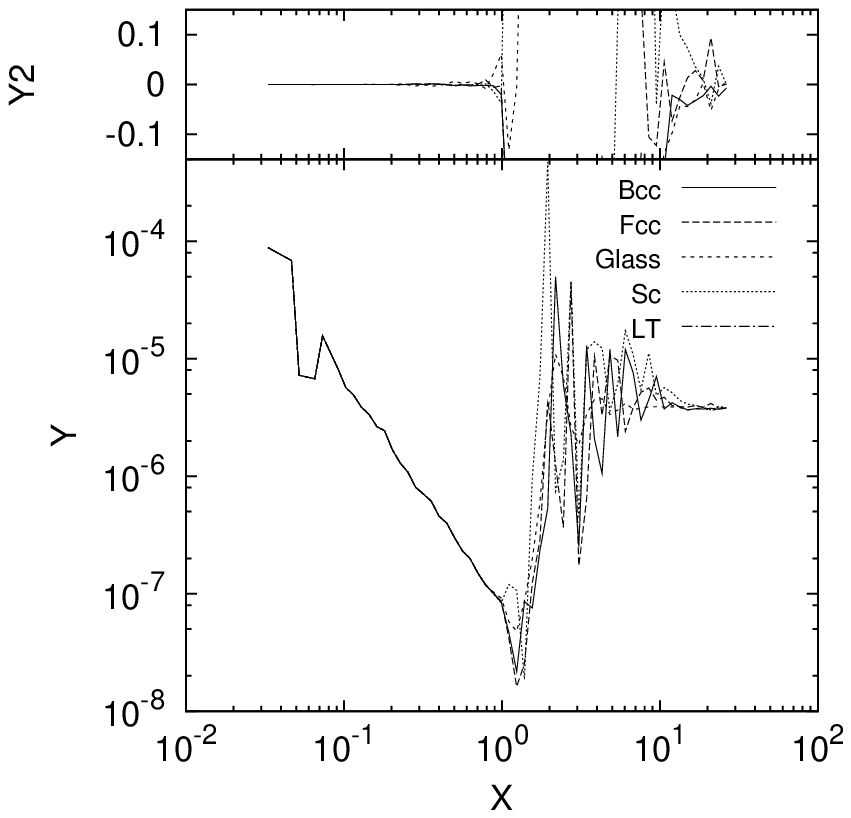}
\psfrag{X}[c]{\large $r$}
\psfrag{Y}[c]{\large $\xi(r)$}
\psfrag{Y2}[c]{\large $\delta \xi^I(r)$}
\includegraphics[width=0.45\textwidth]{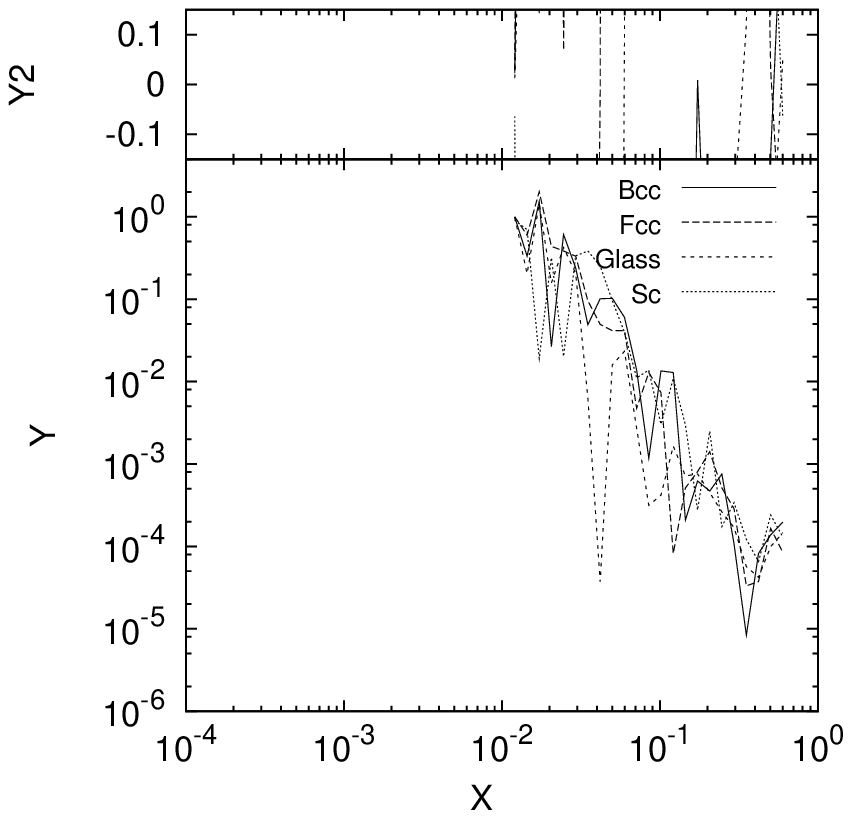} \\
\psfrag{X}[c]{\large $k/k_N$}
\psfrag{Y}[c]{\large $P(k)$}
\psfrag{Y2}[c]{\large $\delta  P^I(k)$}
\includegraphics[width=0.45\textwidth]{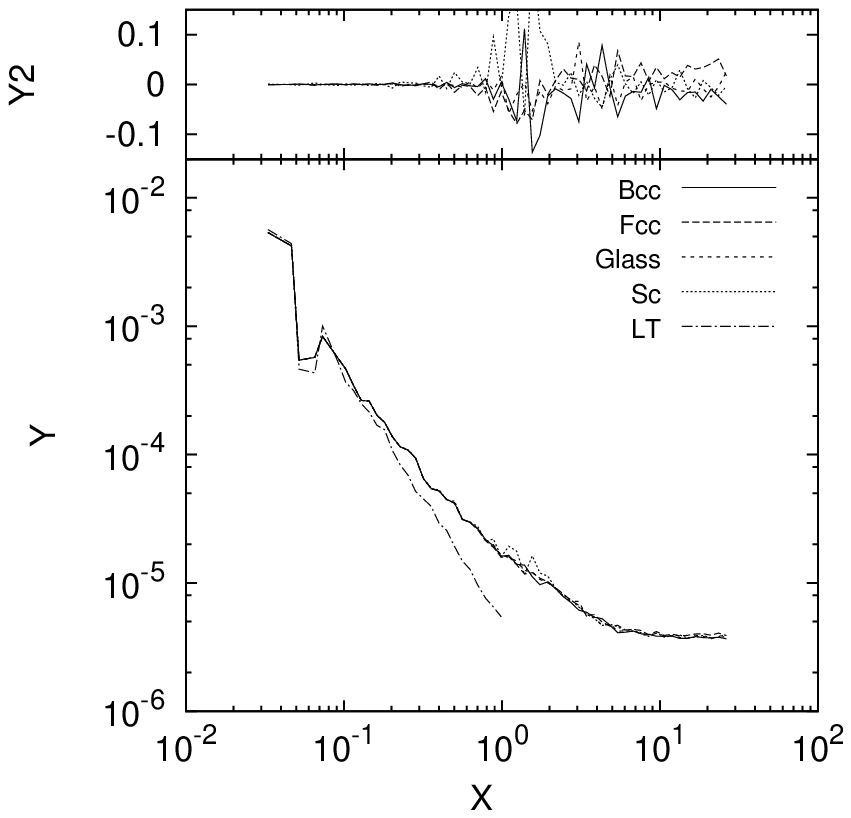}
\psfrag{X}[c]{\large $r$}
\psfrag{Y}[c]{\large $\xi(r)$}
\psfrag{Y2}[c]{\large $\delta \xi^I(r)$}
\includegraphics[width=0.45\textwidth]{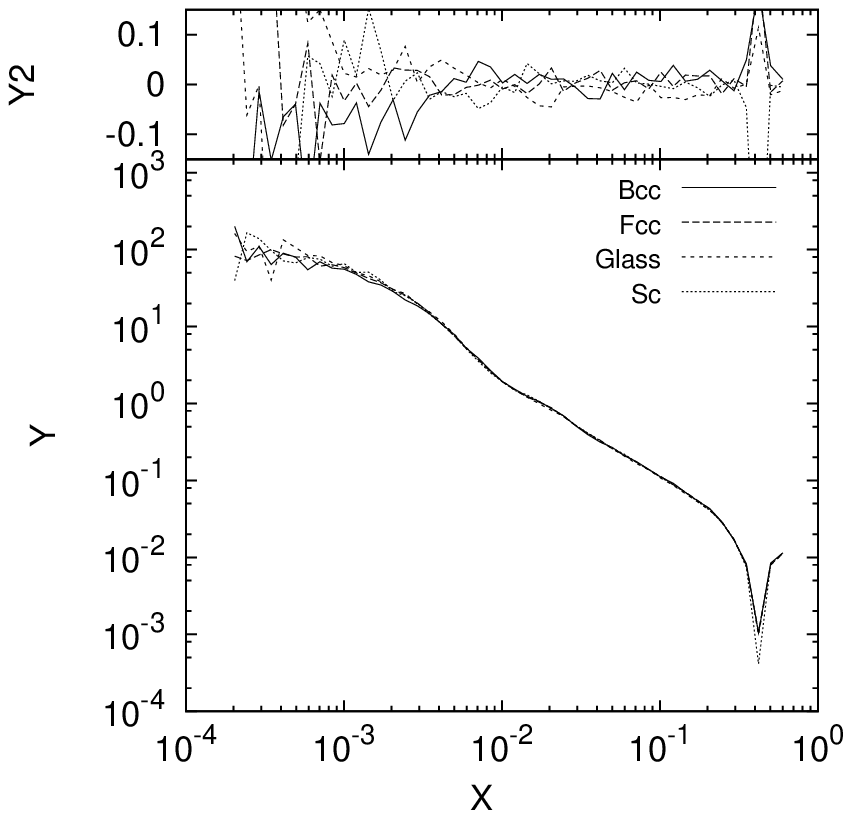} \\
\caption{PS (left column) and correlation function in real space
(right column) for $a=1$ (top row) and $a=2^3$ (bottom row).
 \label{corr0-3}}
\end{figure*}

\begin{figure*}
\psfrag{X}[c]{\large $k/k_N$}
\psfrag{Y}[c]{\large $P(k)$}
\psfrag{Y2}[c]{\large $\delta P^I(k)$}
\includegraphics[width=0.45\textwidth]{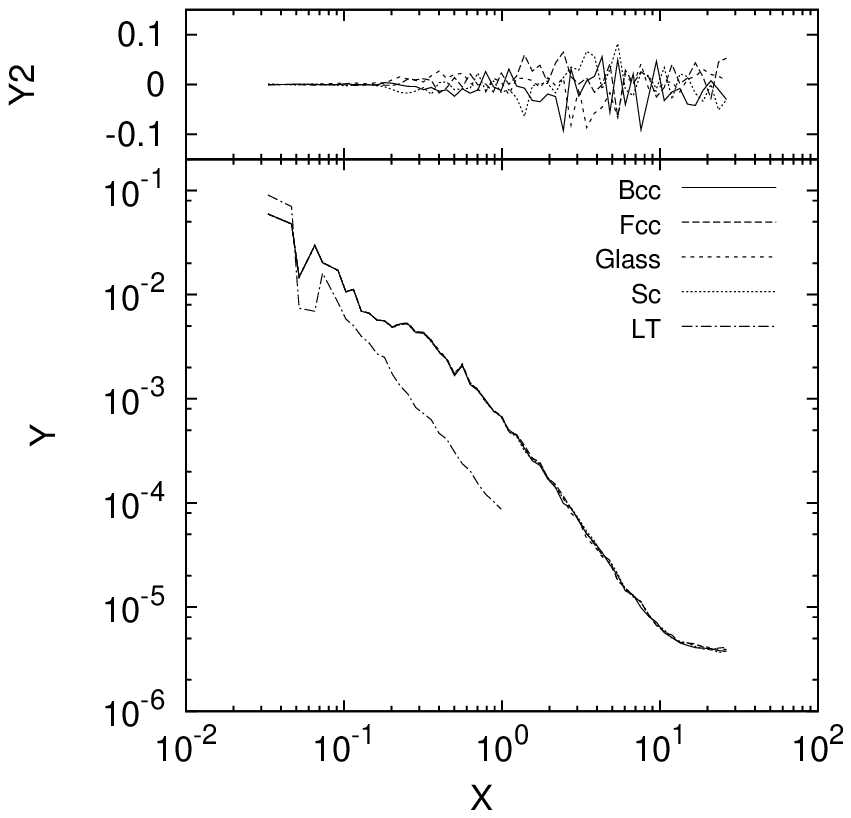}
\psfrag{X}[c]{\large $r$}
\psfrag{Y}[c]{\large $\xi(r)$}
\psfrag{Y2}[c]{\large $\delta \xi^I(r)$}
\includegraphics[width=0.45\textwidth]{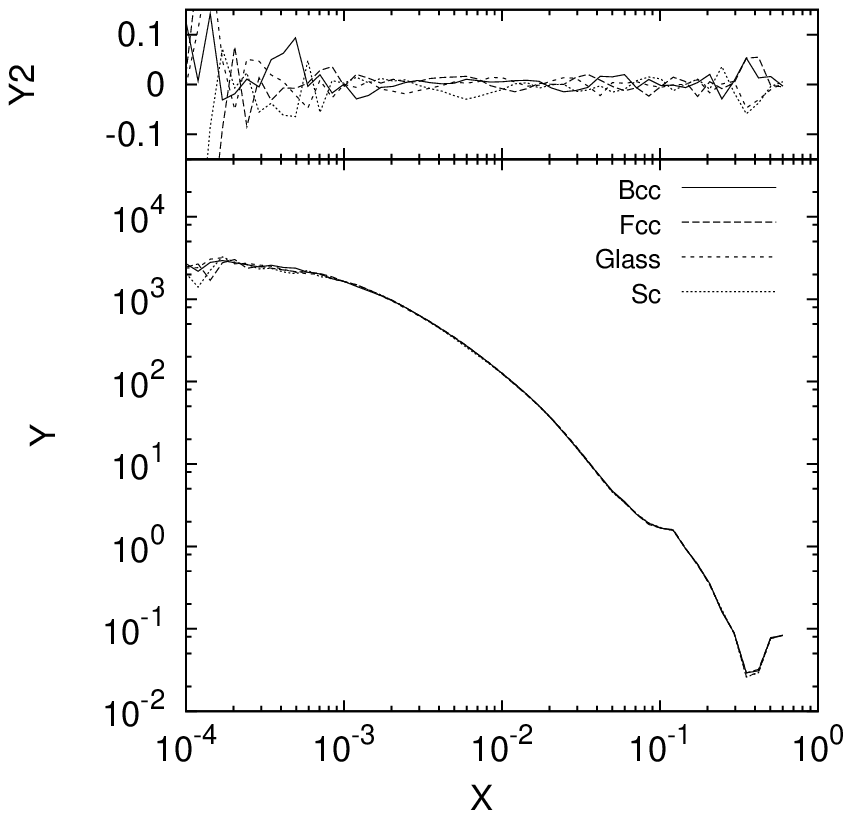}
\\
\psfrag{X}[c]{\large $k/k_N$}
\psfrag{Y}[c]{\large $P(k)$}
\psfrag{Y2}[c]{\large $\delta P^I(k)$}
\includegraphics[width=0.45\textwidth]{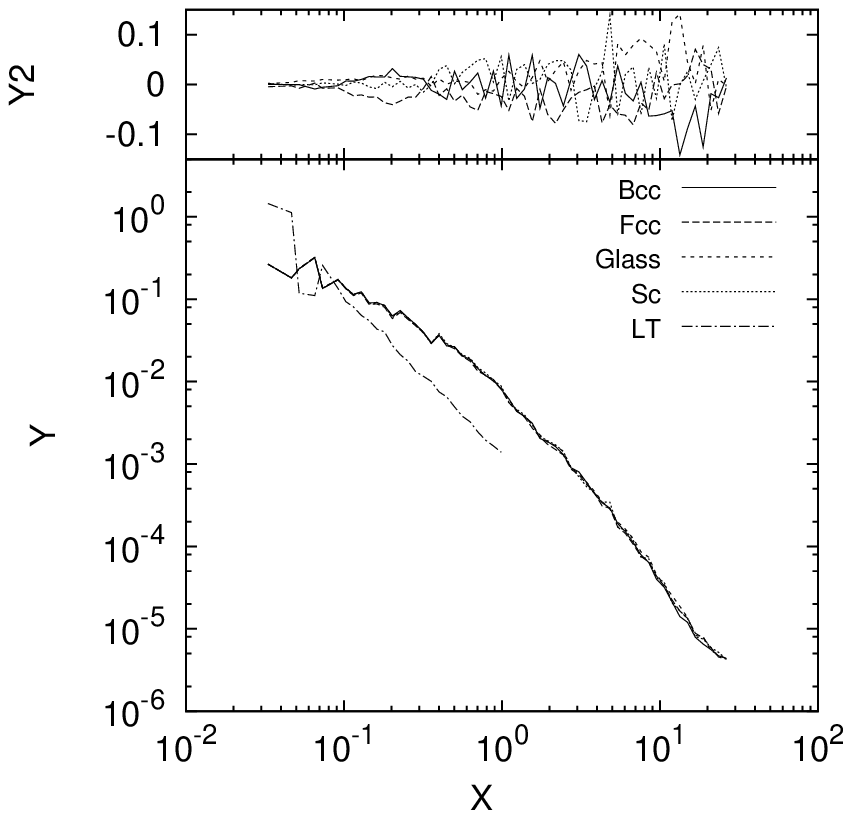}
\psfrag{X}[c]{\large $r$}
\psfrag{Y}[c]{\large $\xi(r)$}
\psfrag{Y2}[c]{\large $\delta \xi^I(r)$}
\includegraphics[width=0.45\textwidth]{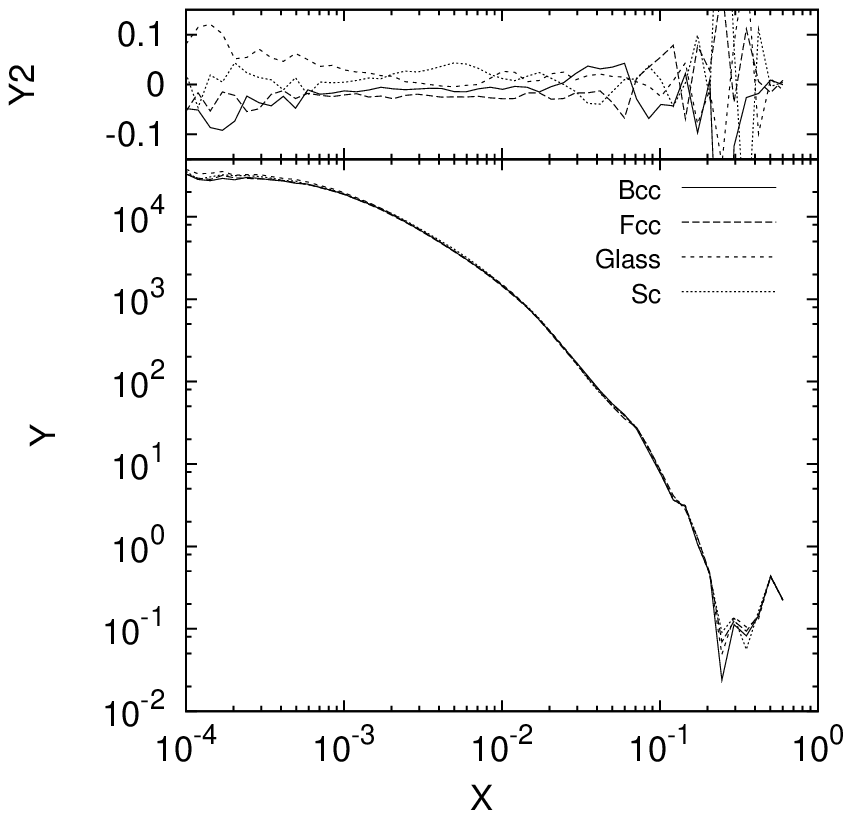}
\\
\caption{PS (left column) and correlation function in real space
(right column) for $a=2^5$ (top row) and $a=2^7$ (bottom row).
 \label{corr5-7}}
\end{figure*}

Some of the differences observed visually in
the earlier time snapshots are manifestly related to the subtle 
differences in the initial conditions, and are therefore clearly
discreteness effects. The differences in the more evolved snapshots
are, however, not necessarily indicative of anything other than 
the intrinsically chaotic dynamics of the non-linear regime 
of the evolution\footnote{For a discussion
of chaos in $N$-body self-gravitating systems see, e.g.,
\cite{sideris+kandrup}.}. What we are interested
in, and will now examine, are differences in 
the {\it statistical} properties of these 
distributions, which are what we use them to infer in cosmology.

\subsection{Power spectrum and correlation function}

Let us consider more quantitatively the differences in the 
two point properties of these distributions.  
In Figs.~\ref{corr0-3} and \ref{corr5-7} 
we plot the reduced two point correlation function 
$\xi(r)$ and the PS $P(k)$, for a series of four different
time slices\footnote{Details of how these quantities have
been estimated are given in Appendix~\ref{app-estimation}.
Note that we write the two point correlation function and
the PS as functions only of the modulus of their arguments
as an average over spherical shells in real and reciprocal
space respectively is performed.}. Also shown, in an inset 
panel in each case, are the normalised residuals of each 
quantity with respect to the average, i.e., for a quantity 
$C^I(i) $ in the $i$-th bin
(of $k$ or $r$) in the simulation of initial conditions
$I$ ($I$=SC, BCC, FCC, glass) the residual is 
\be
\label{residuals}
\delta C^I (i)=\frac{4 \,C^I (i)-\sum_{I} C^I (i)}
{\sum_{I} C^I (i)}\,.
\ee
Also shown in each case is the ``linear theory'' (LT)
prediction for the evolution of the theoretical PS, i.e., the 
initial theoretical PS multiplied by $a^2$.

These results reflect broadly the impression gained by visual
inspection above. In particular inspection of the correlation
function shows four distributions which are apparently very
different at the initial time evolve to closely resemble one 
another already at $a=2^3$. Note indeed that, at this time
and the subsequent ones, the correlation functions are so
similar as to be indistinguishable in the main plot down
to about $r=10^{-3}$, which coincides with our chosen
$\varepsilon$. In reciprocal space the resemblance of 
the initial conditions --- the fact that they represent
exactly the same realisation of the input PS --- can be
seen. Indeed the initial power below $k_N$ agrees in all 
cases to a precision of less than a small fraction of a percent. 
Already in the next time slice shown, at $a=2^3$, the PS in
the main plot of the four distributions
are superimposed almost perfectly over the entire range,
except for a still visible difference for the sc
configuration, corresponding to the difference we identified
by visual inspection. In the last time slices the curves
for the PS are again, as in the case of $\xi(r)$,
effectively indistinguishable over, in the final slice,
more than four orders of magnitude in power. Note that 
in our length units the asymptotic Poissonian behaviour 
(of {\it any} translationally 
invariant stochastic point process) corresponds to 
$P(\bk \rightarrow \infty)=1/n_0=1/64^3 \approx 4 \times 10^{-6}$.
When the PS asymptotes to this value it indicates that
the PS measured is dominated by the intrinsic noise of
the discrete process. We see that this maximal resolved
wavenumber propagates to larger $k$ in time, reaching a final value
of order $\pi/\varepsilon$ (i.e. $k/k_N=\ell/\varepsilon$).

\begin{figure}
\psfrag{Y}[c]{\large $\si_{PI}(k,a)/\overline{P}(k,a)$}
\psfrag{X}[c]{\large $k/k_N$}
\includegraphics[width=0.45\textwidth]{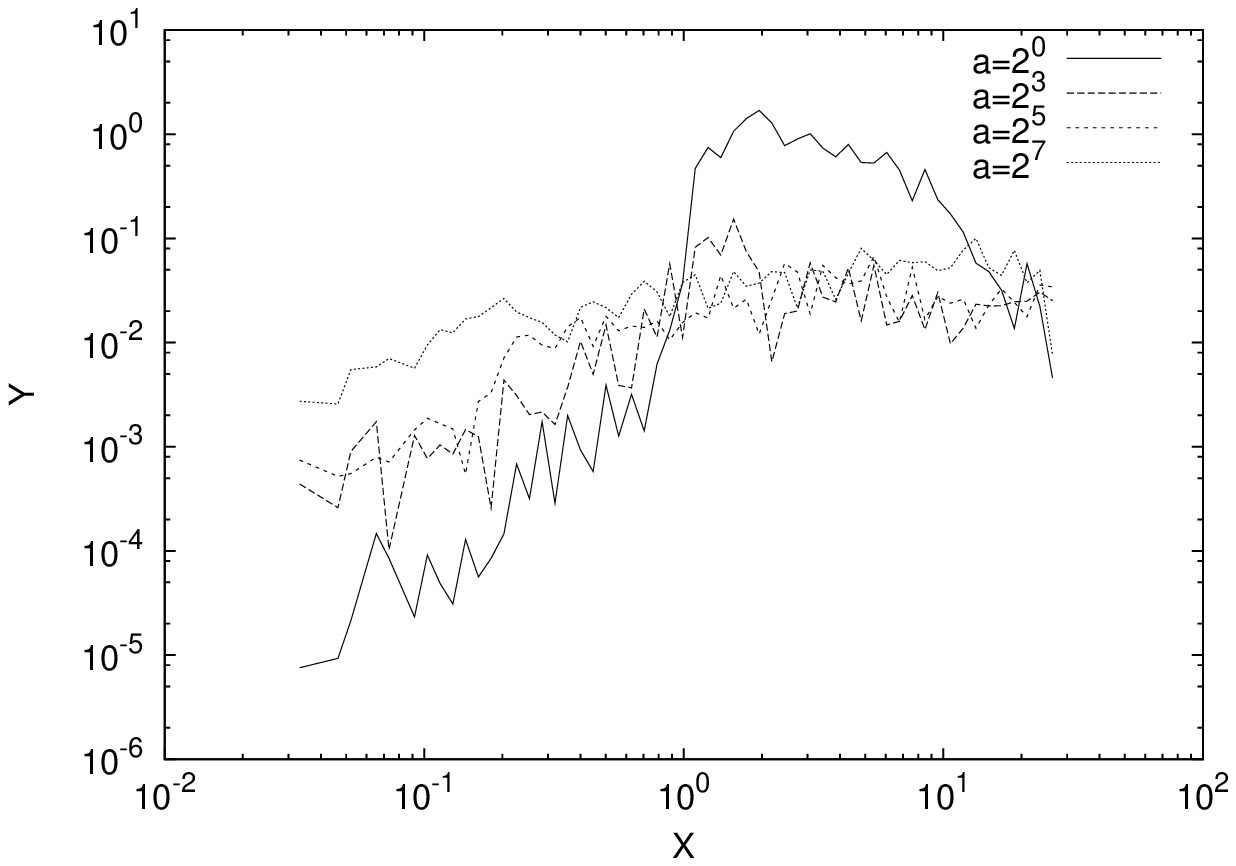}
\caption{Normalized variance of the PS $\si_{PI}(k)$, defined in Eq.~\eqref{variance_pi}.
\label{variance_basic}}
\end{figure}

This first view of these results thus supports very strongly 
the ``common wisdom'': there is very efficient transfer of
power from large to small scales which wipes out memory
of the differences between the initial conditions at small
scales. Rapidly results converge down to a scale characterised
by $\varepsilon$. There is no significant dependence on $preIC$,
and therefore the associated discreteness effects are wiped out. 

A more careful analysis of the evolution of the residuals
in each plot shows, however, some behaviours which are 
unexpected according to this common wisdom. Firstly,
{\it for $k<k_N$} we see differences in the PS which 
appear and {\it grow monotonically with $a$}, i.e., the gravitational 
evolution appears to produce some small differences at 
large scales which were not present initially. 
While such an effect is predicted by PLT, as documented 
in detail in \cite{discreteness2_mjbm} and described
briefly above, this is valid only at sufficiently early
times\footnote{Note that since the residuals are normalised, 
they are constant under fluid LT.}. Thus the 
expectation that non-linear transfer
of power from larger scales may wipe out the effects of
PLT at these times appears not to be correct. For
$k>k_N$, on the other hand, such an effect is indeed
apparent, but appears only to be operative at the very 
earliest times when the non-linear structures first 
develop at smaller scales. Indeed, between $a=2^3$ and 
the final plot, at $a=2^7$, there appears to be little 
evidence for any further washing out of the residuals 
in the power. On the contrary they appear to grow, giving 
a dispersion of order several percent at the final time.

These results are clearly illustrated in Fig.~\ref{variance_basic},
which shows the square root of the variance of PS calculated at each
$k$, and each of the four times as indicated, over the four
realizations, i.e.,
\be
\label{variance_pi}
\si_{PI}(k,a)=\left\{\frac{1}{m-1}\sum_{I=1}^m \left[P_I(k)-\overline{P}(k)\right]^2\right\}^{1/2}
\ee
where $I$ labels the $m=4$ different $preIC$, and we have defined
\be
\label{averaged_P}
\overline{P}(k)=\frac{1}{m}\sum_{I=1}^m P_I(k).
\ee
Given that the amplitude of the effects are so small ---
at small $k$ and early times in particular --- we evidently need 
to be careful in interpreting these differences as resulting
from the physical discreteness effect we set out to measure.
We could envisage that such a time and space dependent 
dispersion could be the produced, in particular, by numerical
effects in the evolution or by statistical effects in the
estimators. For example, it is conceivable that there is
an interplay between the numerical errors relating to the
calculation of the force and each particular initial condition,
or that the variance measured is simply a statistical variance
which would decrease if we took more particles (i.e. a larger
box at the same particle density). In the rest of this section
we examine this question carefully, establishing --- we 
believe very convincingly --- that, at least up to the 
slice at $a=2^5$, this measured dispersion is a discreteness
effect. 

\subsection{Numerical convergence}

Let us first consider the stability of the results with respect to
variation of the {\it numerical} parameters, i.e., those controlling the
accuracy of the numerical integration of the $N$-body equations 
at given values of the discreteness parameters ${\cal D}_\alpha$.
In the GADGET2 $N$-body code, there are two sets of such parameters:
a first set controlling the time-stepping and a second one the 
resolution in the calculation of the force. In Appendix~\ref{app-gadget}
we give the full details of two sets of parameter choices for
which we now compare results: a ``low resolution'' (LR) simulation,
corresponding to the values used in obtaining the results given
above and subsequently in the paper, and a ``high resolution'' (HR)
simulation. As we discuss in further details in Appendix~\ref{app-gadget},
the LR are typical choices for large cosmological simulations in 
the literature (e.g. those of the VIRGO consortium, as described in
\cite{VIRGO}), while our HR values are even more stringent choices
than typically used in similar convergence tests reported in the literature
(e.g. \cite{crocce_06}). 

\begin{figure}
\psfrag{Y}[c]{\large $\frac{2|P_{bcc}(k)-P_{sc}(k)|}{(P_{bcc}(k)+P_{sc}(k))}$}
\psfrag{X}[c]{\large $k/k_N$}
\includegraphics[width=0.45\textwidth]{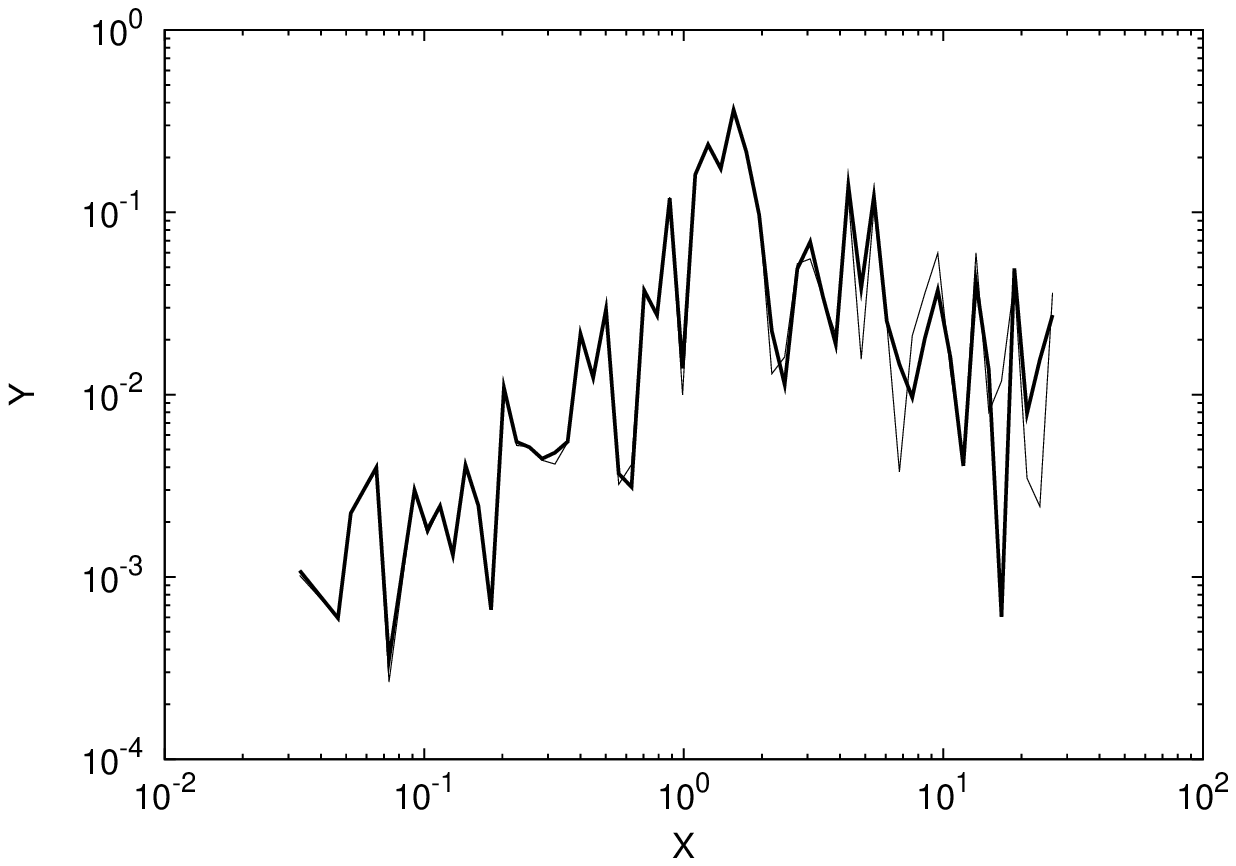}\\
\includegraphics[width=0.45\textwidth]{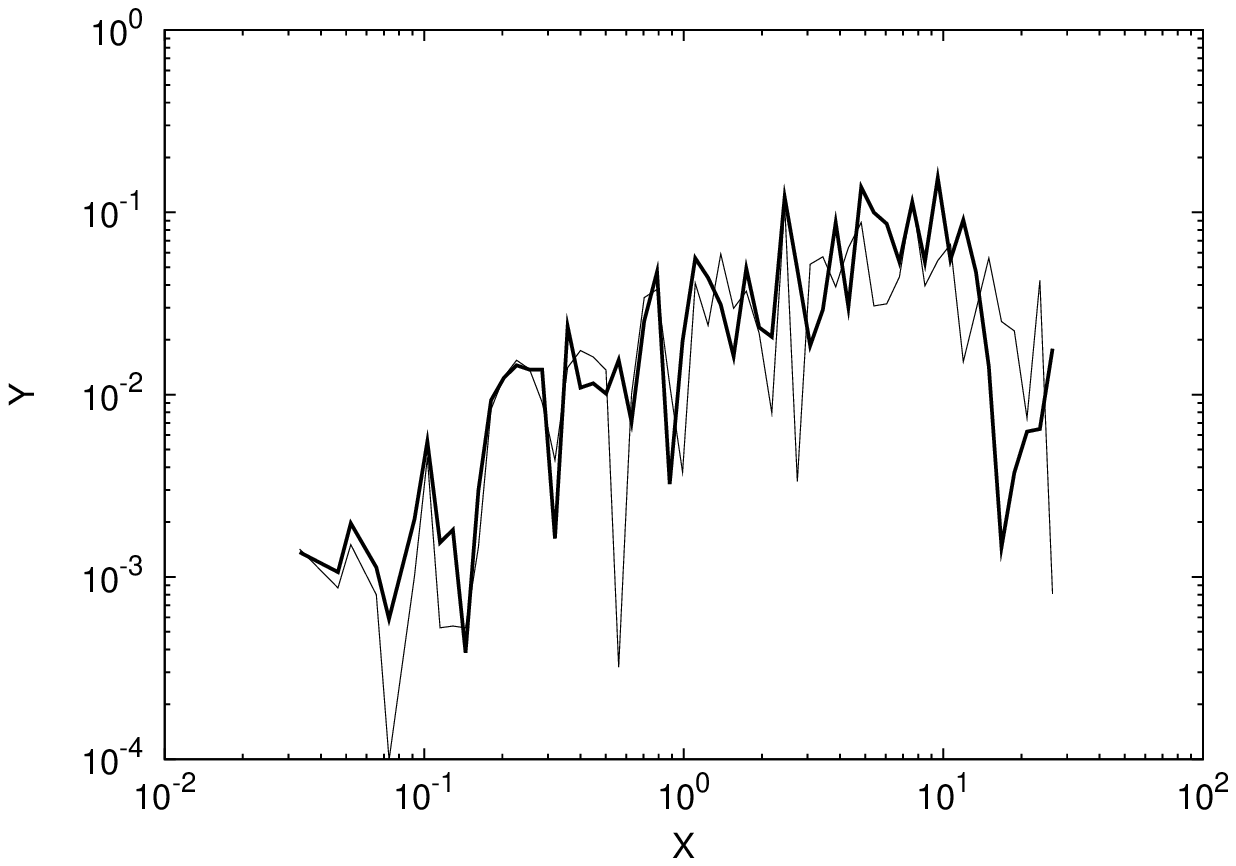}\\
\includegraphics[width=0.45\textwidth]{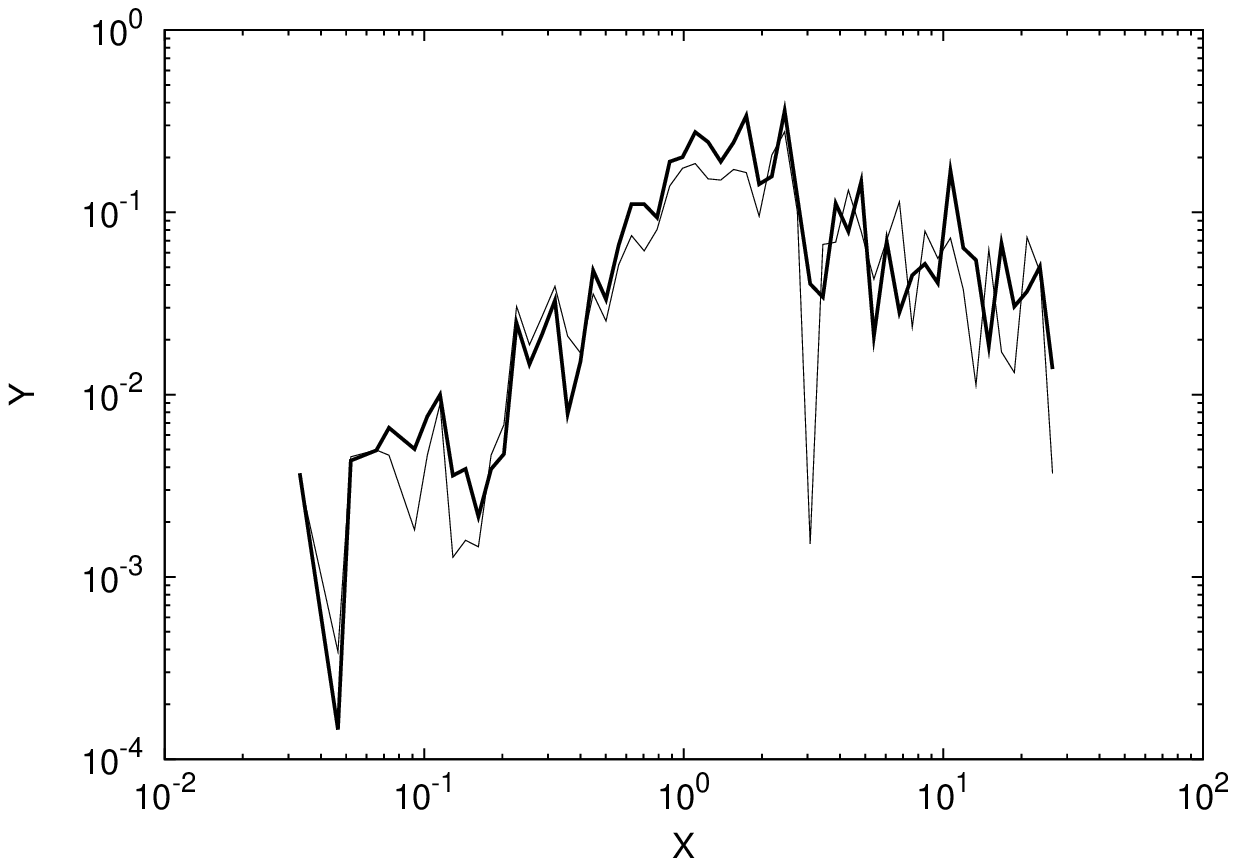}
\caption{Normalized differences of the PS for the BCC and
SC $preIC$ configuration, for high resolution simulations (thick
lines) and low resolution simulations (thin lines) at $a=2^3$, $a=2^5$ and $a=2^7$.
\label{fig_resol1}}
\end{figure}

\begin{figure}
\psfrag{Y}[c]{\large $\frac{2|P_{I}^{\rm HR}(k)-P_{I}^{\rm LR}(k)|}{(P_{I}^{\rm HR}(k)+P_{I}^{\rm LR}(k))}$}
\psfrag{X}[c]{\large $k/k_N$}
\includegraphics[width=0.45\textwidth]{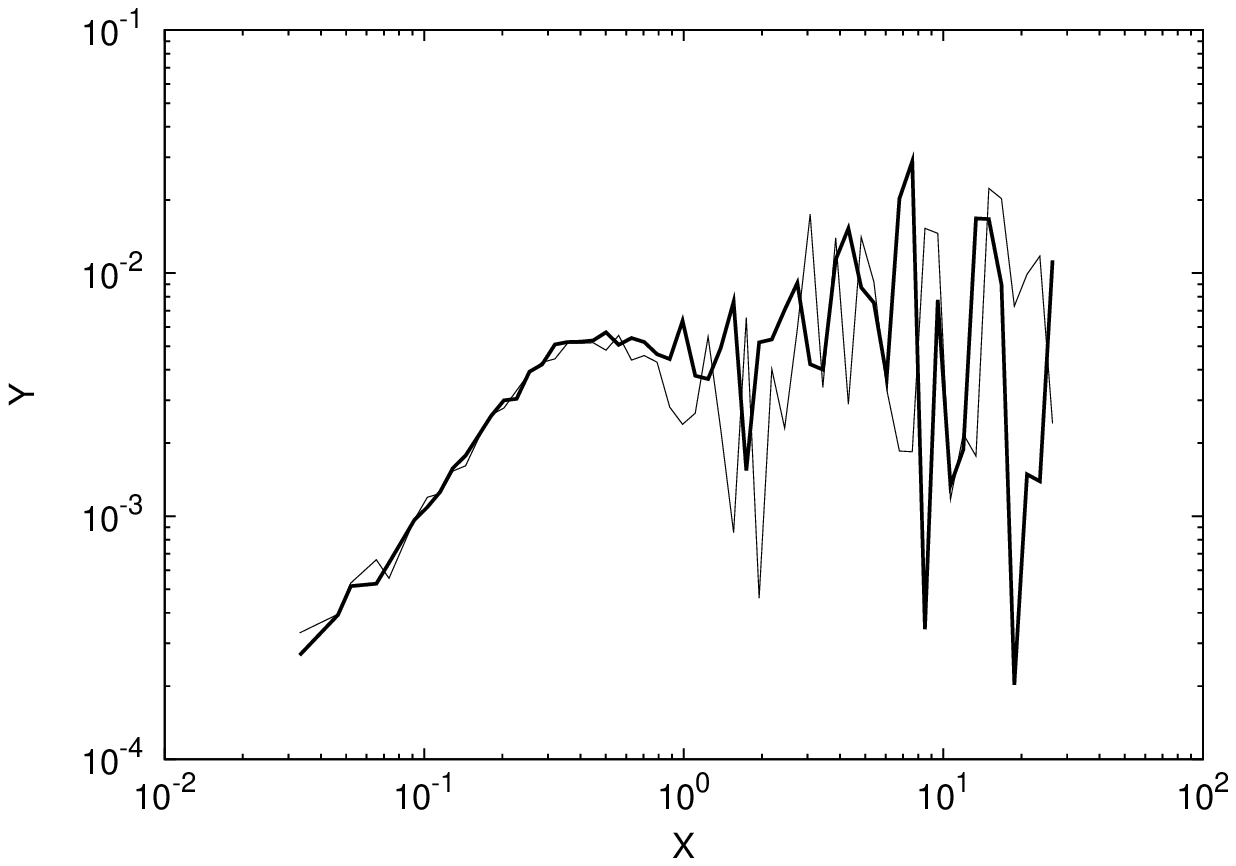}\\
\includegraphics[width=0.45\textwidth]{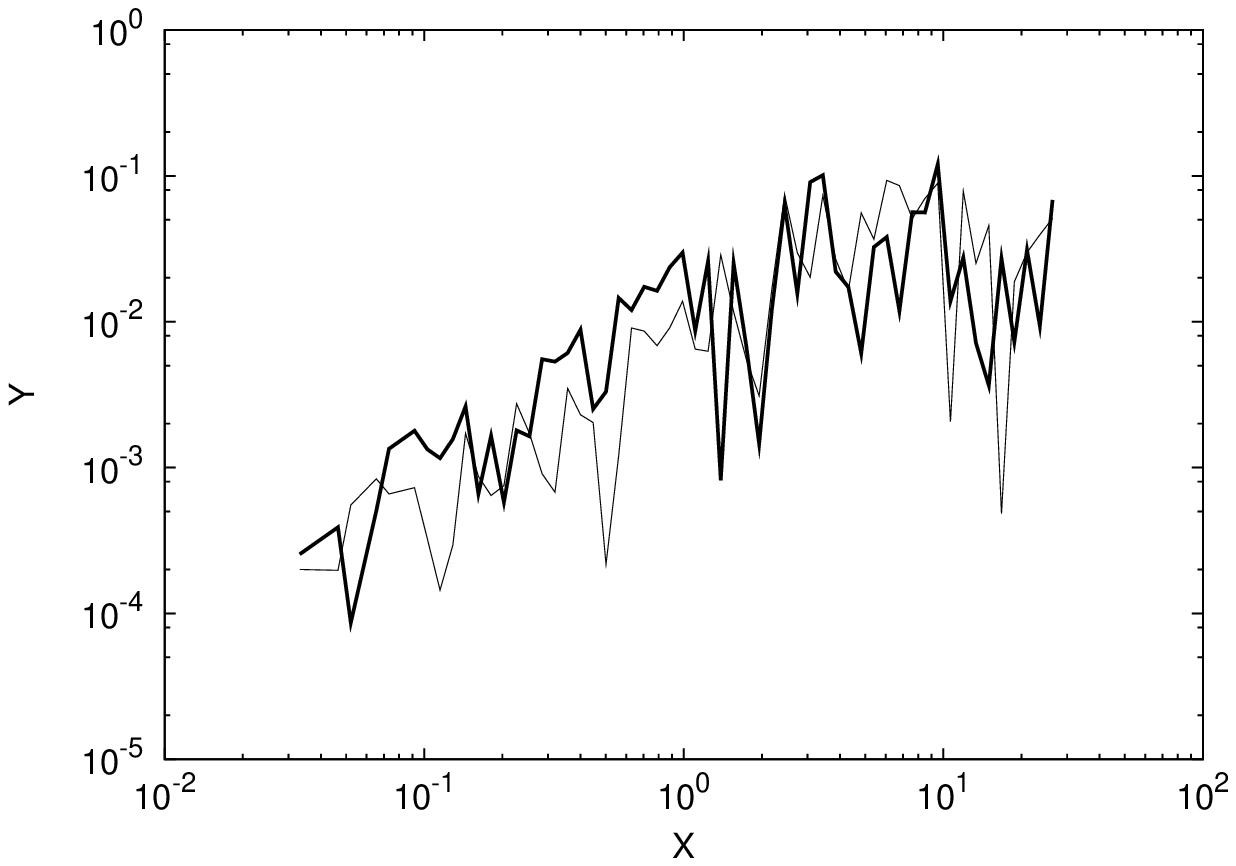}\\
\includegraphics[width=0.45\textwidth]{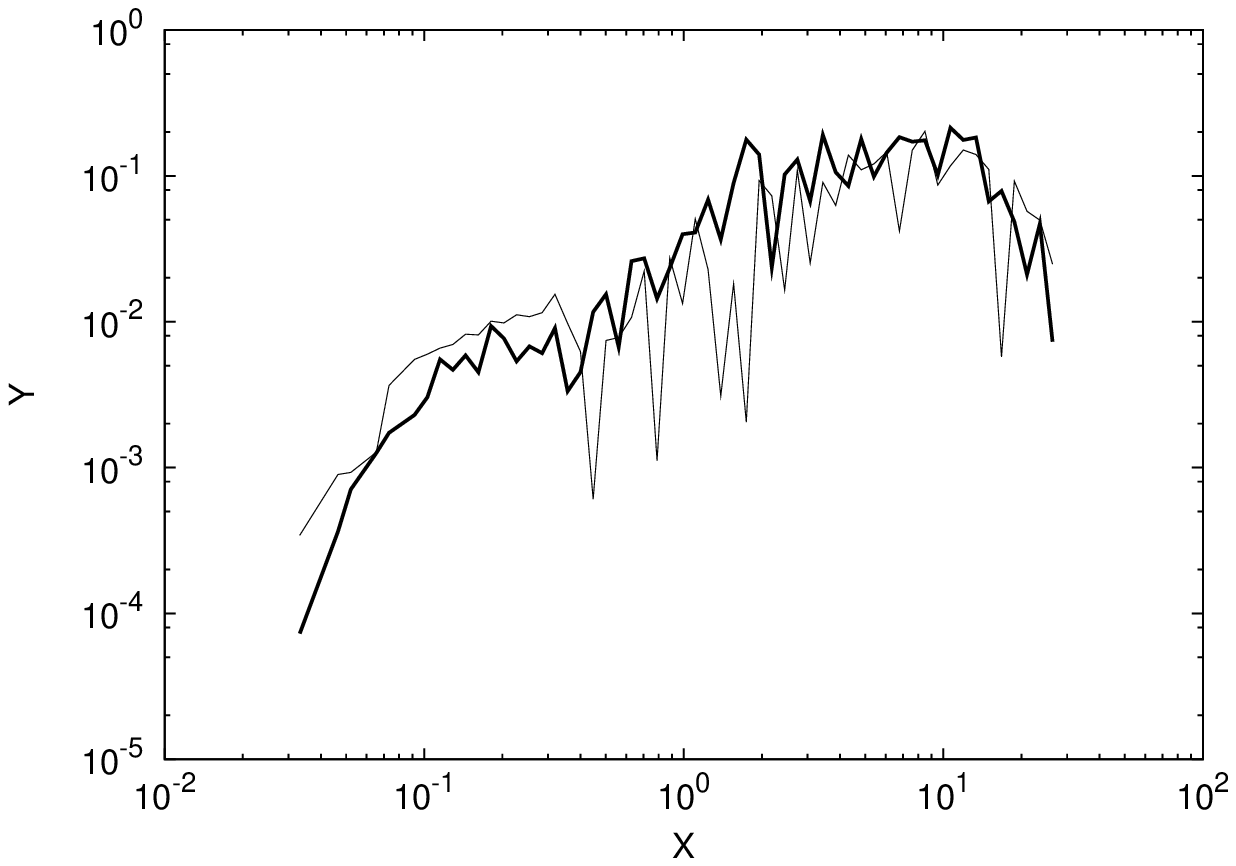}
\caption{Normalized differences of the PS of HR and LR
 runs, at $a=2^3$, $a=2^5$ and $a=2^7$. In each panel the thick lines 
corresponds to the BCC $preIC$ configuration and the thin one to 
the SC one. 
\label{fig_resol2}}
\end{figure}

We illustrate the degree of convergence between the LR and HR
simulations in Figs.~\ref{fig_resol1} and \ref{fig_resol2}. The former
shows the excellent stability of normalized differences like those we
have considered above --- and will focus on in the rest of the paper
--- in the PS for the two $preIC$ indicated. The latter figure shows,
on the other hand, the differences of the results of the LR and HR
simulations for the full PS in each of two $preIC$ taken separately.
We see that these differences are, at the two later times, comparable
in magnitude to the differences we measure (in the previous figure),
over a part of the range of $k$. Thus the full PS measured in each of
the two $preIC$ simulations changes as a function of the numerical
parameters in this range by as much as the differences between them
which we are studying here (and which we have just seen to be well
converged numerically). This means simply that the numerical errors
associated with these changes in parameters are correlated strongly
with the full PS, which is very close to the same in the two cases,
and so cancel out when we take the difference.  This suggests that, in
more general, it may be easier to place this kind of lower bound on
discreteness effects than to attain a comparable level of numerical
convergence on other quantities (such as the full PS).

\subsection{Comparison with PLT}

The PLT formalism for the evolution of the displacements off
the lattice, developed explicitly in \cite{joyce_05, marcos_06} for 
the SC lattice, has been generalized in \cite{marcos07} to both
BCC and FCC lattices. We exploit these analytical results here,
for the case of the SC and BCC lattice, as a control on the
accuracy of our numerical simulations at sufficiently early 
times when PLT is a valid approximation. Conversely this comparison
can be seen --- given the results just shown above on the
numerical convergence of our results --- as a check on the range
of applicability of PLT. We will see that this range turns out to 
be considerably greater than that which was established in the
studies in \cite{marcos_06}, making PLT a very 
useful tool for calibrating numerical results. 

To compare our numerical results with PLT we simply generate,
for each set of BCC and SC initial conditions, the configurations 
given by PLT evolution of Eq.~(\ref{eigen_evol}), where the
eigenvalues and eigenvectors are those for the corresponding
lattice. The details of these latter calculations may be
found in \cite{marcos07}.  

\begin{figure*}
\psfrag{Y}[c]{\large $\si_{PI}(k,a)/\overline{P}(k,a)$}
\psfrag{X}[c]{\large $k/k_N$}
\includegraphics[width=0.45\textwidth]{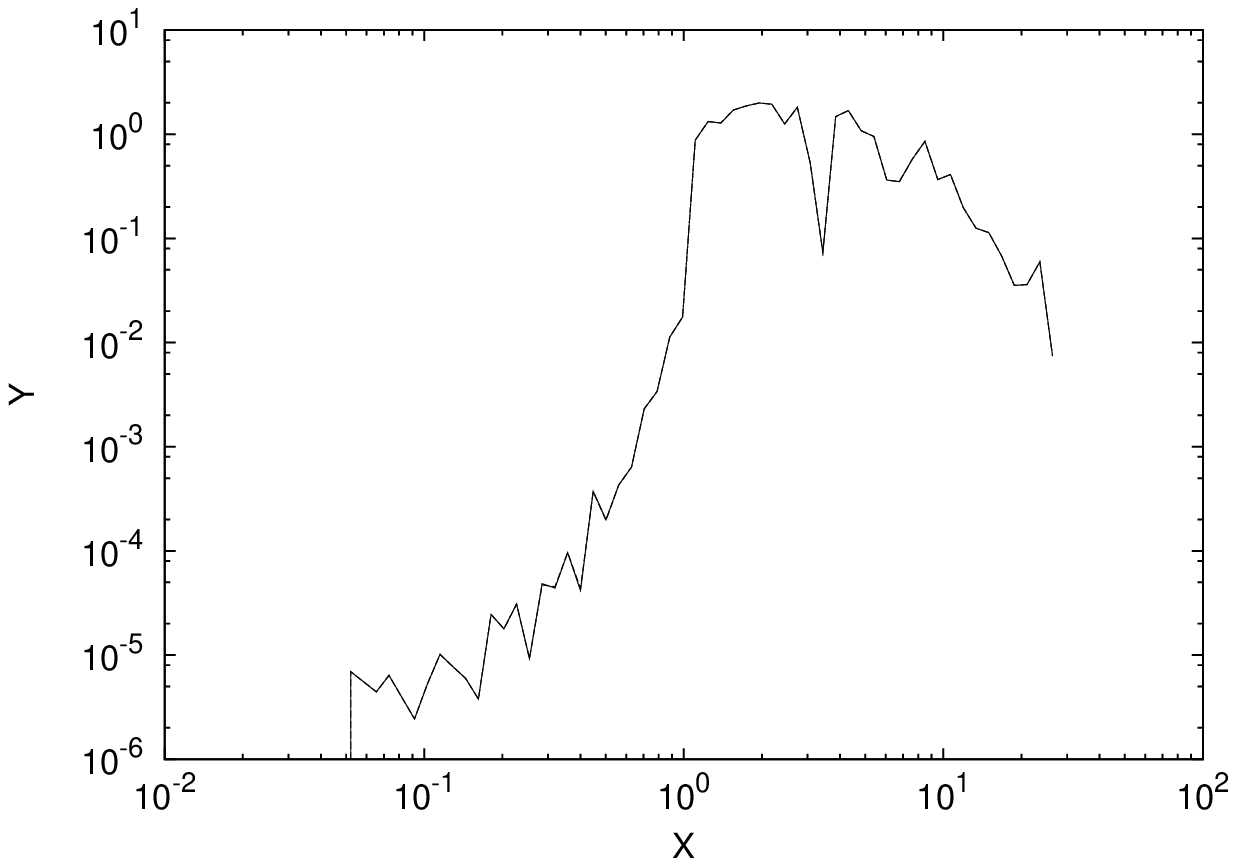}
\includegraphics[width=0.45\textwidth]{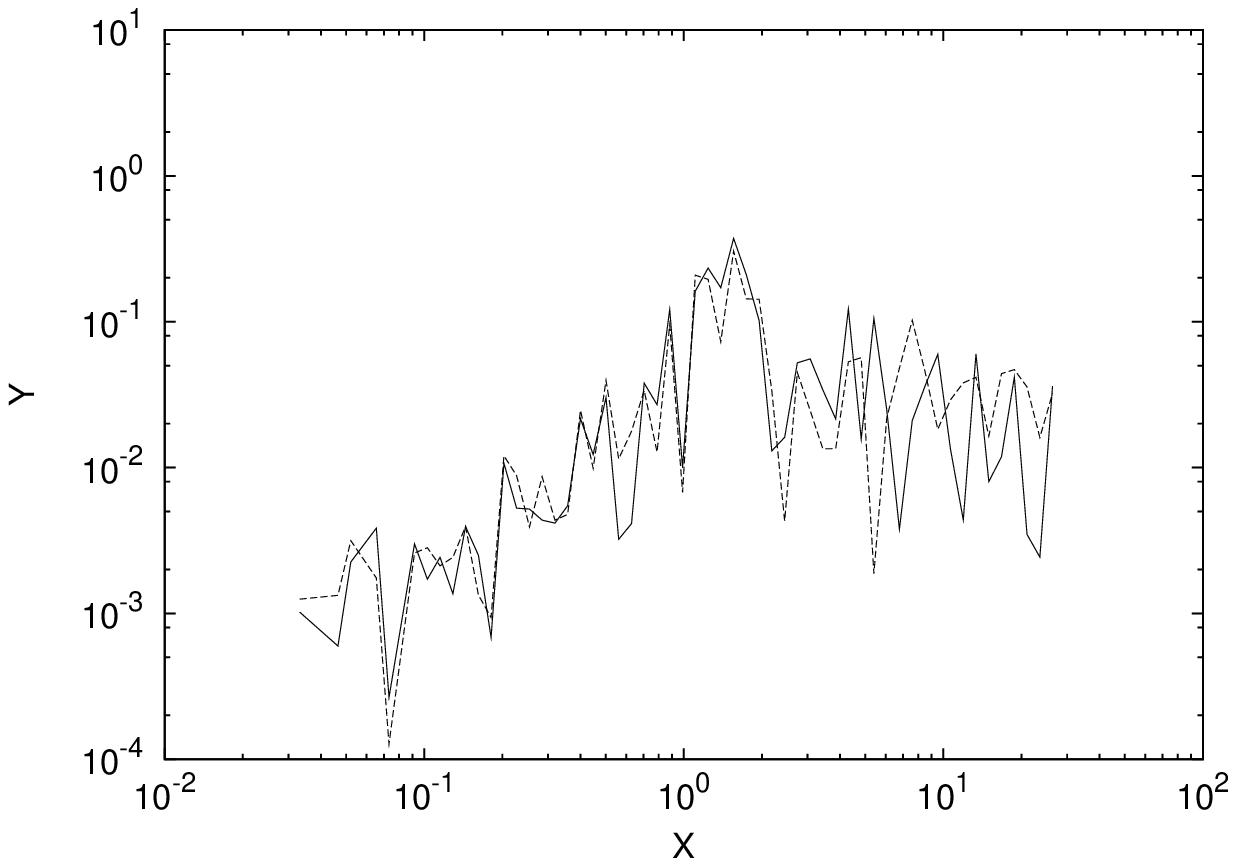}\\
\includegraphics[width=0.45\textwidth]{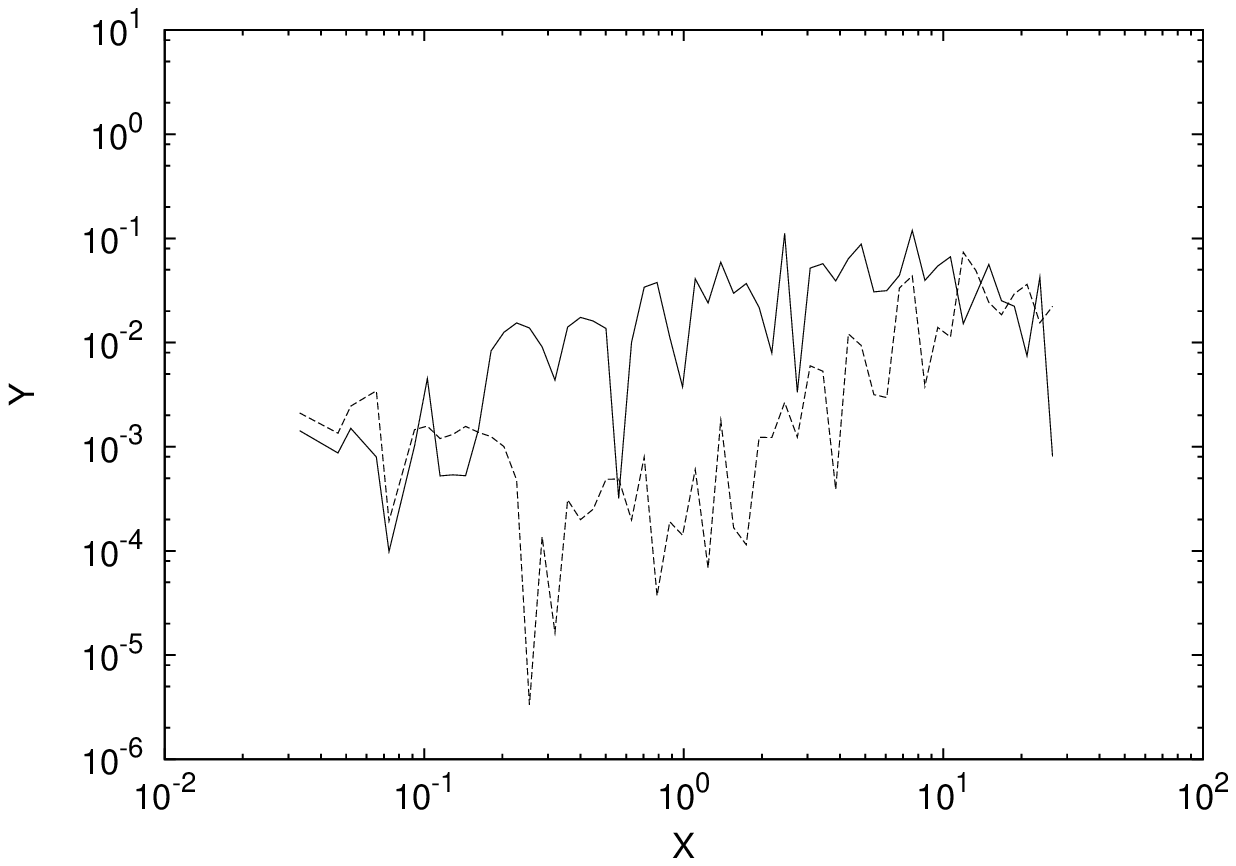}
\includegraphics[width=0.45\textwidth]{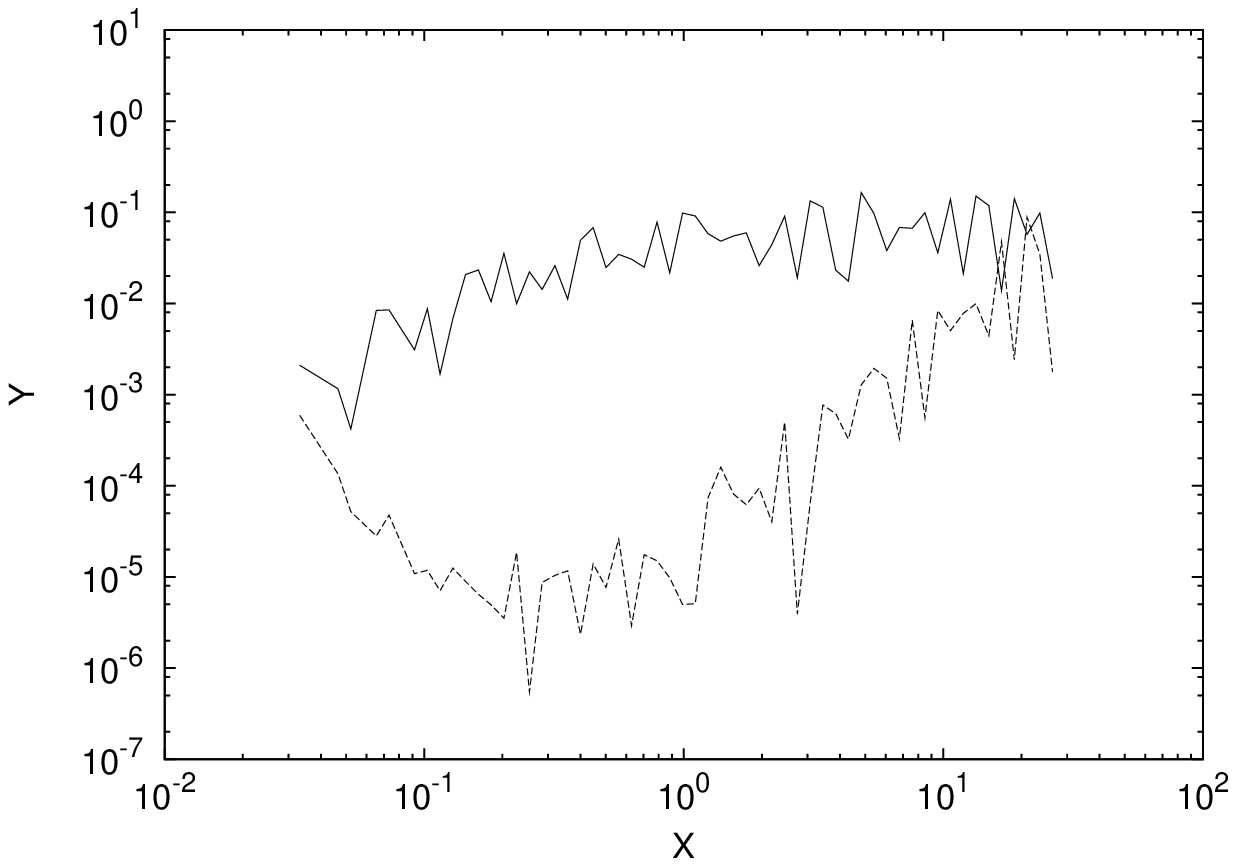}
\caption{Normalized variance of the PS of the BCC and SC, computed from our
GADGET simulations of full gravity (continuous lines) and simulations 
evolved using PLT (dashed lines) for, from top to bottom and left to right, $a=2^0$, $a=2^3$, $a=2^5$ and $a=2^7$. All the curves are normalized to
the PS for the full gravity (FG) case.
\label{variance_basic_PLT}}
\end{figure*}

\begin{figure*}
\psfrag{Y}[c]{\large $\si_{PI}(k,a)/\overline{P}(k,a)$}
\psfrag{X}[c]{\large $k/k_N$}
\includegraphics[width=0.45\textwidth]{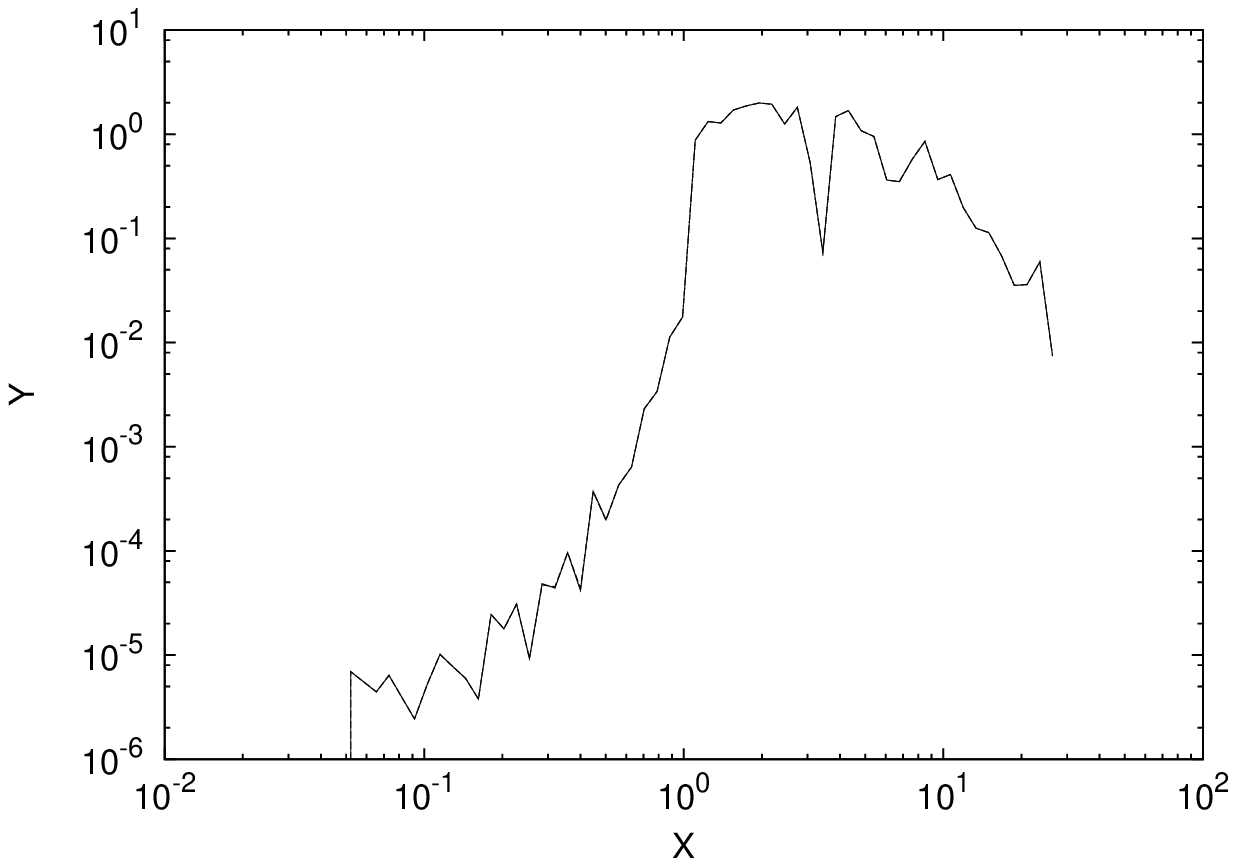}
\includegraphics[width=0.45\textwidth]{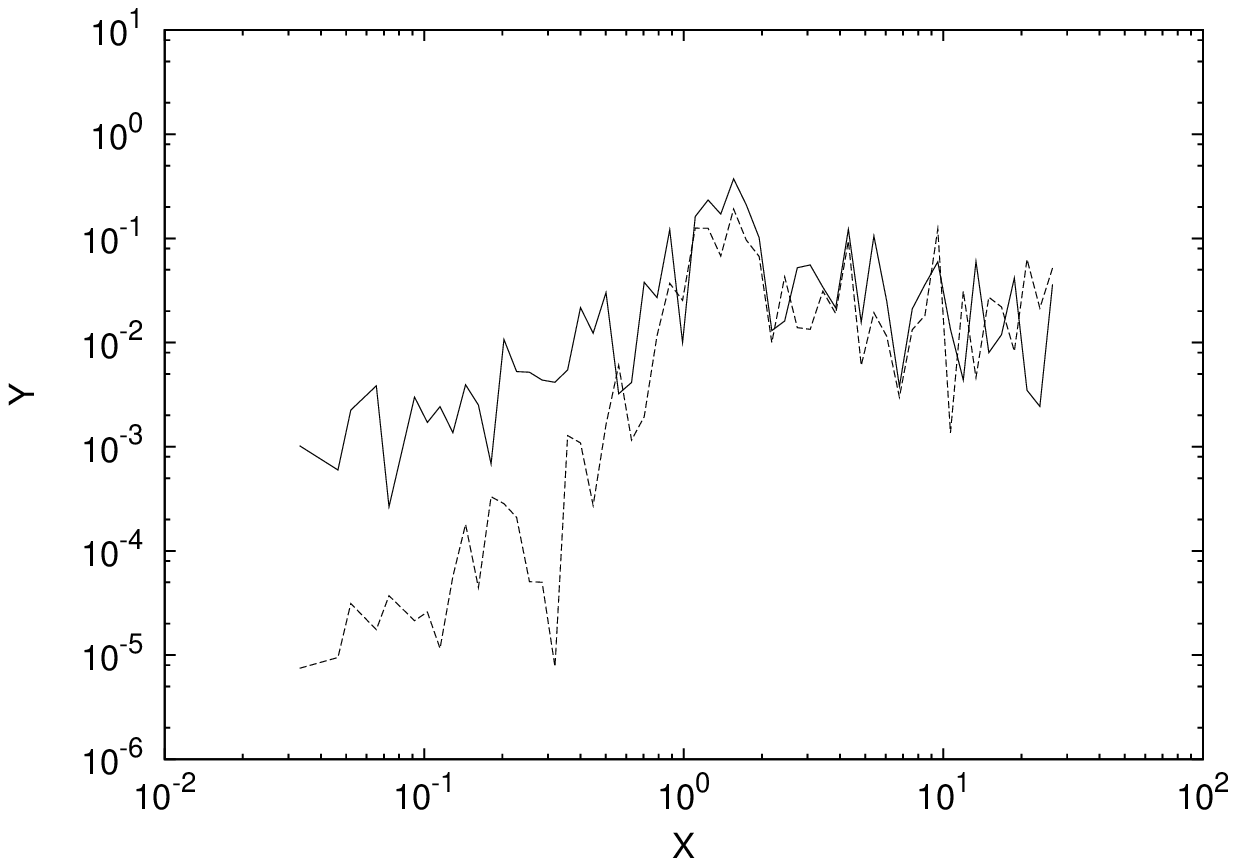}\\
\includegraphics[width=0.45\textwidth]{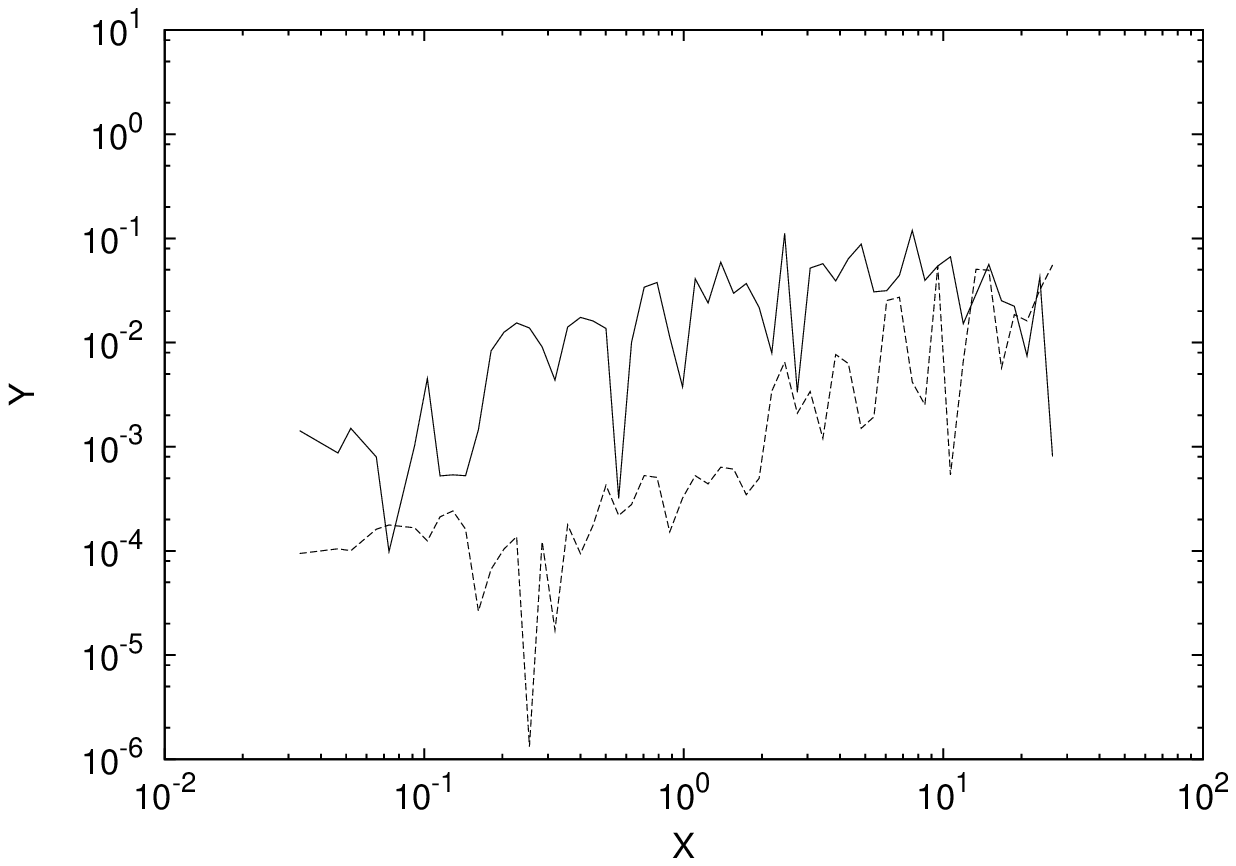}
\includegraphics[width=0.45\textwidth]{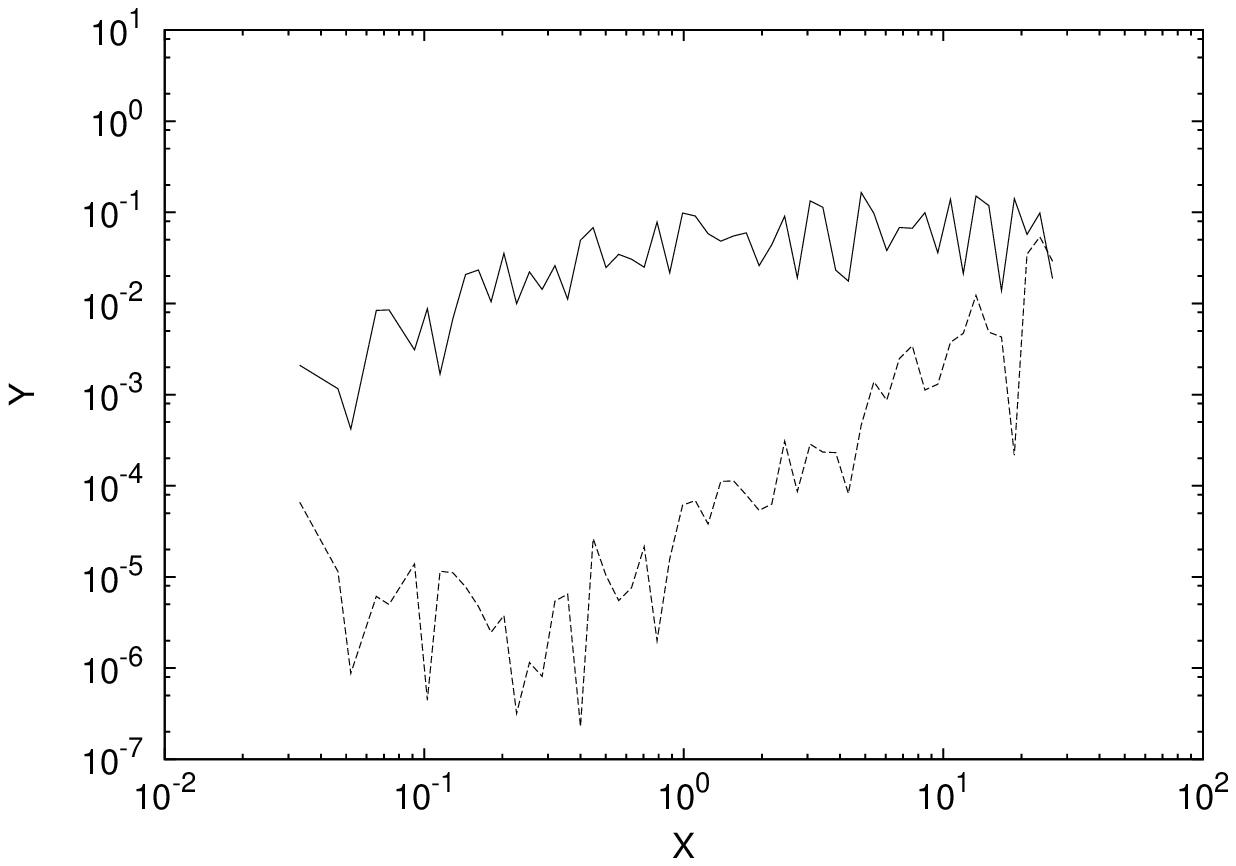}
\caption{Normalized variance of the PS of the BCC and SC PI, computed from
FG simulations (continuous lines) and from simulations using FLT (dashed lines) for, from top to bottom and left to right, $a=2^0$, $a=2^3$, $a=2^5$ and $a=2^7$. All the curves are normalized with the PS obtained with FG.
\label{variance_basic_FLT}}
\end{figure*}

In Fig.~\ref{variance_basic_PLT} are shown, for the 
GADGET2 simulations and the PLT evolved configurations, the 
normalised differences between the PS for the BCC and SC, i.e., 
\be
\frac{P_{\rm bcc}  (k,a) - P_{\rm sc} (k,a)}
{\frac{1}{2}\left[P_{\rm bcc} (k,a)+ P_{\rm sc} (k,a)\right]}
\label{def-normalized-diff}
\ee
where the subscript indicates the $preIC$. 
We also show for comparison in Fig.~\ref{variance_basic_FLT} 
the same quantities, except that the PLT evolved configurations
are replaced by those evolved with its fluid limit (which we will
denote by FLT, for ``fluid linear theory''). 
For initial conditions set up, as done here, with the Zeldovich 
approximation, this is simply the extrapolated evolution in
this same approximation. 

The agreement with PLT at $a=2^3$ is extremely good for all the 
measured $k$, while at $a=2^5$ it is restricted only to
the very longest wavelength modes in the box. FLT, on
the other hand, traces the observed differences well
until $a=2^3$, but only the $k$ larger than $k_N$.

These different ranges of agreement for PLT and FLT are simple to
understand, using the results quoted above in
Eqs.~(\ref{discrete-continuous}-\ref{iofk}). These formulae relate, at
sufficiently early times and small $k$, the theoretical PS of density
fluctuations $P_{\mathrm{th}} (k)$ to the full PS of density
fluctuations in the generated point distributions. FLT gives a linear
amplification of the displacement fields, independent of $\bk$, and
therefore a linear amplification of the terms $P_{\mathrm{c}}(\bk)$
and $P_{\mathrm{al}}(\bk)$. Outside the range of $k$ where
$P_{\mathrm{PI}}(\bk)$ contributes, i.e., inside the FBZ, FLT thus
simply describes a linear amplification of the full initial PS, which
leaves the normalized quantity in Eq.~(\ref{def-normalized-diff})
strictly invariant. Outside the FBZ, on the other hand,
the term $P_{\mathrm{PI}}(\bk)$ becomes important.  When this is the
case the full evolution is well approximated by the FLT evolution
because (see \cite{discreteness2_mjbm}) the evolving term, $P_{\mathrm
{al}} (\bk)$, is in fact dominated by initial power at small $k$ for
which the evolution is very well approximated by FLT.

The regime in which PLT traces the differences
very well, but FLT does not, corresponds to the
$k$ inside the FBZ which are, in PLT, amplified 
linearly in slightly different ways on each lattice.
In this case the physical discreteness effect
arises thus from the modification with respect to
FLT of the {\it dynamical evolution} of the same
initial power. In the regime in which FLT
gives a good approximation, on the other hand, the 
corresponding discreteness effects arise from the 
power associated with the slightly different 
initial samplings on the different lattices
of modes which evolve approximately in the 
same way.

\begin{figure}
\psfrag{Y}[c]{\large $|1-P^{\rm PLT}(k,a)/P^{\rm FG}(k,a)|$}
\psfrag{X}[c]{\large $k/k_N$}
\includegraphics[width=0.45\textwidth]{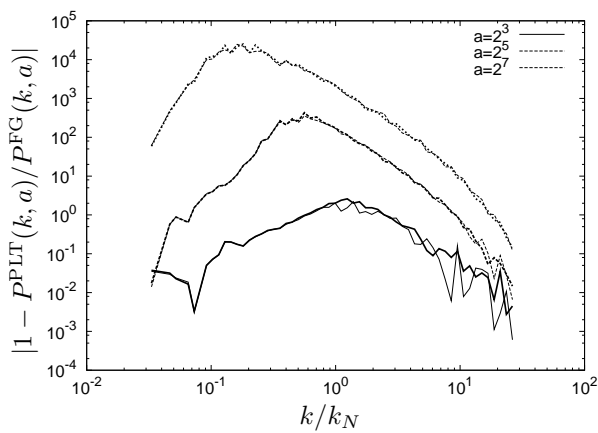}
\caption{Normalized difference  between the PS computed with FG and PLT for the BCC (thick lines) and SC (thin lines) configurations.
\label{ratio_FG-PLT}}
\end{figure}

Let us consider further the range of validity of PLT in these
plots. The perturbative expansion underlying PLT as developed 
in \cite{marcos_06} and \cite{marcos07} is strictly valid, as 
we have discussed above, only when the relative separation
of {\it all} particles is small compared to their initial 
separation. In \cite{marcos_06} it has been shown that it
gives a very good approximation to the evolution of 
the PS (and significantly better than FLT) at least until 
the time when a significant fraction of the particles have 
come close to another particle for the first time (which corresponds
approximately to
shell crossing in the fluid limit). However, its possible
validity beyond this time has not been established. What
the results in these plots show is that its validity indeed 
extends considerably longer, as there has already been
very significant shell crossing already at $a=2^3$, 
and clearly at $a=2^5$ the evolution is well beyond this
point. In Fig.~\ref {ratio_FG-PLT} we show the 
modulus of the ratio 
\be
\frac{P_{I}^{\rm FG} (k,a)- P^{PLT}_{\rm I} (k,a)}
{P_{\rm I}^{\rm FG} (k,a)}
\label{def-normalized-diffPLTFG}
\ee
for both $I=$SC and $I=$BCC, i.e., the fractional
deviation of the power at each $k$ in the PLT evolved initial
conditions simulation from that in the full gravity (FG) 
simulation of the same initial conditions. Comparing with
the results of Fig.~\ref{variance_basic_PLT}, we see 
that the range in which PLT correctly describes the
differences between the SC and BCC simulations extends
in fact to when the plotted quantity is of order two, i.e.,
into a regime in which PLT no longer follows well the full 
power in each PS accurately. This is evidently possible 
only if the deviation from the PLT evolved initial conditions 
is essentially the same for both $preIC$, i.e.,
this additional non-linear power itself has smaller
discreteness corrections than those given by PLT. We note 
that this is very coherent with our observations 
above concerning the numerical integration: we observed in
that case much better numerical convergence of the measured 
differences than in each of the PS individually. 
Thus these numerical residuals are strongly 
correlated with the non-linear power which
is the same in both simulations, and so cancel 
out when we take the difference. We will discuss
briefly in our conclusions these observations about
the regime of validity of PLT.

\begin{figure}
\psfrag{Y}[c]{\large ${\rm Dev}(k,a)$}
\psfrag{X}[c]{\large $k/k_N$}
\includegraphics[width=0.45\textwidth]{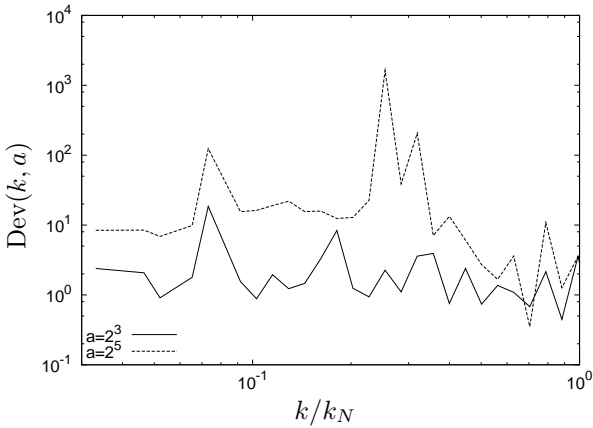}
\caption{Real deviation normalized by the estimated deviation computed
 using Eq.~\eqref{real-div-est}. 
\label{fig_real-div-est}}
\end{figure}

It is instructive also to examine, in the range in which PLT 
traces accurately the differences in the evolution, what the
relation is between this measured difference, and the true
discreteness error, which can also be calculated in PLT.
Indeed, in the FBZ, it is simply the difference between the PLT 
evolved power and FLT evolved power. A simple qualitative
measure is thus: 
\bea
\label{real-div-est}
&{\rm Dev}(k,a)=\\\nonumber
&\left[\frac{\left(P_{bcc}^{PLT}(k)-P_{bcc}^{FLT}(k)\right)^2+\left(P_{sc}^{PLT}(k)-P_{sc}^{FLT}(k)\right)^2}{\left(P_{bcc}^{PLT}(k)-P_{sc}^{PLT}(k)\right)^2}\right]^{1/2},
\eea
where the $a$-dependence on the right hand side is left implicit. Limiting
ourselves to the modes for which PLT furnishes a good approximation
to the full evolution of the individual PS, i.e., to the regime in
which the quantity plotted in Fig.~\ref{ratio_FG-PLT} is less than
or order one, we see that ${\rm Dev}(k,a)$ shows a clear tendency 
to increase with $a$, particularly for smaller $k$. Indeed at $a=2^5$,
for the very smallest $k$ for which PLT is still approximately valid,
our lower bound on the discreteness error is {\it one order of 
magnitude smaller} than the real discreteness
error. The reason is simply that the difference in the exponents
characterising, in PLT, the growth of the displacement fields in
these two different lattices at these values of $k$ is considerably 
smaller than the difference between these exponents and the FLT behaviour
(giving growth in proportion to $a^2$). Even if, taking this factor
into account, we arrive at a real discreteness error of order of one 
percent, this result shows that, for all our results here, we should
bear in mind that {\it the measured differences provide only lower
bounds on the discreteness error which may be very different from
the full discreteness error}.

\subsection{Variation of particle number $N$}

\begin{figure*}
\psfrag{Y}[c]{\large $\si_{PI}(k,a)/\overline{P(k,a)}$}
\psfrag{X}[c]{\large $k/k_N$}
\includegraphics[width=0.45\textwidth]{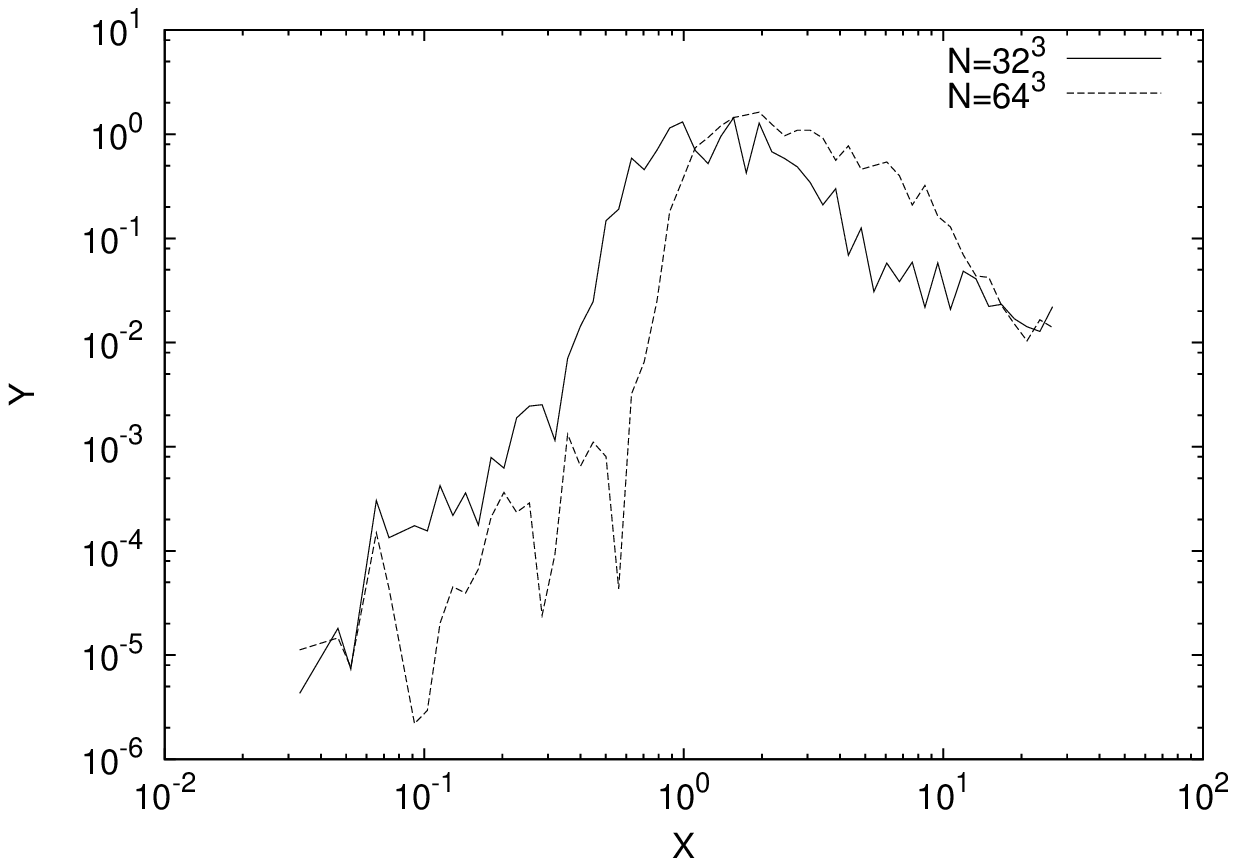}
\includegraphics[width=0.45\textwidth]{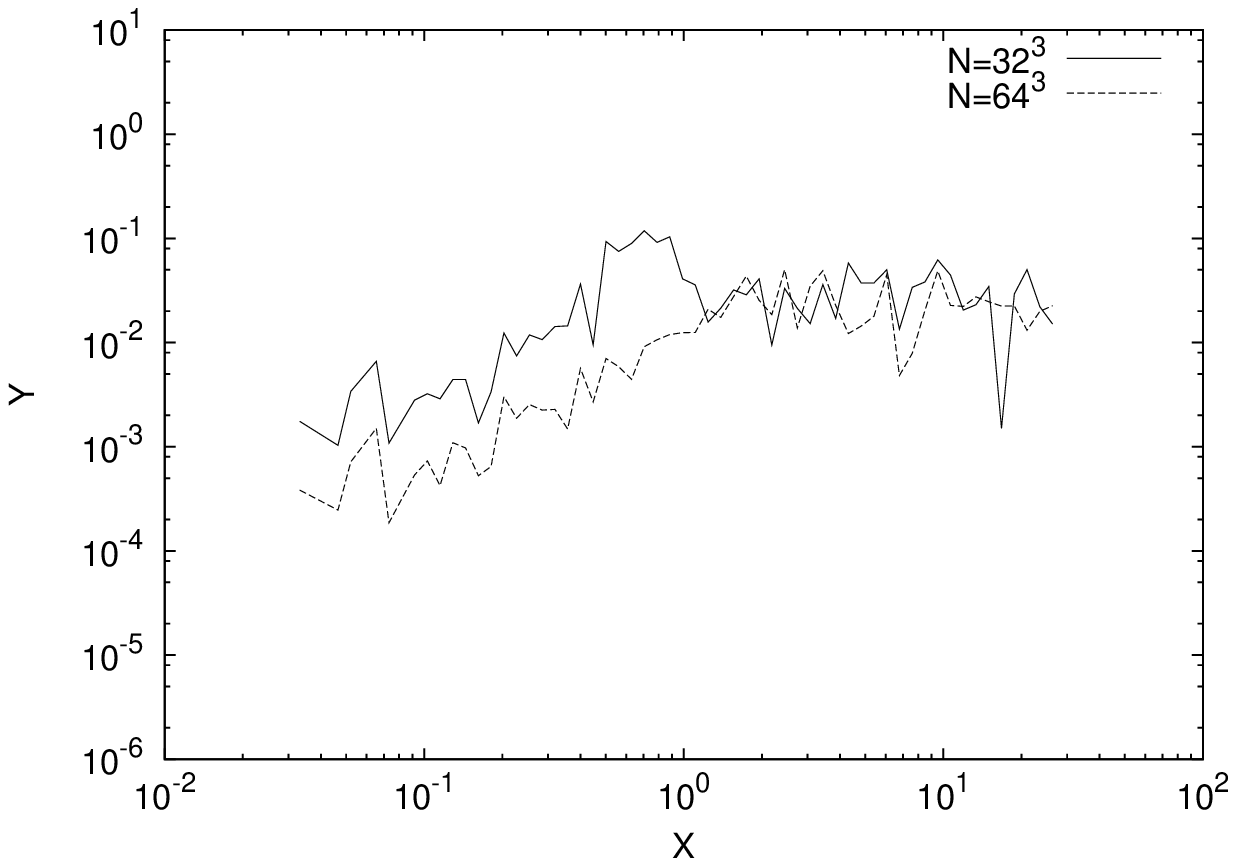}
\\
\includegraphics[width=0.45\textwidth]{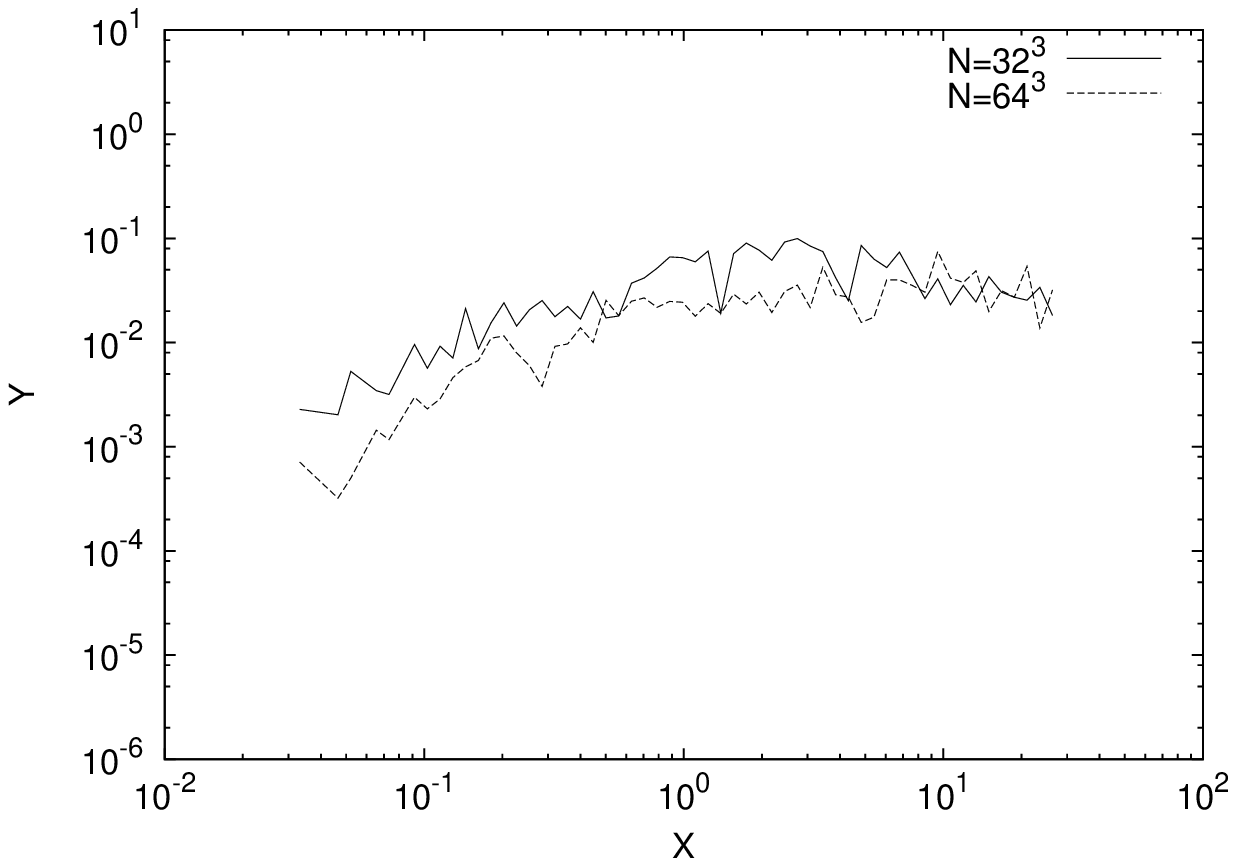}
\includegraphics[width=0.45\textwidth]{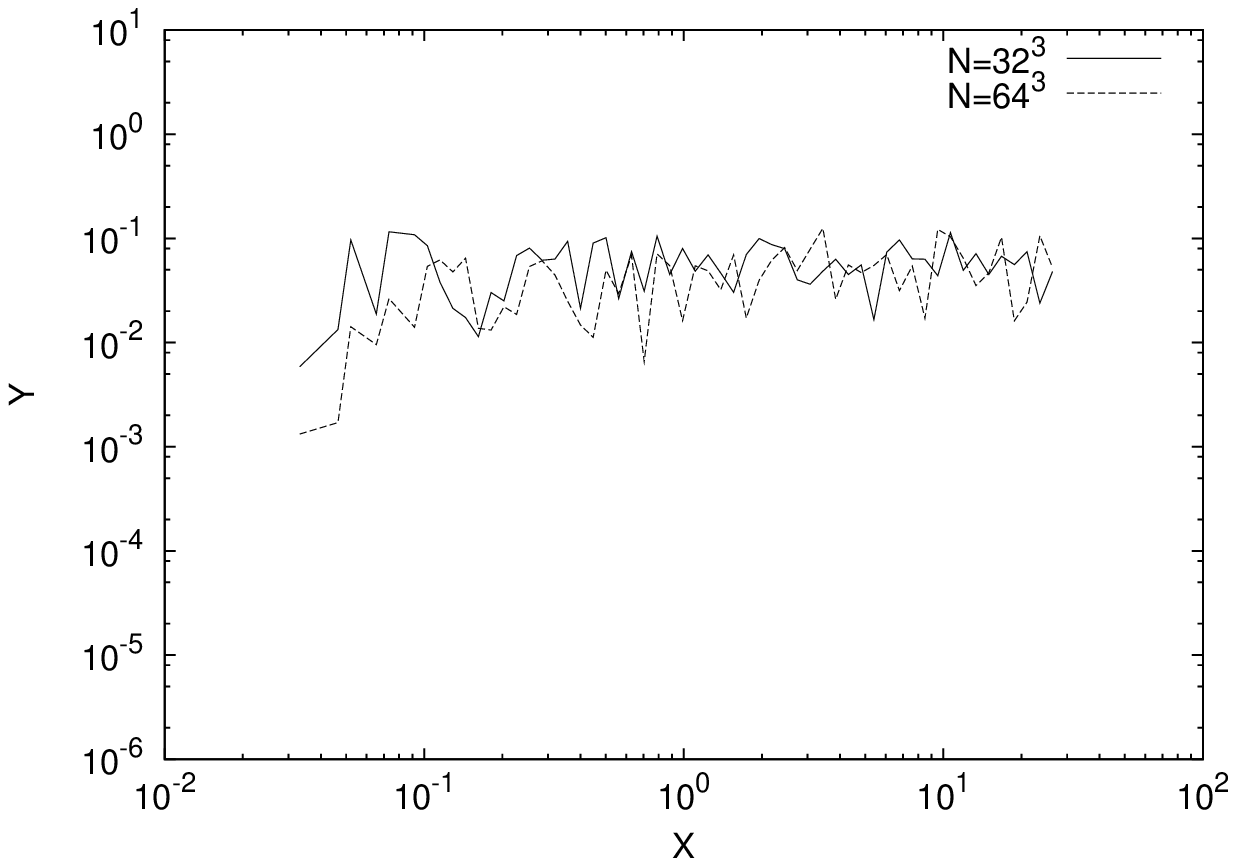}
\caption{Normalized variance of the PS for $N=32^3$ and $N=64^3$ particles
at, from top to bottom, $a=1$, $a=2^3$ and $a=2^7$. The horizontal axis is normalized at the Nyquist frequency of the $N=64^3$ distribution.
 \label{variance_res}}
\end{figure*}

A further direct numerical check on our interpretation of the
differences we have identified as discreteness errors may be 
given by looking at their dependence on particle number $N$
(i.e. effectively, given that we work at fixed box size, 
on the particle density parametrized by $\ell$). We thus
consider varying $N$ while keeping all other discreteness
parameters fixed (and, again, checking also the stability of 
results considered to the variation of numerical parameters).
Shown in  Fig.~\ref{variance_res} are the results for the 
normalized variance $\si_{PI}(k)$ [as defined in 
Eq.~\eqref{variance_pi}] on a set of four simulations of
identical initial conditions (i.e. identical modes of the
displacement fields on the four $preIC$) for 
$N=32^3$ and $N=64^3$ particles\footnote{As described in 
Sect.~\ref{Initial conditions} above, the number of 
particles in the non-SC configurations are chosen as
close as possible to these numbers (Table~\ref{table1}).}. 
Note that, to have identical
displacement fields in the two cases, we have cut the initial
PS at the Nyquist frequency of the $N=32^3$ 
distribution\footnote{More precisely, as described in 
Sect.~\ref{Initial conditions} we sample exactly the modes 
in the FBZ of the SC lattice.}. We see clearly the explicit
$N$ (or $\ell$) dependence of the results in all but the very strongly
non-linear regime. In the PLT regime the 
difference in power depends parametrically on $N$ as 
$N^{-2/3}$ [see Eq.~(\ref{PS-differences-PLT}) below].
Interestingly one can observe between $a=2^3$
and $a=2^5$ an apparent ``spreading'' of this explicit dependence
to larger $k$, a behaviour which is naturally interpreted as 
the transfer of the discreteness effects accumulated at
$a=2^3$ in the linear regime to larger $k$ modes as the
corresponding scales go non-linear. On the other hand, we
see no clear evidence for a dependence on $N$ at the 
strongly non-linear scales --- and most notably over the 
entire range at the final time, $a=2^7$ --- and so we will
not assume here that the measured differences are discreteness
effects. It is to be noted, however, that this is a very
conservative assumption: the differences even at $a=2^7$ 
may quite consistently be, and indeed are naturally, 
ascribed to those at the previous time, $a=2^5$, without
the latter having to show the same explicit dependence
on $\ell$. Indeed in the 
strongly non-linear regime \footnote{We recall that, at 
$a=2^7$, the whole box has gone non-linear with most of the 
matter in only a few halos (see Fig.~\ref{snapshot_evol}).}
we do not expect a simple dependence of the final power
on the amplitude of the power in the preceeding 
weakly non-linear phase, and therefore the dependence on
$\ell$ inherited from this phase could quite possibly be 
much weaker than that observed 
in the preceeding phase.  

\subsection{Variation of $\varepsilon$}

\begin{figure*}
\psfrag{Y}[c]{\large $\si_{PI}(k,a)/\overline{P(k,a)}$}
\psfrag{X}[c]{\large $k/k_N$}
\includegraphics[width=0.45\textwidth]{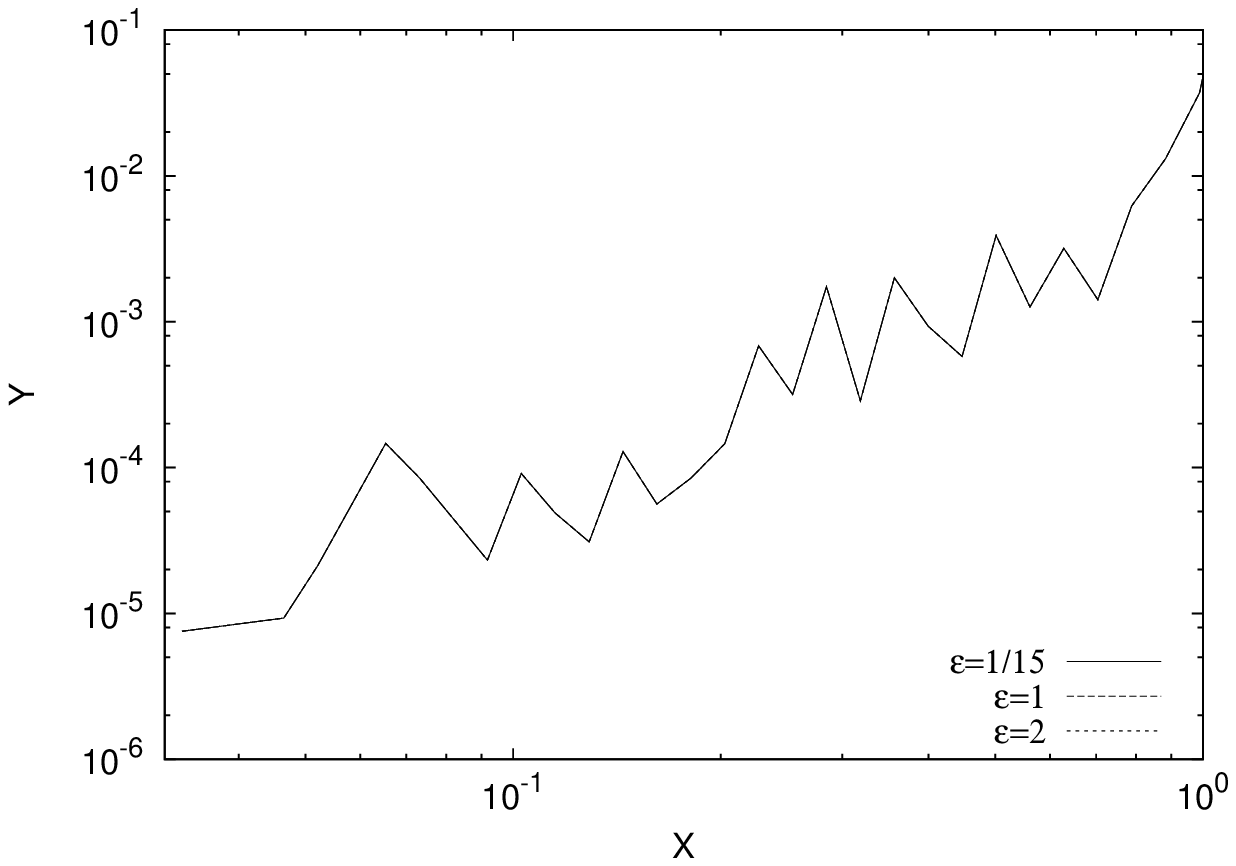}
\includegraphics[width=0.45\textwidth]{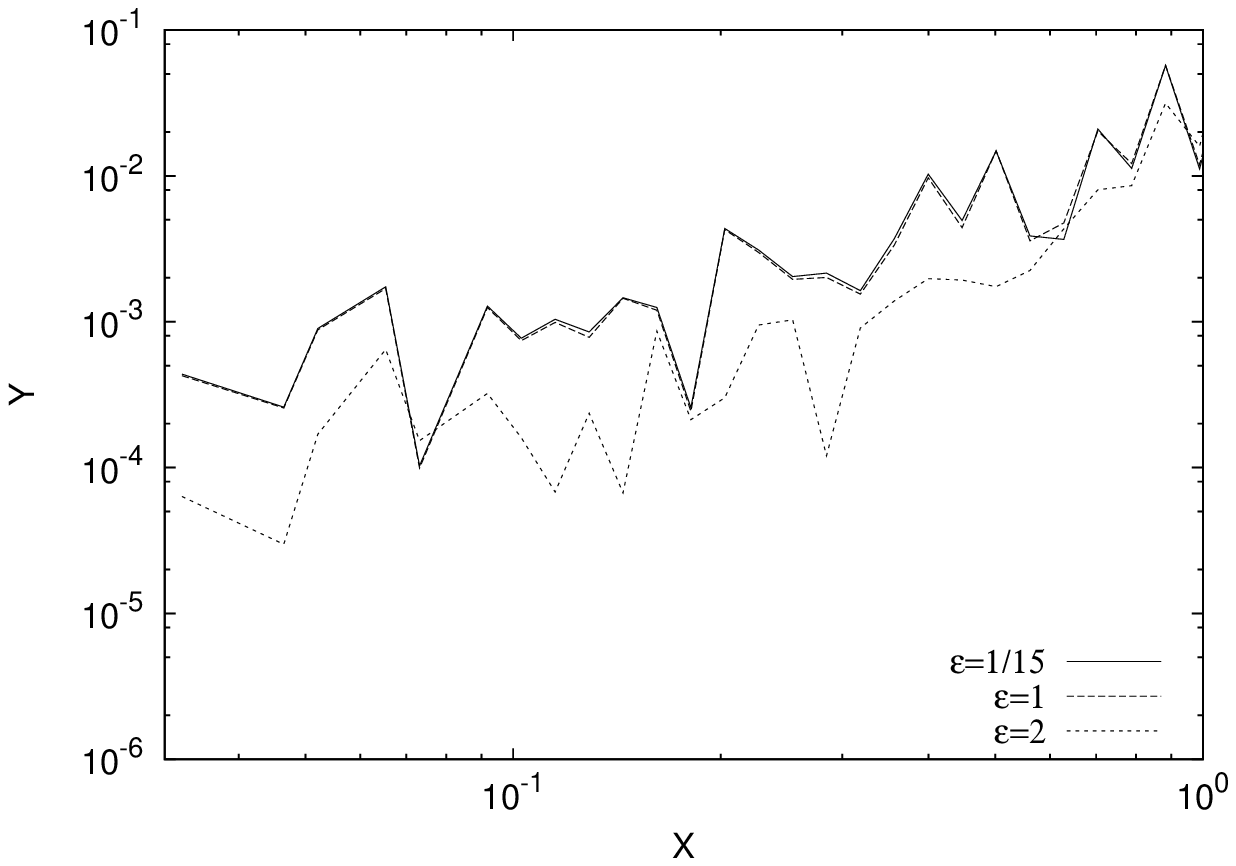}
\\
\includegraphics[width=0.45\textwidth]{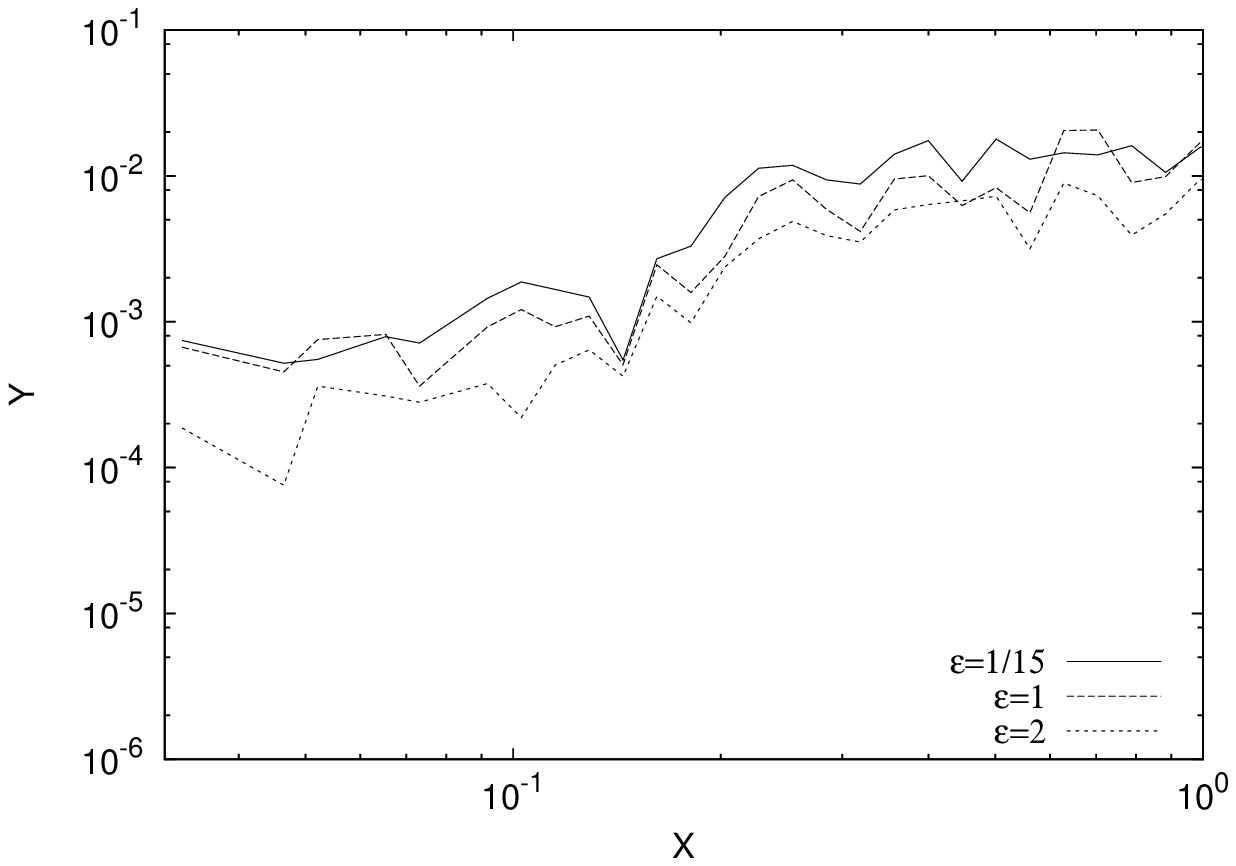}
\includegraphics[width=0.45\textwidth]{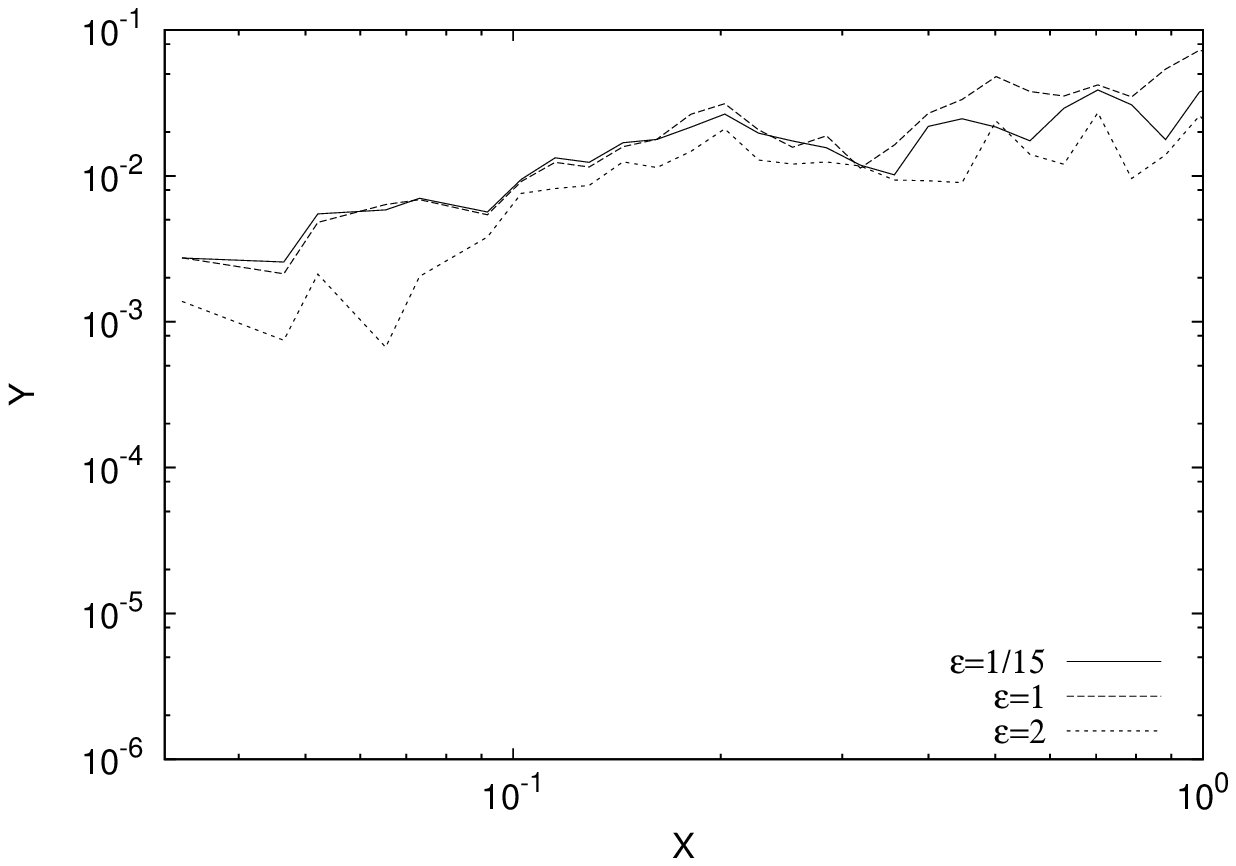}
\\
\caption{Normalized variance of the PS for $\ep=1/15$, $\ep=1$ and $\ep=2$
at, from top to bottom, $a=1$, $a=2^3$ and $a=2^7$. 
 \label{variance_soft}}
\end{figure*}

Another check on our results is given by considering the effect
of varying $\varepsilon$, keeping all the other discreteness
parameters fixed. In Fig.~\ref{variance_soft} we show again
the normalized variance $\si_{PI}(k)$, now again for four
different simulations with $N=64^3$, for three different values
of $\varepsilon$: the same one as used in the results reported
until now ($\varepsilon=\ell/15$), and now also for simulations
(from exactly the same initial conditions) with $\varepsilon=\ell$
and $\varepsilon=2\ell$. We show only the range of $k$ below
the Nyquist frequency as this is the regime of physical interest, i.e., 
in which results are expected to converge to those for 
(unsmoothed) gravity, fixed approximately by the mode inverse to 
the largest value of $\varepsilon$. (We do not show results 
for smaller $\varepsilon$ as they are negligibly different
in this range from those at $\varepsilon=\ell/15$). The 
behaviour observed  at $a=2^3$ is completely 
consistent with what is expected given that we have seen 
that PLT provides an excellent description of these 
differences at this time: the exponents for growth of the
modes of the  displacement field calculated in PLT (which may be calculated
for any two-body potential) only begin to change significantly
when $\varepsilon \sim \ell$, simply because PLT is an expansion
about the particles placed at their lattice sites. As $\varepsilon$
increases the deviation from the fluid evolution becomes in fact
more and more significant (see \cite{discreteness2_mjbm}), but 
this deviation does not manifest itself as a difference 
between evolution on the different lattices as the 
smallest scales on which they differ are then smoothed over. 
Thus the differences
we measure decrease (in the FBZ, where they are due to the
difference in the exponents relative to their FLT values).
At $a=2^5$ we see essentially the same behaviour for the
modes for which PLT was valid, while for the larger modes
there is also some more marked decrease already for 
$\varepsilon=\ell$. At $a=2^7$ we see a larger spread,
with an apparent tendency for the largest $\varepsilon$
to lead to the smallest differences, which would certainly
be consistent with the hypothesis that these errors could 
also be interpreted as due to discreteness. It is important 
to note that, in all these figures, 
the reduction of the differences measured as $\varepsilon$ is 
increased does not imply a convergence of the simulations towards the physical
(VP) limit, but at most towards a smoothed version of it, which 
may be further from the physical limit than the results 
obtained with the smallest $\varepsilon$. Indeed in the PLT regime
we have shown in \cite{discreteness2_mjbm} that increasing 
$\varepsilon$ at fixed $\ell$ does indeed increase the deviation
of the growth exponents of modes from their fluid value.

\subsection{Variation of initial red-shift}

The initial red-shift $z_i$ is the remaining parameter
in the list of discreteness parameters ${\cal D}_\alpha$
we gave in our discussion in Sect.~\ref{The problem of discreteness}.
As the dependence of varying it while keeping the other parameters
fixed can be understood analytically using PLT, in the regime in
which we know it to be valid (of small relative displacements), we 
do not report here numerical results\footnote{See also 
\cite{discreteness2_mjbm} for quantitative results. Some numerical
results for the effect of varying $z_i$ only are given in \cite{mcdonald05},
but only for very specific quantities (ratios of PS for
different dark energy models).}. Quite simply
we note that, in the EdS cosmology, the evolution of the PS
in PLT can be written (\cite{marcos_06}) to a very good 
approximation as  
\be 
P(\bk,a)= a^{2+\delta_i(\hat{\bk}) k^2\ell^2 } P(\bk, a=1)
\label{linear-general}
\ee 
where $\delta_i(\hat{\bk})$ is a function of the orientation
of the vector which depends on the $preIC$. It follows that
the normalized difference in the power, averaged in a bin
of wavevectors centred at wavenumber $k$, scales 
approximately as
\be
\sim \left[\bar{\delta}_{\rm sc}(k) - \bar{\delta}_{\rm bcc}(k)\right]
k^2 \ell^2 \log a
\label{PS-differences-PLT}
\ee
where $\bar{\delta}_{\rm i}(k)$ are appropriate effective values of
the parameter ${\delta}_{\rm i}(\hat{\bk})$ over the bins of wavevectors.
The differences we have measured thus increase without limit
as $z_i$ does, with a logarithmic depedence on the latter.

\section{Discussion and conclusions}

We now summarize our main findings and conclusions, as well as
indicating some directions for further study to clarify these
issues.

\begin{itemize}

\item Cosmological simulations should evidently be tested for
discreteness effects by an appropriate, and well controlled,
extrapolation of the relevant parameters. To recover the continuum VP
limit, we have noted that the simplest such extrapolation is to
increase the particle density (i.e. $\ell \rightarrow 0$) keeping the
other relevant parameters introduced by the discreteness fixed ---
specifically the force smoothing $\varepsilon$, initial red-shift
$z_i$ and $preIC$.  While this may seem rather evident, this kind of
procedure is not systematically applied in the literature, apart from
the few isolated studies we have mentioned (notably those of Melott
and collaborators). More specifically many of the (relatively few)
convergence studies in the literature adopt a different approach,
typically decreasing $\varepsilon$ in proportion to $\ell$, keeping
always $\varepsilon \ll \ell$.  While such an extrapolation is not
necessarily wrong, i.e., it may allow one to arrive at conclusions
which are correct concerning discreteness effects, it has the
intrinsic problem that it does not converge to the VP
limit. Physically this means that such an extrapolation {\it does not
remove the non-VP effects} in the dynamics (e.g. two body
collisionality, or the effects described by PLT) but simply moves them
to smaller scales. Given that the interplay of different scales in the
fully non-linear regime of gravity is not understood, this is not a
solid procedure. In this respect we note also that in this approach,
additional power in the initial conditions --- corresponding to the
extra modes which may be sampled as $\ell$ is decreased --- is usually
added. This means that structures do indeed form first at the smallest
scales, where discreteness is manifestly important. Further such
modification of the initial conditions makes it difficult to identify
with precision, as in the present study, variations which are due to
discreteness. We note, however, that using wavelet techniques
\cite{romeo08} have recently claimed to detect numerically
discreteness effects embedded in the scatter of a set of cosmological
simulations using different realizations of the initial conditions
(and extrapolated power).
   
\item There has been some controversy in the literature about 
the widely used practice of taking results to be physical
(i.e. representative of the VP limit) at scales below $\ell$,
in simulations with $\varepsilon < \ell$. We have addressed
this issue with a controlled numerical study of such a simulation
(with $\varepsilon=\ell/15$). Our conclusion is that such a 
procedure appears to be reasonable, {\it to a first approximation}: 
efficient transfer of power from large to small scales does
indeed tend to make the results on scales below $\ell$ converge,
``wiping out'' the significant differences on these scales in
the initial conditions (see e.g. \cite{Little+weinberg+park_1991},
and \cite{bagla_discreteness08}). {\it However} this mechanism is by no
means perfect and we have demonstrated with our study beyond
doubt that there are indeed {\it measurable} residual effects of 
discreteness {\it at all scales}, at a level relevant to the precision
(of order a percent) now set as a target for such simulations.
Considering, very conservatively, only our results up to $a=2^5$ in 
Fig.~\ref{variance_basic} as indicative of what one would find
in a typical cosmological simulation (i.e. starting at an initial 
red-shift $z_i=32$), one infers a lower bound on discreteness effects 
which reach about one percent at the Nyquist frequency. We emphasize
that these measures are only {\it lower bounds}, which may be very
much below the full discreteness error. Indeed we have seen that
in the regime (of validity of PLT) in which we can calculate this 
full error, the lower bound is (at small $k$, at $a=2^5$) one order
of magnitude larger than the estimated error (i.e. about one percent
rather than the measured lower bound of a tenth of a percent at these scales).
Most importantly the only way to attain greater precision, and indeed
{\it the only way to firmly establish the convergence to the physical
limit}, is to extrapolate to $\ell \ll \varepsilon$. Thus, while the
``common wisdom'' is probably reasonable for the modest precision
required for many uses of the results of these simulations, the
criticisms formulated by some groups (notably Melott and collaborators)
are fundamentally correct and further, relevant for the levels of
precision required for some applications (e.g. future weak lensing 
observations). In this respect we note also that we have analysed 
here solely two point properties (essentially the PS), while Melott 
has emphasized that the numerically measured effects of discreteness 
are more important in other (phase-sensitive) quantities. The 
methods used here to establish ``calibrated'' lower bounds on
discreteness error can easily be generalized to study such quantities.  
Such a study, as well as more extended numerical studies of controlled
extrapolations  to the regime $\ell \ll \varepsilon$ like those of
\cite{splinter}, using possibly also the methods of analysis
employed in \cite{romeo08}, would provide further insight 
into these issues.

\item An important element in our numerical study is the use of 
the PLT formalism. It allows us to fit analytically the measured 
dispersion in results for the PS (or, in principle, any quantity) 
due to discreteness, at sufficiently early times. This allows
us not only  to ``calibrate'' our numerical results, establishing that
the method does really indeed measure discreteness effects (rather
than other numerical or finite-size effects), but also gives us
an understanding of the physical origin of these effects: a finite 
sampling of a fluctuation modifies its evolution with respect to
the smooth limit. This is a physical effect of discreteness
which has not been previously envisaged, and it illustrates very
clearly that the widely made assumption that the effects of
discreteness are solely those which arise from (i) missing initial
power, and (ii) two-body collisionality, is indeed just an
assumption, which can at best be approximately correct.  
Indeed PLT describes explicitly the effect of small scales on larger 
scales which, albeit not the dominant one in the evolution
of the gravitational clustering, is not zero when the ratio 
of these scales is finite. Such effects at large scales (i.e.
significantly larger than $\ell$)
have until now escaped detection in cosmological N-body simulations, 
even in studies which looked for them\footnote{See, e.g., 
\cite{Little+weinberg+park_1991} and both the recent studies
of \cite{bagla_discreteness08} and \cite{romeo08}.}.
Further we have noted that our results indicate that, apart 
from the very early non-linear evolution which ``fills in'' the 
missing power at large $k$, the discreteness errors {\it at any scale} 
continue to grow monotonically in time, as in PLT, throughout 
the whole simulation. Such behaviour would naturally be explained
if the physical effects of PLT continue to act in the non-linear
regime, and indeed it is very plausible that this should be the
case: one would expect that the evolution at any scale will be
affected by the discreteness of the sampling, as in PLT, even 
if this sampling is not uniform in space as in PLT. We underline,
however, that understanding of discreteness effects in the 
fully non-linear regime is completely lacking, and it is quite 
possible that other effects also come into play. For this reason 
alone it is important that carefully controlled extrapolations 
are systematically undertaken.

\item We have seen also in our numerical study that PLT provides an 
excellent fit to the evolved power at a wavenumber $k$, until the 
time that this wavenumber goes non-linear, and indeed describes 
the differences between simulations on different $preIC$ for even 
slightly longer. This extends its validity considerably beyond that
established by the numerical study in \cite{marcos_06}, which 
showed that it extended only to the time when the typical
relative displacement of nearest neighbour particles becomes 
of order the interparticle distance $\ell$. While this is what is expected
from a naive analysis of the validity of PLT --- requiring that
the linearization in the relative displacements of the force 
be valid --- it is not in fact surprising that its regime of
validity extends to the non-linearity of any given mode: to
obtain a good approximation to the evolution of the displacement
fields at a given scale the breakdown of PLT in describing 
the force due to particles at smaller scales is not relevant.
The regime of validity observed is what results if one assumes
that one needs the PLT linearization of the force on a particle
to be valid only for particles at separations of order 
$k^{-1}$ or larger. The fact that PLT does even better in
tracing the differences between evolution from identical
initial conditions sampled on different $preIC$ than in
following the full evolution on an individual $preIC$
indicates that the leading non-linear corrections have
discreteness corrections which are smaller than those 
in PLT at linear order. A full study of the extension of
PLT to next order (i.e. to second order in the Taylor expansion 
of the forces) should be able to explain this behaviour in  
detail. More generally, we underline that the success of PLT
in fitting analytically the quantities we have measured shows
that it can be a very useful instrument for controlling
analytically the results of numerical simulations. Indeed,
to our knowledge, the data in Fig.~\ref{variance_basic_PLT}
are by far the most stringent analytic controls which have
been placed on an $N$-body code, showing that GADGET can
trace correctly, to a precision of as great as one in a 
thousand, differences in the PS from slightly different initial
conditions. Thus, interestingly, the measurement of discreteness 
effects in simulations can be seen as a way of controlling the 
numerical accuracy of codes. Indeed, in cosmological $N$-body
simulation, a reasonable goal for the numerical accuracy of 
any code is that it should measure such effects, as it is
not of physical interest to do better than reach this level
of {\it systematic} error in the $N$-body method.

\item The numerical study presented was for the case of an initial
power law PS $P(k) \propto k^n$ with exponent $n=-2$. We have also 
analysed fully the cases $n=0$ and $n=2$, for which, starting from
similar amplitudes of fluctuations at the scale $\ell$ with the
same number of particles, the range of $a$ prior to that at which
the box goes non-linear is much greater. We have observed qualitatively
the same behaviours, and in particular, the monotonic growth of the
measured lower bounds on discreteness as a function of $a$. The method 
can of course be used for any initial conditions, and in particular
for the current standard $\Lambda$CDM model. The precise results for
this case will depend of course, in particular, on what physical scale 
is identified with $\ell$. The use of PLT as a ``calibrator'' in this
case would require its generalisation to this cosmology, which, as
noted in \cite{discreteness2_mjbm} should be straightforward.
We note that the recent study by 
\cite{romeo08} of this case reaches conclusions very consistent
with those found here (and those of Melott et al over a decade ago): 
using a wavelet analysis of a set of simulations a positive detection 
of discreteness errors is made for spatial scales smaller
than of order the interparticle spacing. It would be interesting
to combine in future studies these methods of numerical analysis 
with the analytical and numerical methods used here. 

\item We have considered only numerical simulations with fixed 
$\ell$ and $\varepsilon$, and our conclusions are valid of course
therefore only for this case (i.e. $PM$ or $P^3M$ simulations).  
One possibility, discussed by \cite{romeo08} in their 
conclusions, and briefly by Melott in a 
comment (\cite{melott_comment2008}, see also
the reply of \cite{romeo_replymelott_08}) on this paper, is that 
the intrinsic limitations on accuracy imposed by discreteness
might be addressed with numerical efficiency using AMR type codes,
with the mesh defining the resolution of the force 
(i.e. effectively $\varepsilon$) being adapted in higher density
regions so that the condition that the number of particles
per cell is always significantly larger than unity. Therefore, the
idea is, one would have always a local interparticle distance
smaller than the effective force resolution scale, thus satisfying
locally the condition apparently necessary to control discreteness 
effects ($\ell \ll \varepsilon$) while allowing a greater 
spatial resolution, in denser regions, than that fixed by
the interparticle distance $\ell$ of the initial grid.
While such an approach would be expected to reduce greatly
certain physical effects of discreteness ---  specifically any 
effects due to deviations from the mean field force 
acting on particles due to particles in their immediate neighbourhood 
(e.g. by two body collisions) --- our findings here lead to us
be very cautious about this conclusion about AMR: we have 
emphasized that the discreteness effects which we have been 
able to understand physically and quantify here (using 
the PLT formalism) are dynamical effects 
induced {\it at any scale} by the coupling to smaller
scales at which particle sampling noise becomes dominant.
When the smoothing scale is changed one does not undo
these effects, but simply modifies them by modifying
the evolution of the fluctuations at small scales.
Indeed in PLT, as has been shown explicitly 
in \cite{discreteness2_mjbm}, 
{\it increasing the force resolution scale $\varepsilon$ 
at fixed $\ell$ does not make the evolution of the $N$-body
system approximate better the physical limit}. Put simply,
the only way to reduce these kinds of effects of discreteness 
at any given scale is to increase the particle density.
Thus we do not consider that it is clear, in general, that
an AMR type code can give a more accurate result 
(i.e. closer to the physical model) than a standard $P^3M$ 
code (with $\varepsilon \ll \ell$) when both codes
use the same particle number. On the other hand,
we would expect that an AMR code may indeed do better
for many quantities than a simple PM code (with
an effective $\epsilon \geq \ell $) at the same 
particle density. In any case, as 
remarked by \cite{melott_comment2008}, careful tests of 
this or any alternative strategy to reduce discreteness 
effects should themselves of course be subjected to 
controlled tests for convergence.

\end{itemize}

\section*{acknowledgements}
We thank Andrea Gabrielli and Francesco Sylos Labini for many
discussions on various issues related to this work. MJ also thanks
O. Agertz, E. Bertschinger, B. Moore, S. Paduroui, A. Romeo,
R. Scoccimarro, U. Seljak and J. Stadel for useful discussions and/or
suggestions. We thank the ``Centro E. Fermi'' (Rome) for the use of
computing resources.

\bibliographystyle{mn2e}

\appendix

\section{Details of numerical integrations}
\label{app-gadget}

Listed in Tables~\ref{lr-runs} and \ref{hr-runs} are the parameter
values in GADGET2 we have used in our ``low resolution'' (LR) and 
``high resolution'' (HR) runs.

GADGET2 uses adaptative time-steps, which are chosen, for each particle, using the formula:
\be
\Delta t={\rm min}\left[\Delta t_{\rm max}, \left(\frac{2\eta\varepsilon}{|\ba|}\right)^{1/2}\right],
\ee
where $\Delta t_{\rm max}$={\tt MaxSizeTimestep}, $\eta=${\tt ErrTolIntAccuracy} and $|\ba|$ is the
acceleration of the particle in the previous time-step
(and $\varepsilon$ is the softening length). In our runs we
have chosen the parameter {\tt MaxSizeTimestep} sufficiently large
compared to {\tt ErrTolIntAccuracy} so that in practice only
the latter parameter is relevant. GADGET2 computes the gravitational 
force using a tree-mesh technique. At large scales, the force is 
computed with a $PM$ algorithm on a grid with $128^3$ cells.
At small scales, the force is computed using a tree algorithm. 
A node is opened (i.e. the force between a particle and a node of 
the tree is computed using the
monopole moment of the gravitational force) if
\be M l^2 > |\ba| r^4, \ee where $M$ is the mass of the node of
extension $l$ at a distance $r$ of the particle of which the force is
computed, $\ba$ the total acceleration in the last time-step and
$\alpha=${\tt ErrTolForceAcc}. We set the option {\tt
TypeOfOpeningCriterion}$=1$ and therefore the parameter {\tt
ErrTolTheta} is used only in the first force computation, and 
is therefore irrelevant.

Our LR simulations use the same range of parameters
usually used in the literature. For example, for
the VIRGO consortium, the parameter which controls the time accuracy
taken as {\tt ErrTolIntAccuracy}$=0.01$ and the one which controls the
calculation of the force as {\tt ErrTolForceAcc}$=0.005$ are
considered the fiducial ones \cite{VIRGO}, and are the ones which are
effectively used (e.g. \cite{stoehr_03,stoehr_06}), for a softening
length $\varepsilon\in[0.0002,0.02]\ell$. Other works
(e.g. \cite{crocce_06}) divide their runs like us in
``low-resolution'' and ``high-resolution'' ones. Their
``low-resolution'' runs have similar resolution than our
``low-resolution'' ones ({\tt ErrTolIntAccuracy}$=0.025$ and {\tt
ErrTolForceAcc}$=0.005$), but our ``high resolution'' ones take more
stringent parameters ({\tt ErrTolIntAccuracy}$=0.01$ and {\tt
ErrTolForceAcc}$=0.002$), for a value of the softening length  $\varepsilon\in[0.02,0.4]\ell$.

\begin{center}
\begin{table}

\begin{tabular}{@{}lr}
\hline
Accuracy of time integration \\
\hline
{\tt ErrTolIntAccuracy}   &   0.025 \\
{\tt MaxRMSDisplacementFac} &  0.2\\
{\tt CourantFac}      &       0.15 \\
{\tt MaxSizeTimestep}   &    0.025\\
{\tt MinSizeTimestep}   &    0.0\\
\hline
Tree algorithm and force accuracy\\
\hline
{\tt ErrTolTheta}     &       0.7\\
{\tt TypeOfOpeningCriterion}  &  1\\
{\tt ErrTolForceAcc}    &      0.005\\
{\tt TreeDomainUpdateFrequency}  &  0.1\\
\hline
Softening lenght\\
\hline
{\tt SofteningHalo}  &   0.00037202380952381\\
\hline
\end{tabular}
\caption{Numerical parameters for our ``low resolution'' runs.}
\label{lr-runs}
\end{table} 
\end{center}

\begin{table}
\begin{tabular}{@{}lr}
\hline
Accuracy of time integration \\
\hline
{\tt ErrTolIntAccuracy}   &   0.001 \\
{\tt MaxRMSDisplacementFac} &  0.2\\
{\tt CourantFac}      &       0.15 \\
{\tt MaxSizeTimestep}   &    0.025\\
{\tt MinSizeTimestep}   &    0.0\\
\hline
Tree algorithm and force accuracy\\
\hline
{\tt ErrTolTheta}     &       0.7\\
{\tt TypeOfOpeningCriterion}  &  1\\
{\tt ErrTolForceAcc}    &      0.0001\\
{\tt TreeDomainUpdateFrequency}  &  0.1\\
\hline
Softening length\\
\hline
{\tt SofteningHalo}  &   0.00037202380952381\\
\hline
\end{tabular}
\caption{Numerical parameters for our ``high resolution'' runs.}
\label{hr-runs}
\end{table}

\section{Definition and estimation of two point statistics}
\label{app-estimation}

\subsection{Real space}
The reduced two-point correlation function $\tilde\xi(\br)$ is 
defined (see e.g. \cite{book}) as
\be
\label{xi-full}
\tilde\xi(\br)=\langle\delta(\br+\bx)\delta(\bx)\rangle,
\ee
where $\langle...\rangle$ means ensemble average over all the possible
realizations of the system. For particle distributions $\tilde\xi(\br)$ 
has a singularity at $\br=\mathbf 0$, and it is therefore convenient 
to divide it as 
\be
\tilde\xi(\br)=\frac{1}{n_0}\delta_D(\br)+\xi(\br).
\ee
where $n_0$ is the mean number density.
The quantity we give results for in the paper, and denote by
$\xi(r)$, is a direct real space angle-averaged
estimator of $\xi(\br)$:
\be
\xi(r)+1=\frac{1}{n_0V(r,\delta r)N_c}\sum_{i=1}^{N_c} N_i(r),
\ee
where $N_i(r)$ is the number of particles in the spherical shell of
radii $r,\,r+\delta r$, volume $V(r,\delta r)$, centred on the
$i^{\mathrm{th}}$ particle of a subset of $N_c\le N$ particles 
randomly chosen from the $N$ particles of the system.

\subsection{Reciprocal space}

Because we consider distributions with periodic boundary conditions we
can write the density contrast as a Fourier series:
\be
\delta(\bx)=\frac{1}{V}\sum_{\bk} \exp(i\bk\cdot\bx)\tilde\delta(\bk)
\ee
with $\bk\in \{(2\pi/L)\bn|\bn\in\mathbb Z^3\}$. The coefficients
$\tilde\delta(\bk)$ are given by
\be
\tilde\delta(\bk)=\int_V \delta(\bx)\exp(-i\bk\cdot\bx)d^3x.
\ee
The PS is defined as 
\be
P(\bk)=\frac{1}{V} \langle |\tilde\delta(\bk)|^2\rangle,
\ee
which we estimate with  
\be 
P(k) = \frac{1}{N(k)} \sum_{k\le|\bk'|\le k+\delta k}
|\tilde\delta(\bk')|^2 
\ee
where $N(k)$ is the number of vectors $\bk'$ considered in the sum. To
speed up the computation we perform a sampling at larger $k$ on the 
vectors $\bk'$. We have checked that our results are robust to this
choice. 

\end{document}